# Spontaneous exciton dissociation enables spin state interconversion in delayed fluorescence organic semiconductors

*Alexander J. Gillett[1]\*, Claire Tonnelé[2], Giacomo Londi[3], Gaetano Ricci[4], Manon Catherin[5], Darcy M. L. Unson[1], David Casanova[2], Frédéric Castet[6], Yoann Olivier[4], Weimin M. Chen[7], Elena Zaborova[5], Emrys W. Evans[1,8], Bluebell H. Drummond[1], Patrick J. Conaghan[1,9], Lin-Song Cui[1], Neil C. Greenham[1], Yuttapoom Puttisong[7]\*, Frédéric Fages[5]\*, David Beljonne[3]\* and Richard H. Friend[1]\*.*

[1]Cavendish Laboratory, University of Cambridge, JJ Thomson Avenue, Cambridge, CB3 0HE, UK.

[2]Donostia International Physics Centre (DIPC), Donostia, Euskadi, Spain.

[3]Laboratory for Chemistry of Novel Materials, Université de Mons, Place du Parc 20, 7000 Mons, Belgium.

[4]Unité de Chimie Physique Théorique et Structurale & Laboratoire de Physique du Solide, Namur Institute of Structured Matter, Université de Namur, B-5000 Namur, Belgium.

[5]Aix Marseille Univ, CNRS, CINaM UMR 7325, AMUtech, Campus de Luminy, 13288 Marseille, France.

[6]Institut des Sciences Moléculaires, Université de Bordeaux, 33405 Talence, France.

[7]Department of Physics, Chemistry and Biology (IFM) Linköping University, Linköping, Sweden.

[8]Department of Chemistry, Swansea University, Singleton Park, Swansea, SA2 8PP, UK.

[9]ARC Centre of Excellence in Exciton Science, School of Chemistry, University of Sydney, NSW 2006, Australia.

*Corresponding authors: Alexander J. Gillett: E-mail: ajg216@cam.ac.uk; Yuttapoom Puttisong: E-mail: yuttapoom.puttisong@liu.se; Frédéric Fages: E-mail: frederic.fages@univ-amu.fr; David Beljonne: E-mail: david.beljonne@umons.ac.be; Richard H. Friend: E-mail: rhf10@cam.ac.uk.

**Engineering a low singlet-triplet energy gap ($\Delta E_{ST}$) is necessary for efficient reverse intersystem crossing (rISC) in delayed fluorescence (DF) organic semiconductors, but results in a small radiative rate that limits performance in LEDs. Here, we study a model DF material, BF2, that exhibits a strong optical absorption (absorption coefficient =$3.8\times10^5$ cm$^{-1}$) and a relatively large $\Delta E_{ST}$ of 0.2 eV. In isolated BF2 molecules, intramolecular rISC is slow (260 μs), but in aggregated films, BF2 generates intermolecular CT (inter-CT) states on picosecond timescales. In contrast to the microsecond intramolecular rISC that is promoted by spin-orbit interactions in most isolated DF molecules, photoluminescence-detected magnetic resonance shows that these inter-CT states undergo rISC mediated by hyperfine interactions on a ~24 ns timescale and have an average electron-hole separation of ≥1.5 nm. Transfer back to the emissive singlet exciton then enables efficient DF and LED operation. Thus, access to these inter-CT states resolves the conflicting requirements of fast radiative emission and low $\Delta E_{ST}$.**

In organic light-emitting diodes (OLEDs), the spin statistical recombination of injected electrons and holes leads to the formation of a 3:1 ratio of spin-triplet to spin-singlet excitons[1]. Since triplet excitons in organic emitters are optically dark, the internal quantum efficiency (IQE) of an OLED is restricted to 25% if no further steps are taken to utilise triplet excitons for light emission. One approach has been to develop organic semiconductors that can use a reverse intersystem crossing (rISC) process to convert these dark triplet states into bright singlet excitons, assisted by ambient thermal energy[2]. This class of materials, known as delayed fluorescence (DF) emitters, enable OLEDs with IQEs approaching 100%[3]. To obtain a singlet-triplet energy gap ($\Delta E_{ST}$) small enough for an efficient rISC process (typically <0.1 eV), it is necessary to weaken the electron exchange interaction. To achieve this, DF emitters generally comprise of spatially separated electron donor (D) and acceptor (A) moieties, resulting in excitons with intramolecular charge transfer (intra-CT) character[4]. Consequently, rISC can proceed through spin-orbit interactions, often also involving triplet states localised on either the D or A ($^3$LE) that are in close energetic proximity to the intra-CT excitations[5–8]. However, DF emitters with intra-CT-type excitons suffer from a low oscillator strength and can only operate efficiently if competing non-radiative processes are highly suppressed[9]. As a result, it is challenging to create DF materials that achieve the necessary balance between the conflicting requirements of a small $\Delta E_{ST}$ for effective rISC and a large oscillator strength for rapid and efficient emission. Therefore, it is desirable to design new classes of organic semiconductors that can combine the strong absorption and emission of localised excitons with the weak exchange interaction of electronic excitations with spatially separated electrons and holes.

We report here on observations of spontaneous exciton dissociation into intermolecular CT (inter-CT) states in neat films of DF emitters and observe ISC processes taking place on these loosely bound excitations. We study four DF emitters (Fig. 1), selected to provide a range of exchange energies and intra-CT transition oscillator strengths, from large (BF2) to small (TXO-TPA), see Table S1. We begin by focussing on BF2, which supports efficient OLED operation in the near infra-red region[10]. BF2 comprises of two triphenylamine D units and a boron difluoride-based A core, forming a DAD structural motif. Consequently, the excitons of BF2 possess a partial intra-CT character, though the electron-hole overlap is still sufficient to retain a high absorption coefficient of $3.8\times10^5$ cm$^{-1}$ (Fig. S1) and a moderate $\Delta E_{ST}$ of 0.2 eV (Fig. S3)[10].

To investigate the excited state dynamics of BF2, we have performed ultrafast transient absorption (TA) spectroscopy on a neat BF2 film (Figs. 2a, 2b). In the TA spectra at 0.2-0.3 ps, we observe two positive bands centred at 630 and 740 nm and a negative photo-induced absorption (PIA) feature peaking at 1000 nm. We assign the positive bands at 630 and 740 nm to the ground state bleach (GSB) and stimulated emission (SE) of BF2, respectively (see Fig. S4 for steady-state absorption and photoluminescence). The PIA at 1000 nm is present immediately after excitation and is due to the spin-singlet intra-CT exciton (intra-$^1$CT) of BF2. The SE and intra-$^1$CT PIA decay on picosecond timescales, demonstrating that the bright intra-$^1$CT excitons are being lost, but there is very little decrease in the GSB region on these same timescales. In the kinetic taken from the high-energy GSB edge (570-630 nm) to avoid overlap with the decaying SE, the GSB intensity remains unchanged at 10 ps, whilst the SE has decreased to 30% of peak intensity. Thus, the process quenching intra-$^1$CT excitons on these timescales is efficient but does not result in decay back to the ground state. As the forward ISC rates for organic DF emitters are typically ~$10^8$-$10^7$ s$^{-1}$, we can rule out ISC to the triplet manifold as the reason for the loss of the intra-$^1$CT on picosecond timescales[6,11–13]. We notice the formation of a new PIA band at 950 nm; the growth of this species mirrors the loss of the SE, signifying it is being formed from intra-$^1$CT excitons. The new species with a PIA at 950 nm is the hole polaron on BF2, as seen in the TA of a BF2:PC$_{60}$BM 1:1 blend film (Fig. S5), where electron transfer is induced from BF2 to PC$_{60}$BM. We therefore confirm that intermolecular charge transfer takes place in neat BF2 on picosecond timescales. This is consistent with our observations in organic solar cell (OSC) devices fabricated with a neat BF2 active layer (Fig. S6a) where we obtain a moderate short-circuit current density of 0.43 mA/cm$^2$ under AM1.5G 1 sun illumination (Table S2), and a peak photovoltaic external quantum efficiency (EQE$_{PV}$) of 2.1% at 555 nm under short-circuit conditions (Fig. S6c). We find that the EQE$_{PV}$ obtained is limited by the relatively short excited state lifetime in neat BF2 (Fig. S10h), where only a small fraction of the initially photogenerated excitations remain by 1 µs, a timescale that is typical for charge extraction in OSCs[14].

We also find that there can be efficient regeneration of the emissive intra-$^1$CT excitons in BF2 from the spatially separated charge carriers at longer times. This is most easily observed when BF2 is diluted in the non-interacting OLED host material CBP (4,4′-Bis(*N*-carbazolyl)-1,1′-biphenyl), which allows for control over the intermolecular interactions and kinetics. In Figs. 2c, 2d, we present the nanosecond-microsecond TA of BF2 diluted at 10 wt% in CBP. Here, we observe the delayed regrowth of the SE band between 650-725 nm over microsecond timescales, matching closely the blue-shifted SE of dilute BF2 in the corresponding ultrafast TA measurements (Fig. S8c). In addition, we show that the emissive species present on the timescales of SE reformation has the same photoluminescence (PL) spectrum as the intra-$^1$CT of BF2 (Fig. S12b), confirming that these states are being regenerated. These findings indicate that the electronic excited states of BF2 can readily interconvert between localised intramolecular excitations, loosely bound singlet inter-CT (inter-$^1$CT) states and even free charge carriers. As discussed later, the efficient interconversion between the intra-$^1$CT and inter-$^1$CT can be rationalised by electronic structure calculations that suggest strong electronic coupling and a small energy gap between them (Table S8). A full study and discussion of the dilute BF2 blends is reported in the SI.

We consider that this spontaneous exciton dissociation plays a critical role in OLEDs employing BF2 in the emissive layer. The thermally assisted intramolecular rISC process that is used to convert dark triplet excitons into bright singlet states in organic DF emitters is typically promoted by spin-orbit interactions, often involving intermediate triplet states[5–8]. In-line with previous reports[10], our calculations reveal the presence of an intra-$^3$CT (Table S5) in between the lowest-energy intra-$^1$CT and intra-$^3$CT excitations in the BF2 monomer, leading to an endergonic rISC that can rationalise the very long the "delayed" lifetime of isolated BF2 molecules (Fig. S10b). However, the formation of inter-CT states with spatially separated electrons and holes in aggregated BF2 will greatly reduce the strength of the electron exchange interaction, bringing in near-resonance the singlet and triplet states, as shown by electronic structure calculations (Fig. S29). Thus, it becomes possible for the hyperfine interaction (HFI) to mediate the interconversion of loosely bound inter-CT states with singlet and triplet character in aggregated BF2. Typical HFI-ISC rates are on the order of ~$10^8$–$10^6$ s$^{-1}$, enabling efficient spin state interconversion via a periodic oscillation between the $M_S$=0 singlet ($^1$CT$_0$) and the $M_S$=-1,0,+1 triplet sublevels ($^3$CT$_-$, $^3$CT$_0$, $^3$CT$_+$) of an inter-CT state (Fig. 3a), otherwise known as a spin-correlated radical pair[15–18].

We utilise PL-detected magnetic resonance (PLDMR), performed at 293 K to ensure relevance to OLED device operation, to explore whether the HFI mediates the ISC processes in BF2. In the PLDMR of a neat BF2 film (Fig. 3d, full field range data in Fig. S13), we observe a narrow negative signal at ~333 mT, which is attributed to inter-CT states[19]. To understand the relevance of this feature, we note that under an external magnetic field (**B$_0$** ~333 mT), the Zeeman interaction decouples HFI-induced transitions between the inter-$^1$CT$_0$ and either inter-$^3$CT$_+$ or inter-$^3$CT$_-$[20]. Consequently, microwave "spin-pumping" from the inter-$^3$CT$_0$ (formed via HFI-ISC from inter-$^1$CT$_0$) to inter-$^3$CT$_+$ and inter-$^3$CT$_-$ reduces the number of inter-$^3$CT$_0$ states that can couple to the singlet manifold, lowering the PL yield from the intra-$^1$CT (Fig. 3b). Therefore, the appearance of this feature provides strong evidence that: (i) HFI-ISC processes are occurring in loosely bound inter-CT states with a very small exchange energy (on the order of neV)[21]; and (ii) the nominally-dark inter-CT states can readily interconvert with the bright intra-$^1$CT excitons. From the full width at half maximum (FWHM) of the negative signal (1.5 mT), when the magnetic field of the applied microwave radiation (**B$_1$**) is 0.12 mT, we estimate an upper bound of 0.75 mT (21 MHz) for the zero-field splitting **D**-parameter, which is related to the average inter-spin distance[22]. **D** ≤0.75 mT corresponds to a lower bound of ≥1.5 nm for the inter-CT state electron-hole radius ($r_{e-h}$) in neat BF2 (full calculation details in the SI)[23,24], far larger than expected for an intramolecular exciton[22]. We also observe the presence of this same negative signal in BF2 diluted at 10 wt% in CBP (Fig. S14), confirming that HFI-ISC can also take place in the dilute films that are typically employed in OLED devices. Interestingly, the signal is stronger in the 10 wt% blend than the neat BF2 film, which we attribute to the greater number of excited states living long enough to undergo HFI-ISC processes in the 10 wt% film (Fig. S10).

Additional information about the HFI-ISC processes occurring in inter-CT states can be obtained by investigating the response of the PLDMR signal to increased microwave intensities, where **B$_1$** is perpendicular to **B$_0$**. When **B$_1$** exceeds the hyperfine field (**B$_{HF}$**, typically ~1 mT in organic radicals[16,25–27]), the rate of the HFI-mediated inter-$^1$CT$_0$/inter-$^3$CT$_0$ transitions is reduced as the radical spins begin to precess around **B$_1$** with the same frequency, negating their HFI-induced spin precession frequency difference[21,28,29]. In the absence of an effective spin-mixing process, the spin population now becomes trapped in the initially generated inter-$^1$CT$_0$, leading to "spin-locking" (Fig. 3c). Spin-locking results in a reduction of the number of inter-$^3$CT$_0$ states formed, so the spin-pumping in the triplet manifold becomes less efficient. Consequently, the PL yield begins to rise again under a large **B$_1$**, manifesting as the formation of a broader resonance with a characteristic "W"-shaped peak in the PLDMR[21]. We observe the formation of this "W"-shaped signal at a **B$_1$** of 0.38 mT in the neat BF2 film. The effective $B_{HF}$ can be estimated from twice $B_1$ at the onset of spin-locking, and we obtain an effective $B_{HF}$ of ~0.76 mT; this corresponds to a HFI-induced periodic singlet-triplet interconversion time of ~24 ns in the inter-CT states of BF2 (full calculation details in the SI). Thus, due to the large $\Delta E_{ST}$ and slow rISC process of isolated BF2 molecules, we propose that HFI-ISC between inter-CT states is the dominant triplet to singlet interconversion process in OLED devices fabricated from BF2. Furthermore, the singlet-triplet interconversion time of ~24 ns and $r_{e-h}$ of ≥1.5 nm in BF2 suggest that the HFI-ISC processes observed may be in the coherent regime[30].

We have explored three further model DF materials with smaller exchange energies and packing motifs that may or may not allow for the formation of inter-CT states: APDC-DTPA, TXO-TPA, and 4CzIPN[2,31,32]. In the first two materials, we observe a strong negative signal at ~333 mT in both the neat and dilute (10 wt%) films, with spin-locking also present in the neat films (Figs. S15-S18). This finding suggests that the dissociation of excitons is present in well-studied DF emitters in the solid state, with HFI-ISC between inter-CT states contributing to the triplet to singlet conversion in these OLED devices. We believe that this can be attributed to the close intermolecular interactions facilitated by the tendency of many DF materials to aggregate, even when diluted in a non-interacting host at a low wt%[10,33,34]. Additionally, many DF materials, including APDC-DTPA[31], also demonstrate excellent performance in non-doped OLEDs[31,35–42], with a growing number of DF emitters also exhibiting aggregation-induced DF[40–46]. In aggregation-induced DF, the DF properties are only present in environments with significant intermolecular contact, such as neat films; therefore, we propose that HFI-ISC between inter-CT states, which requires close intermolecular contacts between the emitters, may be a significant rISC mechanism in these systems. In contrast, we observe a very weak inter-CT response in a neat 4CzIPN film and no spin-locking (Fig. S19), demonstrating that intermolecular HFI-ISC is not a major singlet-triplet mixing process in this material. This is consistent with the low EQE$_{PV}$ (0.53%) of an OSC device fabricated with a neat 4CzIPN active layer (Fig. S6c), which indicates that the excitons of 4CzIPN do not dissociate as readily as those in BF2. This is confirmed by the TA of a neat 4CzIPN film (Fig. S20), where no evidence for intermolecular charge transfer is observed, and is supported by modelling, see below. As a result, the HFI-ISC channel we observe in the other DF emitters is largely unavailable to 4CzIPN and the ISC processes must proceed via spin-orbit interactions as generally considered in the literature[6,47].

We have carried out computational studies of the lowest energy electronic excitations of the representative cases of BF2 and 4CzIPN, which exhibit a contrasting ability to form inter-CT states. Our theoretical approach combines screened tuned range-separated hybrid functional time-dependent density functional theory (TDDFT) calculations with a (Boys) diabatization scheme that decomposes the electronic eigenstates of the system into a set of pure (diabatic) intra- and inter-CT electronic configurations, see SI for details. We begin by examining BF2 molecules embedded in a dielectric continuum. In line with earlier reports[10], our calculations show that the lowest intramolecular singlet excitation involves in-phase mixing of the D$^+$(AD)$^-$ and (DA)$^-$D$^+$ zwitterionic configurations and carries a very large oscillator strength (Table S3). The lowest-energy intramolecular triplet excitation (intra-$^3$CT) has a greater weighting on the central A unit than the intra-$^1$CT, leading to a relatively large $\Delta E_{ST}$ =0.28 eV (Table S5), in good agreement with our experimental observations. To explore the inter-CT states in BF2, we first note that X-ray diffraction measurements on single crystals of the BF2 parent molecule show the presence of strongly interacting dimers (with an intermolecular separation of ~0.35 nm)[48], driven by electrostatic interactions between the large, short-axis polarized, ground-state molecular dipoles. Because these dimers likely experience different environments in the disordered films, we have performed calculations for a range of intermolecular distances ($d_\perp$) from the crystal value of 0.35 nm to 0.7 nm (Fig. S25-S26).

In Fig. 4a, we summarise the nature of the lowest energy intra- and inter-molecular electronic excitations in BF2, and a full schematic of all the excitations present and their dependence on $d_\perp$ is included in the SI. The most striking result is the appearance of inter-$^1$CT states at a lower energy than the intra-$^1$CT exciton. This energy ordering of the electronic excitations in BF2 can be rationalised through the analysis of the excited-state wavefunctions, namely their corresponding electron-hole radii. For a broad range of intermolecular distances between 0.35-0.5 nm (Fig. S27), the averaged electron-hole distance in the intra-$^1$CT exciton is larger than in the inter-$^1$CT states. This result supports the view that intermolecular charge transfer is thermodynamically favourable in closely interacting BF2 molecular dimers, in line with our experimental observations. Furthermore, we find that the dependence of the inter-$^1$CT state energy with $d_\perp$ is surprisingly weak; a simple fit with a $1/(\varepsilon d_\perp)$ Coulomb law of the inter-$^1$CT state energy would lead an unphysical value of $\varepsilon$ =7.5. This is because the effective electron-hole separation in the inter-$^1$CT state exceeds the intermolecular distance $d_\perp$ owing to a large intramolecular component (at $d_\perp$ =0.35 nm, $r_{e-h}$, is already ~0.6 nm). Thus, the intra-CT character of the exciton effectively crops the short-range part of the Coulomb potential for the inter-$^1$CT states. We expect this to favour the formation of loosely bound inter-CT states, or even free charge carriers, provided continuous pathways that enable long range charge separation are present. In contrast to the relatively large $\Delta E_{ST}$ values calculated for the intra-CT excitons, the $\Delta E_{ST}$ of the inter-CT states is vanishingly small due to the minimal electron-hole overlap. The energy ordering predicted in the triplet manifold is now swapped compared to the singlet, with the lowest energy triplet adiabatic excitation the intra-

$^3$CT (Table S10). This is because the intra-$^3$CT, with an increased contribution from the central A moiety, has an average $r_{e-h}$ of only 0.4 nm and is consequently more strongly bound than the inter-$^3$CT states.

The calculated inter-$^1$CT states in the crystal dimer have a $r_{e-h}$ of ~0.6 nm, which is less than half the value (~1.5 nm) estimated from the FWHM analysis of the PLDMR signal. Though Equation S1 provides only an approximate value for $r_{e-h}$, we conclude that spin conversion occurs either in dimers with intermolecular distances exceeding the equilibrium value in the crystal, or through non-nearest-neighbour interactions in larger aggregates. To investigate the latter hypothesis, we have conducted similar TDDFT calculations on a BF2 tetramer. The results reported in Fig. S29 show that a broad manifold of inter-$^1$CT excitations develops *below* the intra-$^1$CT excitations. Namely, the second-nearest-neighbour inter-$^1$CT states with $r_{e-h}$ ~1 nm are virtually degenerate with their inter-$^3$CT counterparts and in close energy resonance with the intra-$^1$CT states, thus enabling the interconversion between the intra-CT and second-nearest neighbour inter-CT manifolds. We anticipate that more distant (third-, fourth-, … neighbours) inter-$^1$CT pairs will be thermally accessible in larger clusters, on a par with the experimentally measured $r_{e-h}$.

Most importantly, this analysis is consistent with the comparison between the calculated and measured HFI-mediated singlet-triplet conversion rates. We have used DFT calculations to explore the HFI couplings in the BF2 inter-CT states, retaining both the isotropic Fermi contact and the dipolar tensor terms. The nucleus-dependent hyperfine magnetic fields were weighted according to the atomic contributions to the attachment (for anions) and detachment (for cations) densities and arranged to depict a (full) inter-CT excitation, see SI. The calculated HFI couplings, in the range of a few mT (~0.1-0.2 µeV), are in line with expected values for organics and consistent with our experimental findings. We next follow Fay and Manolopoulos to derive rate equations for the HFI-mediated spin conversion from second-order perturbation theory[49,50]. In this limit, the spin conversion rate scales as the square of the HFI coupling and is sensitive to the relative magnitude of the dephasing rate versus energy splitting between the initial and final states (see SI for further information on the model). We use the distance-dependent inter-CT state exchange energy from TDDFT calculations in combination with the computed hyperfine couplings to predict HFI-ISC rates as a function of dephasing in Fig. 3e. The HFI-ISC rate rapidly increases with increasing $r_{e-h}$ because of the lower singlet-triplet exchange energy and, assuming a reasonable range of dephasing times (estimated experimentally to be 40 ns to 1 µs, see SI for details; represented by the dashed vertical white lines), it reaches an order of magnitude approaching the experimental result of $10^7$ s$^{-1}$ at electron-hole pair separations of ≥1 nm. We stress that this occurs in a regime at the edge of validity of the perturbative model (Fig. S35), suggesting the possibility for a coherent process.

In contrast, similar calculations performed on a representative nearest-neighbour 4CzIPN pair (extracted from the single crystal structure[33], see SI for details) yield degenerate singlet and triplet inter-CT states >0.12 eV above the lowest-lying intramolecular excitations (Fig. 4b); a relative energetic ordering opposite to that predicted in BF2. This is expected based on the near spherical symmetry of the 4CzIPN molecules that isolates the central A group of one molecule from the D moieties of neighbouring molecules. This raises the effective intermolecular $r_{e-h}$ in the closest molecular dimer to 0.76 nm: a value significantly larger than the corresponding $r_{e-h}$ of 0.43 nm in the intramolecular exciton. As a result, the intramolecular excitons of 4CzIPN are not expected to readily dissociate into inter-CT states, matching our experimental observations in this material.

From the combination of our experimental and computational results, we can rationalise the spin interconversion processes in BF2 occurring after photoexcitation (Fig. 4c). In the context of OLED operation involving electrical excitation[1], the 25% loosely bound and long-lived dark inter-$^1$CT states initially created can act as a reservoir for the formation of the slightly higher-lying and emissive intra-$^1$CT exciton. The 75% of injected carriers that form inter-$^3$CT states branch between: (i) conversion into inter-$^1$CT via HFI-ISC; (ii) dissociation into free charge carriers that are then recycled in the recombination loop; (iii) recombination into lower-lying intra-$^3$CT excitons. We find little experimental evidence for the latter pathway; in the PLDMR of neat BF2 (Fig. S13) we see a small intra-$^3$CT response, and this is absent in BF2 at 10 wt% in CBP (Fig. S14). Our calculations suggest that intra-$^3$CT/inter-$^3$CT electronic couplings are as strong as the ones obtained in the spin-singlet manifold (Tables S9 and S13), but the intra-$^3$CT/inter-$^3$CT energy gap is much larger (~0.29 eV). Therefore, we speculate that near degeneracy of the intra-$^1$CT and inter-$^1$CT facilitates their interconversion, while this is hindered in the triplet manifold due to the relative stabilization of the intra-$^3$CT. As a result, rISC via HFI or the scrambling of the inter-CT state spins through a few cycles of inter-CT state dissociation and recombination will be favoured. Furthermore, the presence of an intramolecular rISC pathway in BF2, albeit slow, means that recombination to the intra-$^3$CT is also not necessarily a terminal event.

In summary, we have shown that it is possible to reconcile the conflicting requirements of a very strong optical absorption and emission with efficient spin state interconversion in aggregated DF molecules. We achieve this by utilising high oscillator strength intramolecular excitations for interaction with photons, and exploiting loosely bound inter-CT states, with a vanishingly small $\mathit{\Delta E_{ST}}$, for rapid ISC processes mediated by HFI. This enables us to demonstrate a mechanism of OLED operation that combines the desirable optical attributes of conventional fluorescent emitters and the spin manipulation abilities of DF organic semiconductors. Furthermore, the generation of loosely bound charge transfer states, and even free charge carriers, in neat BF2 films without the driving energy provided by offset molecular orbitals in D:A blends is of relevance to OSCs. Thus, our study of BF2 provides an insight into the mechanism for the ultra-low driving energy charge generation seen in non-fullerene acceptor OSCs[51–55], and will also facilitate the development of OSCs that do not require separate electron donor and acceptor materials for efficient free charge carrier generation.

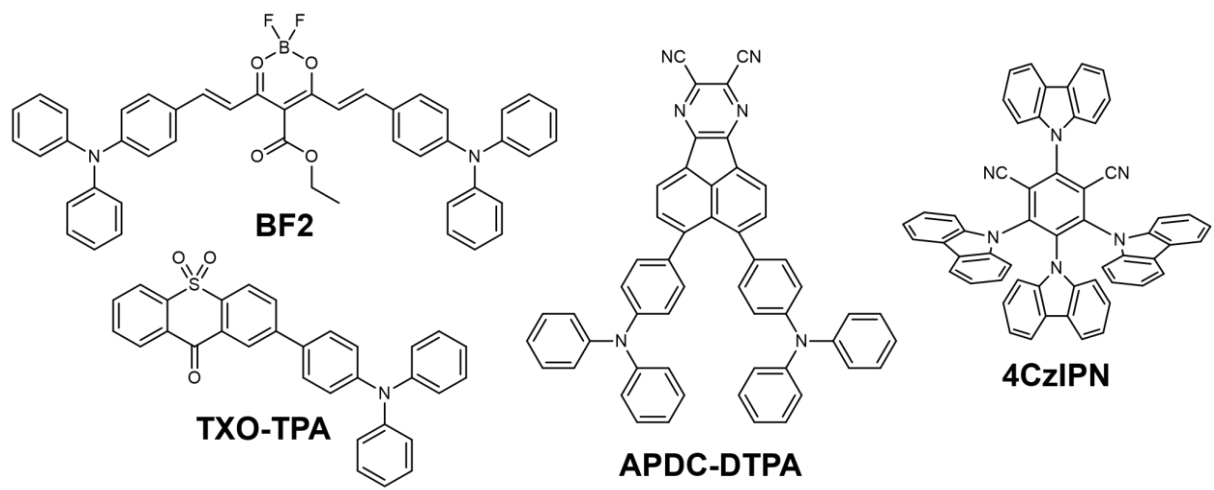

**Figure 1: The molecular structures of the four DF materials investigated in this study.**

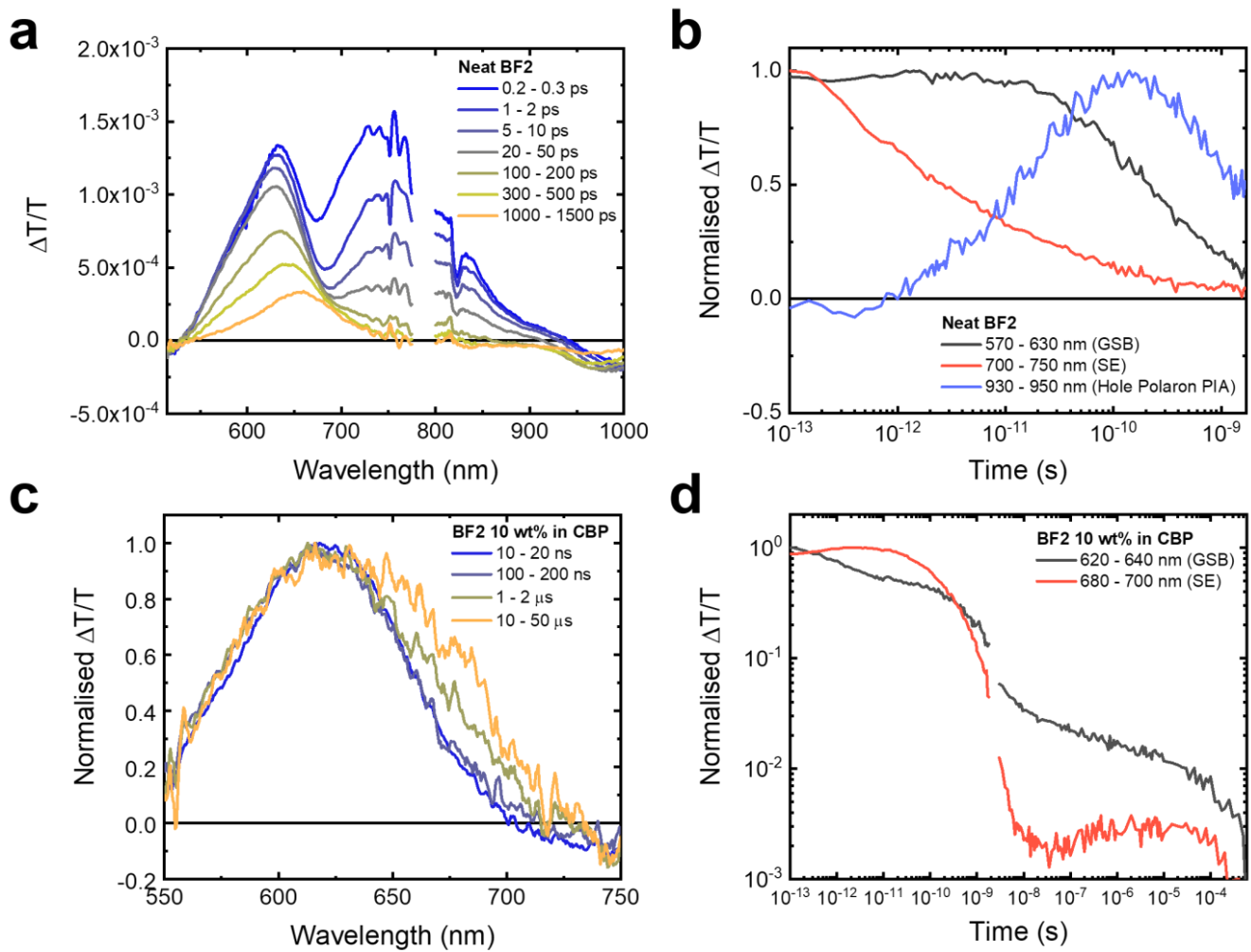

**Figure 2: Transient absorption studies of BF2 films. (a)** The transient absorption spectra of a neat BF2 film, excited at 610 nm with a fluence of 7.0 μJ cm$^{-2}$. **(b)** The transient absorption kinetics of the neat BF2 film, taken from the GSB (570-630 nm), SE (700-750 nm) and hole polaron (930-950 nm) regions. **(c)** The normalised nanosecond TA spectra of BF2 doped at 10 wt% in CBP, excited at 532 nm with a fluence of 15.7 μJ cm$^{-2}$. **(d)** The combined kinetics of the ultrafast and nanosecond-microsecond TA of BF2 doped at 10 wt% in CBP, taken from the GSB (620-640 nm) and SE (680-700 nm) regions. 532 nm excitation was used for both measurements, with a fluence of 15.7 μJ cm$^{-2}$ for the ultrafast TA and 346 μJ cm$^{-2}$ for the nanosecond-microsecond TA. The kinetics were scaled relative to each other by their respective fluences.

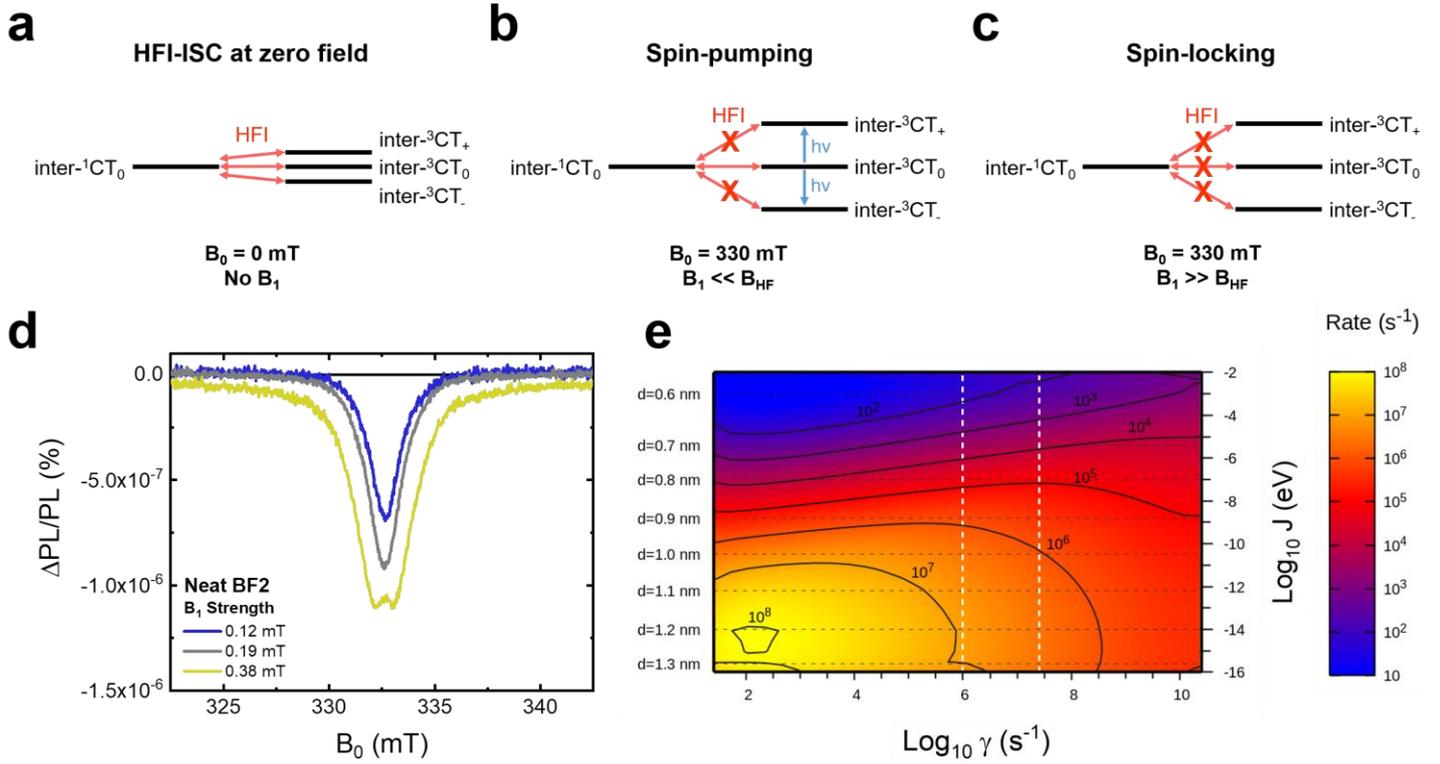

**Figure 3: Magnetic resonance studies of hyperfine couplings in a BF2 film.** (a) A schematic demonstrating the ability of the HFI to mediate spin-mixing processes between the $M_S$=0 inter-$^1$CT and the $M_S$=-1,0,+1 inter-$^3$CT sublevels in the absence of an external magnetic field. (b) Under an applied external magnetic field $B_0$, the Zeeman interaction energetically forbids HFI-induced transitions between the inter-$^3$CT$_+$ and inter-$^3$CT$_-$ and the inter-$^1$CT$_0$. Spin-pumping $M_S$=±1 transitions can then occur between the inter-$^3$CT$_0$ and the inter-$^3$CT$_+$ and inter-$^3$CT$_-$ sublevels; this reduces the inter-$^3$CT population that can ultimately couple to the emissive intra-$^1$CT manifold via the inter-$^1$CT state, resulting in a decreased PL yield from the sample and the sharp, negative PLDMR signal seen at ~333 mT. (c) When the magnetic field of the applied microwaves $B_1$ is perpendicular to $B_0$ and becomes larger than $B_{HF}$, spin-locking occurs. This reduces the rate of the HFI-mediated inter-$^1$CT$_0$ to inter-$^3$CT$_0$, transitions, "locking" the inter-CT state spin-population in the initially generated inter-$^1$CT$_0$. Spin-locking manifests as the formation of a characteristic "W"-shaped peak in the PLDMR as $B_1$ is increased. (d) The PLDMR response of a neat BF2 film at 293 K with 405 nm excitation (30 mW). (e) The calculated HFI-ISC rate in a model BF2 dimer as a function of the dephasing rate, $\gamma$, and the electron exchange energy, J. The dashed white lines represent the experimental dephasing times in BF2 (upper bound =40 ns, lower bound =1 μs), as estimated from the TA measurements of BF2 at 10 wt% in CBP film (see SI for details).

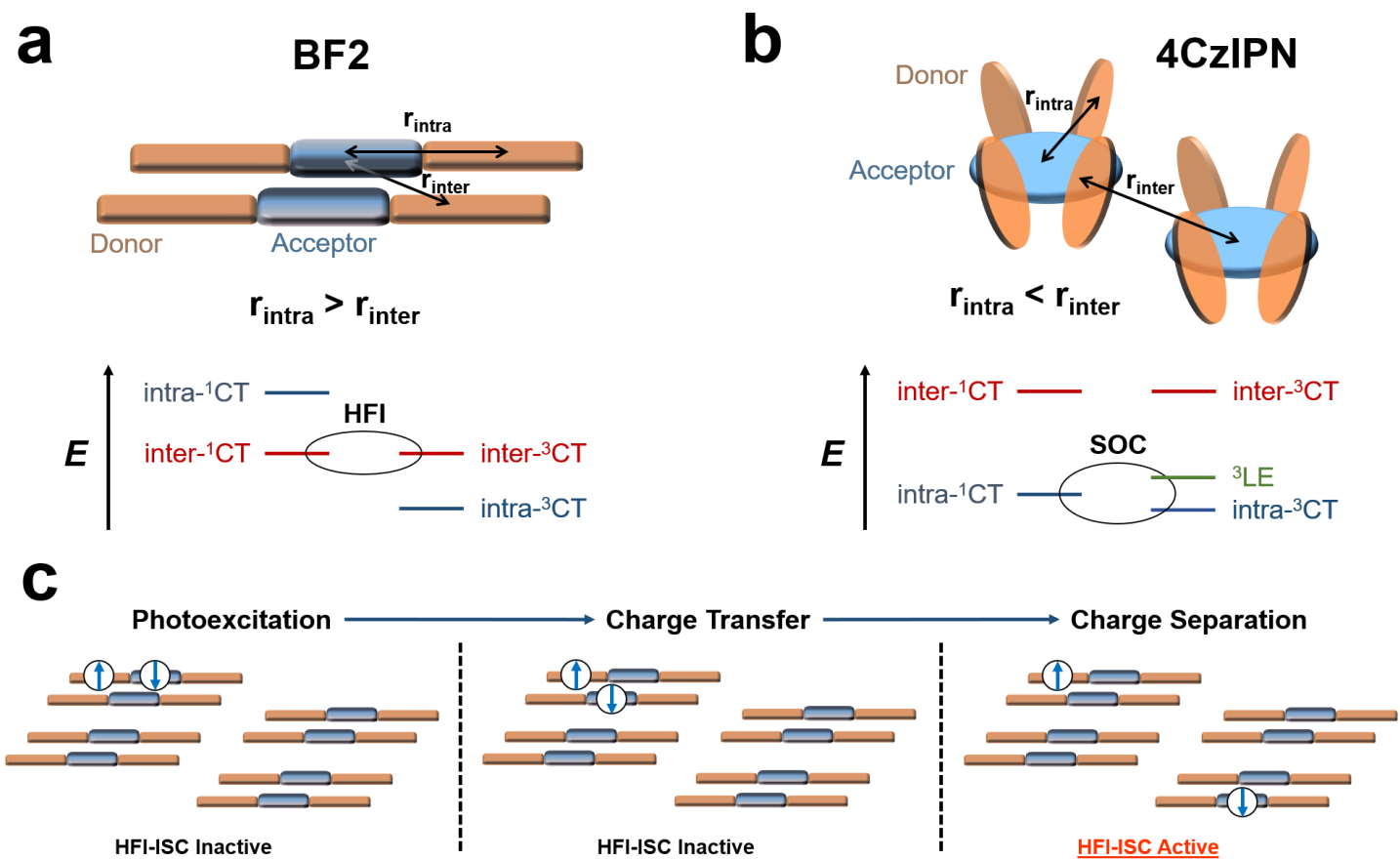

**Figure 4: The role of intermolecular excitations in delayed fluorescence.** (a) A schematic of a representative BF2 dimer and the most relevant electronic excitations (state energies not to scale), demonstrating how the electron-hole separation in the inter-CT state can be less than the intra-CT exciton. This renders the inter-CT states more stable than the intra-$^1$CT, meaning intermolecular charge transfer following photoexcitation is thermodynamically favourable. The loosely-bound inter-CT states can then undergo HFI-ISC processes, enabling efficient spin mixing in BF2, followed by recombination to the intra-$^1$CT for light emission. (b) A schematic of a representative 4CzIPN dimer the most relevant electronic excitations (state energies not to scale), where the electron-hole separation in the inter-CT state is significantly larger than the intra-CT excitons. Consequently, the inter-$^1$CT state is >0.12 eV higher than the intra-$^1$CT excitons and intra-$^1$CT to inter-$^1$CT interconversion does not readily occur. As a result, the ISC processes in 4CzIPN are promoted by spin-orbit interactions, likely involving the intra-$^1$CT, intra-$^3$CT and $^3$LE states. (c) The photophysical processes occurring in a neat BF2 film that enable efficient spin state interconversion via HFI. Following photoexcitation, the intra-$^1$CT dissociates to form an inter-$^1$CT state with the neighbouring molecule in the dimer. Subsequently, the inter-CT state can readily obtain longer range separation, likely enabled by the shallow dependence of the inter-CT state energy on distance, to form loosely-bound inter-CT states where the exchange energy is small enough for HFI to mediate the ISC processes.

# Methods

**TA spectroscopy**

TA samples were fabricated by spin-coating solutions onto quartz substrates using identical conditions to the previously-reported optimised OLED devices[10]. The samples were encapsulated in a nitrogen glovebox environment to ensure oxygen-free measurements.

TA was performed on a setup powered using a commercially available Ti:sapphire amplifier (Spectra Physics Solstice Ace). The amplifier operates at 1 kHz and generates 100 fs pulses centred at 800 nm with an output of 7 W. A non-colinear optical parametric amplifier (NOPA) was used to provide the tuneable ~100 fs pump pulses for the "ultrafast" (100 fs-1.8 ns) TA measurements, whilst the second harmonic (532 nm) of an electronically triggered, Q-switched Nd:YVO$_4$ laser (Innolas Picolo 25) provided the ~1 ns pump pulses for the nanosecond-microsecond (1 ns-700 μs) TA measurements. The probe was provided by a broadband visible (525-775 nm) and near infrared (830-1000 nm) NOPAs. The probe pulses are collected with a Si dual-line array detector (Hamamatsu S8381-1024Q), driven and read out by a custom-built board from Stresing Entwicklungsbüro. The probe pulse was split into two identical beams by a 50/50 beamsplitter; this allowed for the use of a second reference beam which also passes through the sample but does not interact with the pump. The role of the reference was to correct for any shot-to-shot fluctuations in the probe that would otherwise greatly increase the structured noise in our experiments. Through this arrangement, very small signals with a $\frac{\Delta T}{T} = 1\times 10^{-5}$ could be measured.

**Time-resolved (ns-μs) PL**

Time-resolved PL spectra were recorded using an electrically gated intensified CCD camera (Andor iStar DH740 CCI-010) connected to a calibrated grating spectrometer (Andor SR303i). Sample excitation with a 532 nm pump pulse was provided by a NOPA. Temporal evolution of the photoluminescence emission was obtained by stepping the ICCD gate delay with respect to the excitation pulse. The steady-state PL spectra were measured using the "CW" mode of the spectrometer.

**PLDMR spectroscopy**

Samples for PLDMR were prepared by thermal evaporation (APDC-DTPA, TXO-TPA and 4CzIPN) and spin-coating (BF2) onto thin glass microscope cover slides. The substrates were then cut up into ~3 mm thick strips, stacked into quartz EPR tubes and sealed in a nitrogen glovebox with a bi-component resin (Devcon 5-Minute Epoxy), such that all PLDMR measurements were performed without air exposure.

PLDMR was measured at the Swedish Interdisciplinary Magnetic Resonance Center (SIMARC) with a modified Bruker X-band ESR spectrometer (microwave frequency 9.2-9.84 GHz) with optical access for light excitation and detection. A super-high-Q microwave resonator was used with Q = 8000±1000. The microwave power was amplified by AXEPR10 units with maximum power of about 3.3 W at 0 dBm, corresponding to an amplitude of the microwave magnetic field ($B_1$) of about 0.38 mT. $B_1$ was deduced from the calibration sheet provided by Bruker. The samples were excited by a 405 nm solid-state diode laser with excitation power of 30 mW and a beam spot of 2 mm diameter. The PL was detected by a Si photodiode in conjunction with suitable long-wavelength pass filters. PLDMR was registered as a change of PL intensity upon the magnetic resonance condition. Lock-in detection was used by modulating **B₁** with a rectangular waveform at a frequency of 1-100 kHz. The measurements were performed at room temperature (293 K).

**OSC device fabrication**

Indium tin oxide (ITO) patterned glass substrates were cleaned by acetone and isopropanol for 20 minutes each. The substrates were dried using compressed nitrogen, followed by oxygen plasma treatment for 10 minutes. The conventional architecture devices were made by spin-coating a layer of poly(3,4-ethylenedioxythiophene):poly(styrenesulfonate) (PEDOT:PSS, Clevios P VP Al 8043) at 3000 rpm for 40 s onto the ITO substrates in air, followed by annealing in air at 150 °C for 20 minutes. The active layer for BF2 devices was then spin coated on top of the PEDOT:PSS layer inside a nitrogen filled glovebox from a 20 mg/mL chloroform solution at 1500 rpm. For the 4CzIPN devices, the substrates were transferred to a nitrogen filled glovebox and pumped down under vacuum ($<10^{-7}$ torr) for the evaporation of a 70 nm active layer. Finally, a 10 nm thick Ca interlayer followed by a 100 nm thick Al electrode were deposited on top of the active layer for both sets of devices. An Angstrom Engineering Series EQ Thermal Evaporator was used for thermal evaporation. The electrode overlap area was 4.5 mm$^2$. The active area of the device was determined using an optical microscope.

**OSC device testing**

Photovoltaic characteristic measurements were carried out inside a N₂ filled glove box. Solar-cell device properties were measured under illumination by a simulated 100 mW cm$^{-2}$ AM1.5 G light source using a 300 W Xe arc lamp with an AM 1.5 global filter. The irradiance was adjusted to 1 sun with a standard silicon photovoltaic cell calibrated by the National Renewable Energy Laboratory. No spectral mismatch correction was applied. A Keithley 2635A source measurement unit was used to scan the voltage applied to the solar cell between -2 to 1 V at a speed of 0.43 V/s with a dwell time of 46 ms. Scans were performed in both the forward and reverse directions, with no unusual behaviour observed. At least 16 individual solar cell devices were tested for each system reported.

**Steady-state absorption**

Steady-state absorption spectra were measured using an HP 8453 spectrometer.

**Computational details**

BF2 dimers and tetramer and 4CzIPN dimers were optimised at the DFT level with the B3LYP functional[56], using empirical dispersions, and the 6-31G(d) basis set for all the atomic species. Intermolecular distances $d_\perp$ were found to be 0.35 nm after the ground-state optimisation. Singlet and triplet excited state energies of these aggregates were then computed at the TDDFT level, resorting to the Tamm-Dancoff approximation (TDA)[57]. In the dimers, these calculations were performed as a function of $d_\perp$ by keeping fixed one fragment and shifting the other along the long molecular axis. A *screened* range separated hybrid (SRSH)[58,59] approach was employed in such calculations by using the LC-ωhPBE functional[60] and the 6-311+G(d,p) basis set, as implemented in Gaussian16[61].

Diabatization of electronic states in the aggregates was performed with the same functional and the 6-311G(d,p) and 6-31G(d) basis sets for the dimers and tetramer, respectively. Deconvolution of the lowest singlet and triplet TDDFT adiabatic states in terms of diabatic states was done using the Boys localization scheme[62]. These calculations were performed with the Q-Chem package[63].

The HFI calculations have been performed on the BF2 anion and cation using the ORCA 4.2 package[64] using the SRSH approach using the LC-ωhPBE functional and the EPR-III basis set. The HFI couplings for the triplet excited states were obtained by weighting the HFI of the anion and the cation by the attachment (*i.e.* electron) and the detachment (*i.e.* hole) densities computed on the BF2 dimer.

## Acknowledgements

A.J.G. and R.H.F. acknowledge support from the Simons Foundation (grant no. 601946) and the EPSRC (EP/M01083X/1 and EP/M005143/1). Y.P. acknowledges support from the Swedish Energy Agency (EM-48594-1) and The Swedish Research Council (VR-2017-05285). M.C., E.Z. and F.F. acknowledge financial support from Aix Marseille Université and CNRS. C.T. and D.C. thank the Basque Government (PIBA19-0004) and the Spanish Government MINECO/FEDER (PID2019-109555GB-I00) for financial support. G.L and D.B. thank the European Union's Horizon 2020 research and innovation program under Marie Skłodowska Curie Grant agreement No. 722651 (SEPOMO). Computational resources in Mons and Namur were provided by the Consortium des Équipements de Calcul Intensif (CÉCI), funded by the Fonds de la Recherche Scientifiques de Belgique (F.R.S.-FNRS) under Grant No. 2.5020.11, as well as the Tier-1 supercomputer of the Fedération Wallonie-Bruxelles, infrastructure funded by the Walloon Region under Grant Agreement No. 1117545. Y.O. acknowledges funding from the FRS-FNRS under the grant F.4534.21 (MIS-IMAGINE). G.R. acknowledges a grant from the "Fonds pour la formation à la Recherche dans l'Industrie et dans l'Agriculture" (F.R.I.A.) of the F.R.S.-F.N.R.S. P. J. C. and D.M.U. (EP/L01551X/1) acknowledge support from the EPSRC.

## Author contributions

A.J.G., F.F. and R.H.F. conceived the work. A.J.G. performed the TA and trPL measurements. Y.P. conducted the PLDMR measurements. C.T., G.L., G.R., D.C., F.C., Y.O. and D.B. performed the quantum-chemical calculations. M.C. and E.Z. synthesised and characterised BF2. D.M.L.U. and P.J.C. fabricated and tested the OSC devices. A.J.G, E.W.E. and L.-S.C. fabricated the thin film samples used for the measurements. W.M.C., N.C.G. and B.H.D. contributed to the discussion of the PLDMR data. N.C.G., F.F., D.B. and R.H.F. supervised their groups members involved with the project. A.J.G., D.B. and R.H.F. wrote the manuscript with input from all authors.

## Competing financial interests

The authors declare no competing interests.

## Additional information

Supplementary information accompanies this paper at [to be completed in proofs].

Correspondence and requests for materials should be addressed to R.H.F. (rhf10@cam.ac.uk), A.J.G. (ajg216@cam.ac.uk) D.B. (david.beljonne@umons.ac.be), F.F. (frederic.fages@univ-amu.fr) and Y.P. (yuttapoom.puttisong@liu.se).

Reprints and permissions information is available at www.nature.com/reprints.

## Data availability

The data that supports the findings of this study are available from the corresponding authors upon reasonable request.

# Supplementary Information for

# Spontaneous exciton dissociation enables spin state interconversion in delayed fluorescence organic semiconductors

*Alexander J. Gillett[1]\*, Claire Tonnelé[2], Giacomo Londi[3], Gaetano Ricci[4], Manon Catherin[5], Darcy M. L. Unson[1], David Casanova[2], Frédéric Castet[6], Yoann Olivier[4], Weimin M. Chen[7], Elena Zaborova[5], Emrys W. Evans[1,8], Bluebell H. Drummond[1], Patrick J. Conaghan[1,9], Lin-Song Cui[1], Neil C. Greenham[1], Yuttapoom Puttisong[7]\*, Frédéric Fages[5]\*, David Beljonne[3]\* and Richard H. Friend[1]\*.*

[1]Cavendish Laboratory, University of Cambridge, JJ Thomson Avenue, Cambridge, CB3 0HE, UK.

[2]Donostia International Physics Centre (DIPC), Donostia, Euskadi, Spain.

[3]Laboratory for Chemistry of Novel Materials, Université de Mons, Place du Parc 20, 7000 Mons, Belgium.

[4]Unité de Chimie Physique Théorique et Structurale & Laboratoire de Physique du Solide, Namur Institute of Structured Matter, Université de Namur, B-5000 Namur, Belgium.

[5]Aix Marseille Univ, CNRS, CINaM UMR 7325, AMUtech, Campus de Luminy, 13288 Marseille, France.

[6]Institut des Sciences Moléculaires, Université de Bordeaux, 33405 Talence, France.

[7]Department of Physics, Chemistry and Biology (IFM) Linköping University, Linköping, Sweden.

[8]Department of Chemistry, Swansea University, Singleton Park, Swansea, SA2 8PP, UK.

[9]ARC Centre of Excellence in Exciton Science, School of Chemistry, University of Sydney, NSW 2006, Australia.

[*]Corresponding authors: Alexander J. Gillett: E-mail: ajg216@cam.ac.uk; Yuttapoom Puttisong: E-mail: yuttapoom.puttisong@liu.se; Frédéric Fages: E-mail frederic.fages@univ-amu.fr; David Beljonne: E-mail: david.beljonne@umons.ac.be; Richard H. Friend: E-mail: rhf10@cam.ac.uk.

# Table of Contents

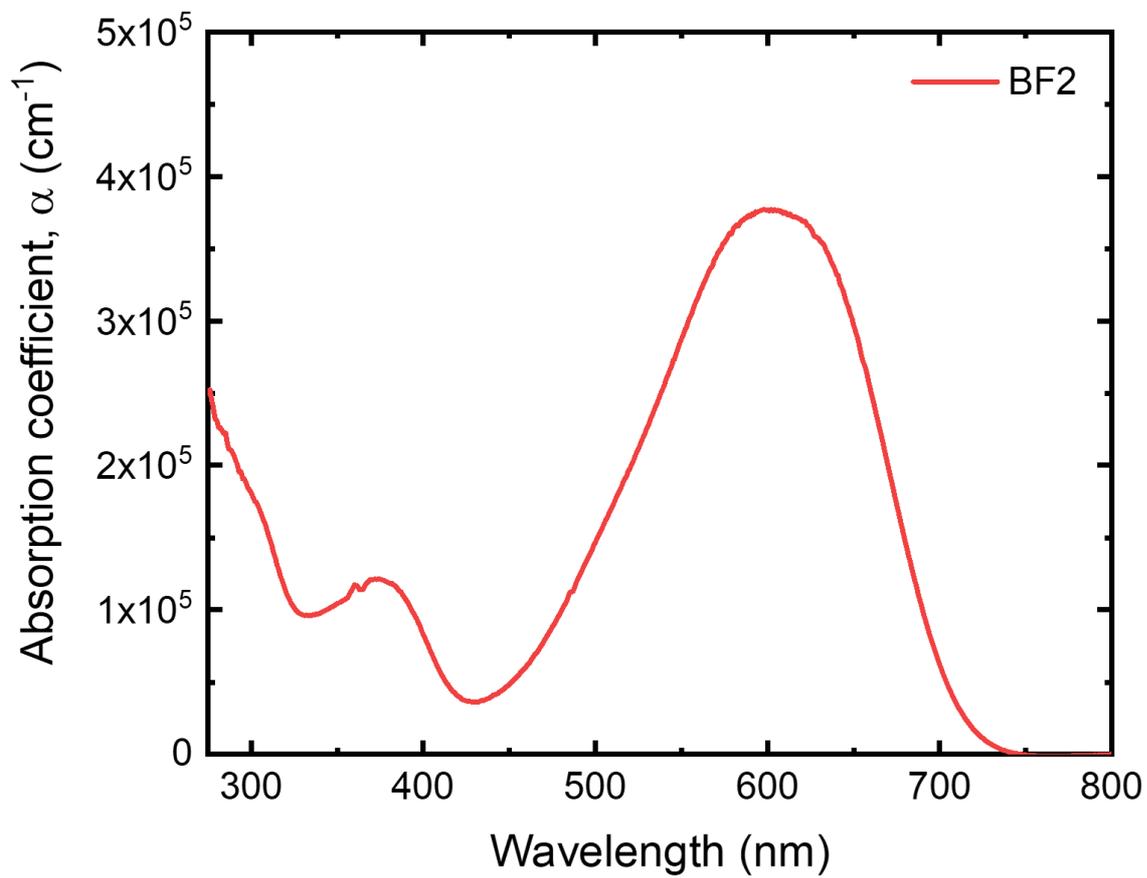

**Figure S1:** The absorption coefficient of a neat BF2 film.

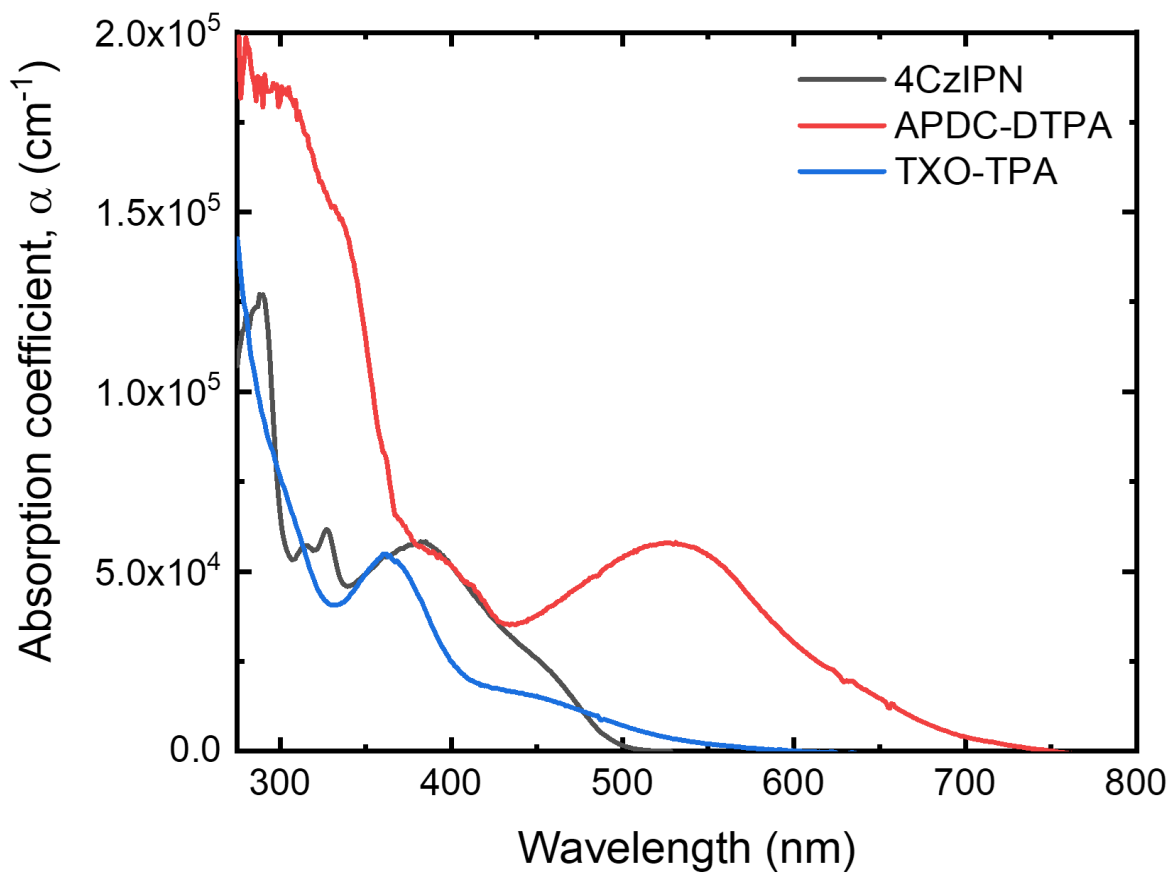

**Figure S2:** The absorption coefficients of neat films of the other organic DF materials investigated.

| Material | Absorption coefficient (cm$^{-1}$) | $\Delta E_{ST}$ (meV) |
|---|---|---|
| BF2 | 3.8x10$^5$ | 200 |
| APDC-DTPA | 5.8x10$^4$ | 140[1] |
| 4CzIPN | 2.5x10$^4$ | 40[2] |
| TXO-TPA | 1.8x10$^4$ | 40[3] |

**Table S1:** The absorption coefficients and $\Delta E_{ST}$ of organic DF materials investigated in this study. The absorption coefficient is taken from the peak of the lowest energy intra-CT absorption band: BF2 (600 nm); APDC-DTPA (530 nm); 4CzIPN (450 nm); and TXO-TPA (425 nm). The $\Delta E_{ST}$ is obtained from low temperature phosphorescence measurements in all cases.

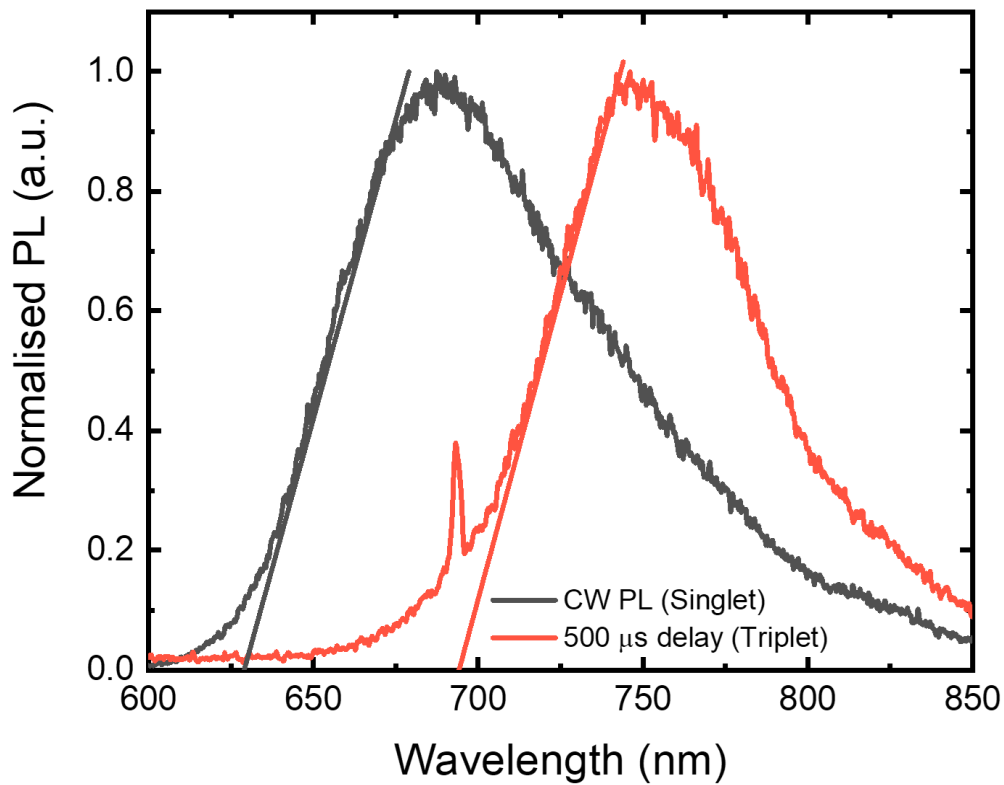

**Figure S3:** The normalised fluorescence and phosphorescence spectra of BF2 doped at 5 wt% in polystyrene, taken at 10 K. Excitation was provided by a 400 nm laser. The onset of the fluorescence is at 1.97 eV, whilst the phosphorescence onset is at 1.77 eV. The difference in onset of the spectra gives a singlet-triplet energy gap of 0.2 eV.

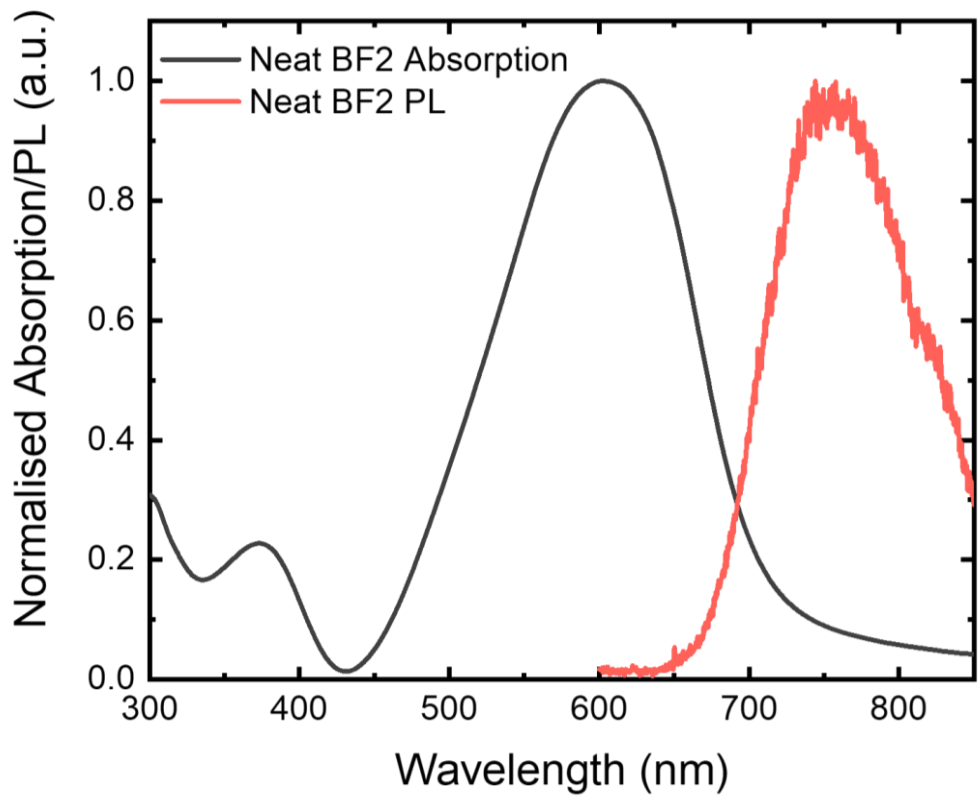

**Figure S4:** The normalised absorption and photoluminescence spectra of a neat BF2 film.

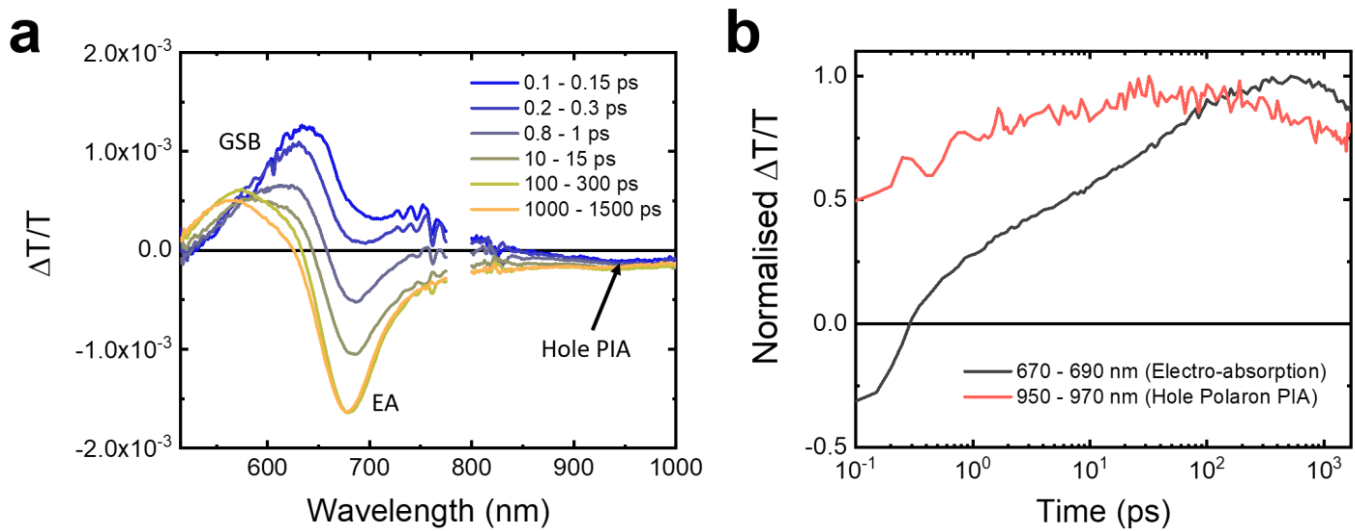

**Figure S5: (a)** The TA spectra of a BF2:PC$_{60}$BM film, excited at 610 nm with a fluence of 3.6 μJ cm$^{-2}$. By 0.1 – 0.15 ps, the BF2 stimulated emission (SE) is already strongly quenched compared to the neat film TA (Fig. 2), suggesting that a significant fraction of the electron transfer from BF2 to PC$_{60}$BM takes place on ultrafast (<0.1 ps) timescales. A PIA centred at 940 nm that is clearly distinct from the BF2 $^1$ICT PIA peaking beyond 1000 nm is also seen at 0.1 – 0.15 ps. Given the rapid quenching of the BF2 $^1$ICT, this new PIA is assigned to the hole polaron photo-induced absorption (PIA) of BF2. As time progresses, a sharp new negative feature forms at the band edge of the BF2 absorption around 680 nm. Due to the spectral location at the edge of the BF2 ground state absorption and distinct kinetics to the hole polaron PIA, this new feature is assigned to the electro-absorption (EA) of BF2, induced by the separation of the charge transfer state into free charges[4–7]. **(b)** The TA kinetics of the hole polaron (950 – 970 nm) and EA (670 – 690 nm). We note that the kinetics of the formation of the EA feature are clearly distinct to the formation of the hole polaron PIA, confirming that these two features have different origins. We note that the EA grows in more slowly than the hole polaron, as would be expected given that charge transfer must take place before the charge transfer state can dissociate into free charges.

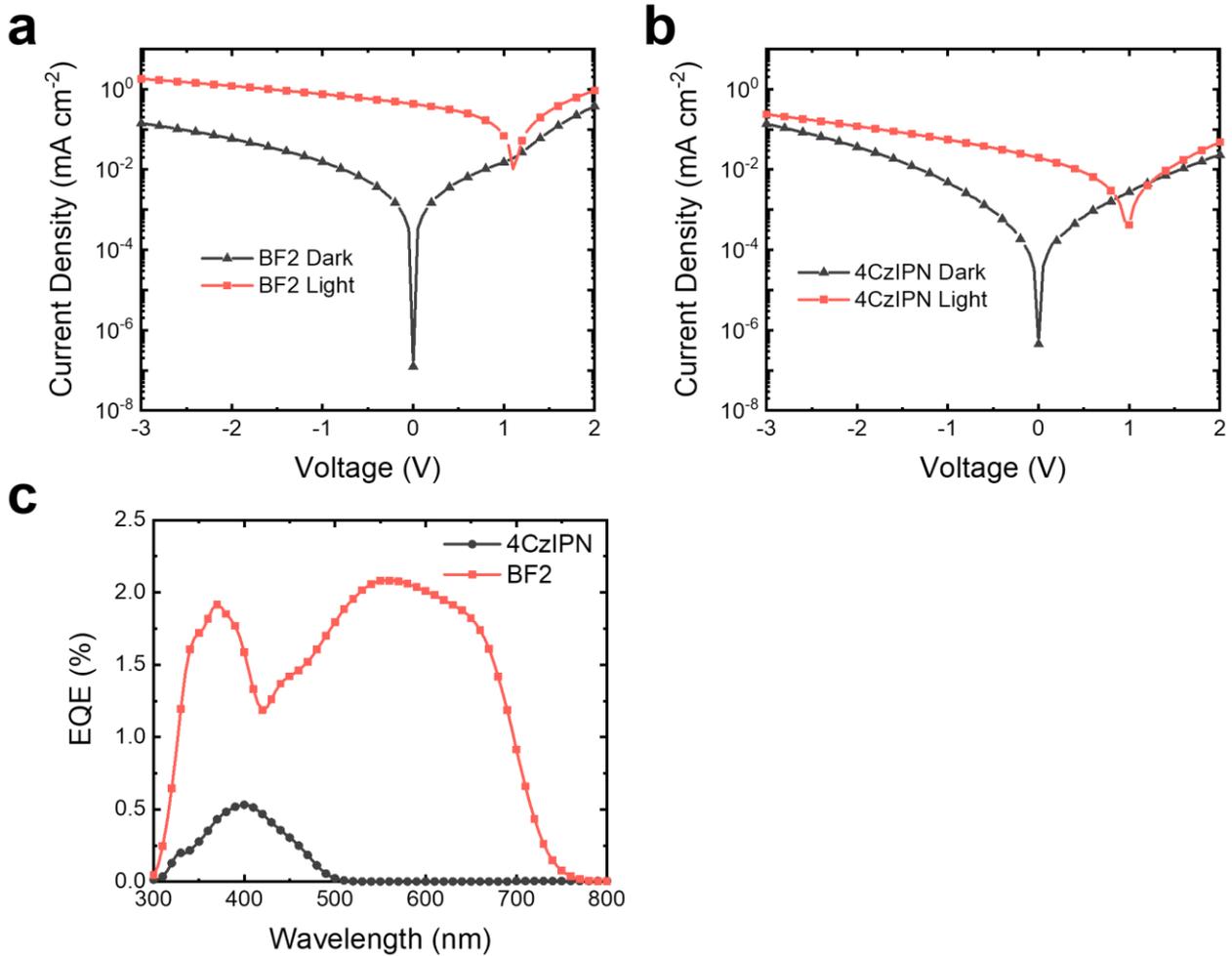

**Figure S6:** (**a**) The dark and light current density-voltage curves under 100 mW cm$^{-2}$ AM1.5G illumination of an OSC device fabricated with a neat BF2 active layer. It can be seen that the current density extracted under reverse bias is greatly enhanced under 1 sun illumination, providing evidence for the photo-generation of bound SCRPs in BF2 that can be dissociated by an external electric field. (**b**) The dark and light current density-voltage curves under 100 mW cm$^{-2}$ AM1.5G illumination of an OSC device fabricated with a neat 4CzIPN active layer. The photocurrent generated from the neat 4CzIPN device is significantly lower than the neat BF2 device. (**c**) The EQE responses of the neat BF2 and 4CzIPN devices taken at short-circuit. The peak EQE of 2.1% at 555 nm in the neat BF2 device demonstrates that BF2 itself can generate a moderate amount of photocurrent without the need for a separate electron or hole acceptor. The lower peak EQE of 0.53% in the neat 4CzIPN device confirms that 4CzIPN alone is not able to generate significant photocurrent.

| Device | $V_{oc}$ (V) | $J_{sc}$ (mA cm$^{-2}$) | FF | PCE (%) |
|---|---|---|---|---|
| Neat BF2 | 1.12 | 0.43 | 0.31 | 0.15 |
| Neat 4CzIPN | 0.98 | 0.02 | 0.22 | 0.004 |

**Table S2:** The summarised performance metrics of the neat BF2 and 4CzIPN OSC devices under 100 mW cm$^{-2}$ AM1.5G illumination.

## BF2 diluted in CBP

To gain a deeper insight into the exciton dissociation process, we have investigated the ultrafast TA of BF2 dispersed in the non-interacting host CBP at 2, 10 and 40 wt% (Fig. S6). We note that the photoluminescence (PL) spectra of BF2 broadens and undergoes a significant red shift as a function of doping concentration (Fig. S5), indicating the presence of interactions between BF2 molecules in the more concentrated films. This can be attributed to the tendency of BF2 to form dimers in low wt% films, followed by larger aggregates at a higher wt%[8]. Thus, we consider the 2 wt% film to represent the behaviour of isolated BF2 molecules, with higher loadings increasingly reflecting the properties of aggregated BF2. In the ultrafast TA of the BF2 dilution series, we observe a dramatic shortening of the SE lifetime with increasing BF2 concentration (Fig. S7); this is consistent with the expectation that closer contact between BF2 molecules will result in faster exciton dissociation. Of particular interest is the nanosecond TA of the BF2 dilution series (Fig. S8). In the 2 wt% film, the SE decays within 10 ns, leaving behind a long-lived GSB signal that represents the small fraction of BF2 singlet excitons that undergo ISC to the $T_1$ state. The kinetics of the GSB decay can be well-described by a bi-exponential function, typical of DF emitters[9,10]. The short "prompt" time constant of 3.3 ns primarily represents the radiative decay of singlet excitons, whilst the long "delayed" time constant of 260 μs confirms that the rISC process of isolated BF2 is particularly slow, as expected for the relatively large $\Delta E_{ST}$ of 0.2 eV. In contrast, the 10 and 40 wt% films exhibit markedly different behaviour. After efficient dissociation of the excitons quenches the SE on picosecond timescales, we observe the regrowth of the SE on microsecond timescales (Figs. 2c, 2d, S8f and S9c). This suggests that the emissive intra-$^1$CT exciton is being reformed from the separated charge carriers. We verify that the emissive state present on microsecond timescales has the same PL spectrum as the steady-state emission (Fig. S10), confirming that the SE regrowth observed in the TA is due to the reformation of the bright intra-$^1$CT exciton.

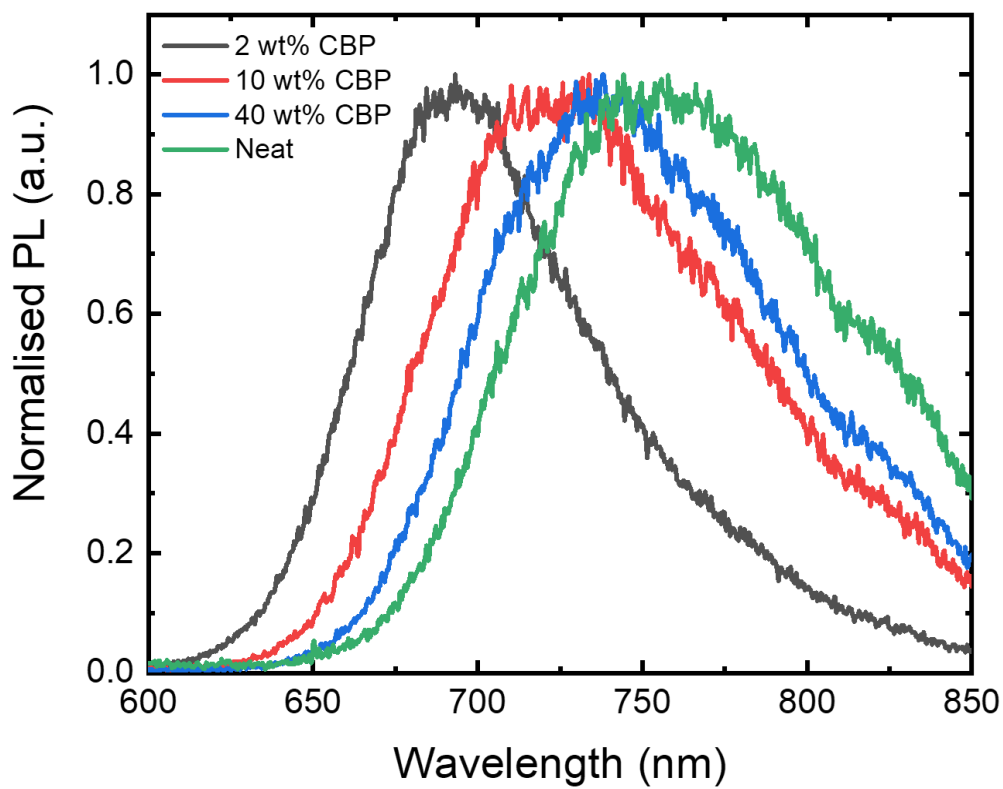

**Figure S7:** The normalised PL spectra of BF2 at 2, 10 and 40 wt% in CBP and a neat film. Excitation was provided by a 532 nm laser. There is a clear red shift of the PL maximum and broadening of the spectra as the doping concentration rises, resulting from the increased levels of interaction between BF2 molecules in the film.

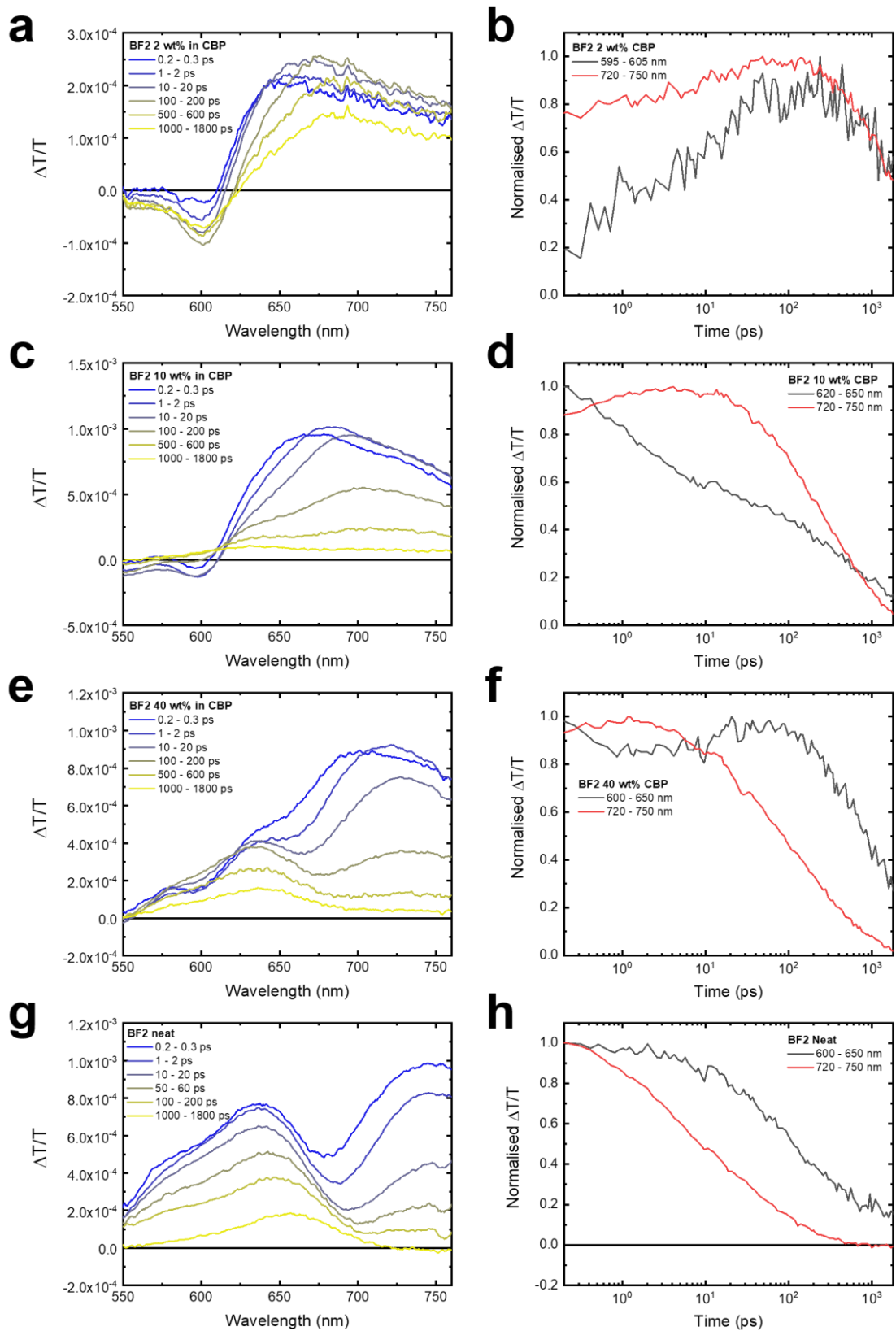

**Figure S8:** The ultrafast TA spectra and kinetics of BF2 at 2, 10 and 40 wt% in CBP and a neat BF2 film. 532 nm excitation with fluences of 15.7, 15.7, 6.3 and 5.3 μJ cm$^{-2}$ were used, respectively.

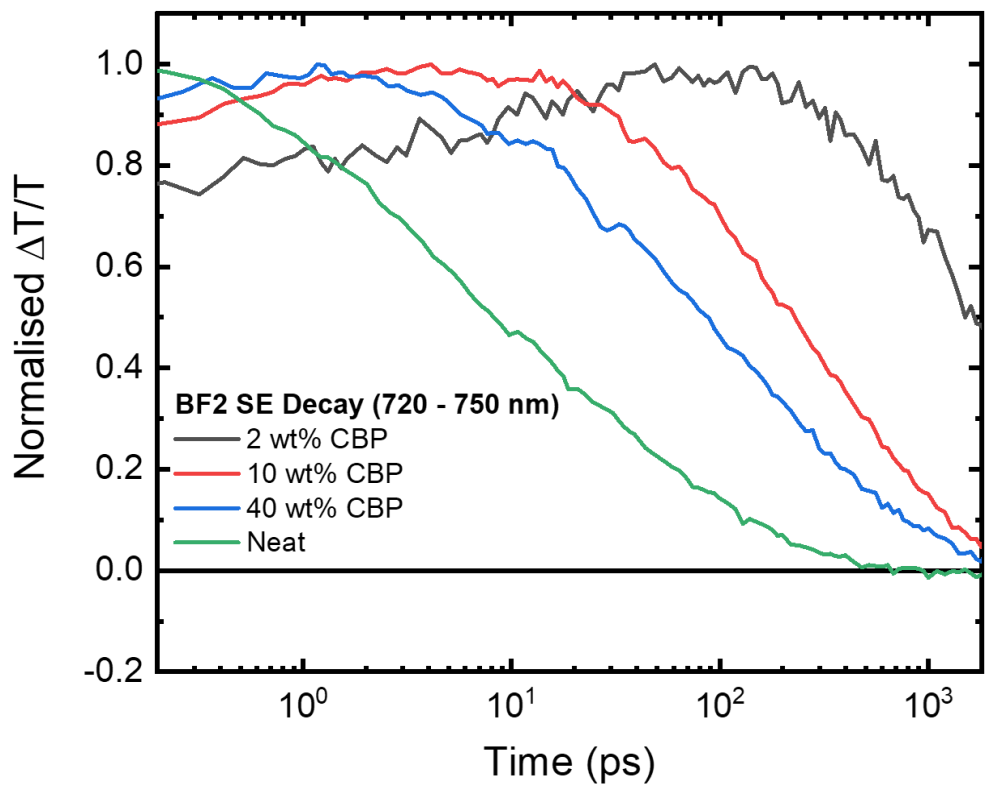

**Figure S9:** The normalised ultrafast TA kinetics of the BF2 SE (720 – 750 nm) for the 2, 10 and 40 wt% films in CBP and the neat BF2 film. The lifetime of the SE falls off rapidly as the doping concentration is increased, confirming that the increased contact between BF2 molecules results in faster charge transfer. The slight increase in the intensity of the SE intensity between 720 – 750 nm in the 2 wt% blend within the first 100 ps is due to a red shift in the SE band over these timescales.

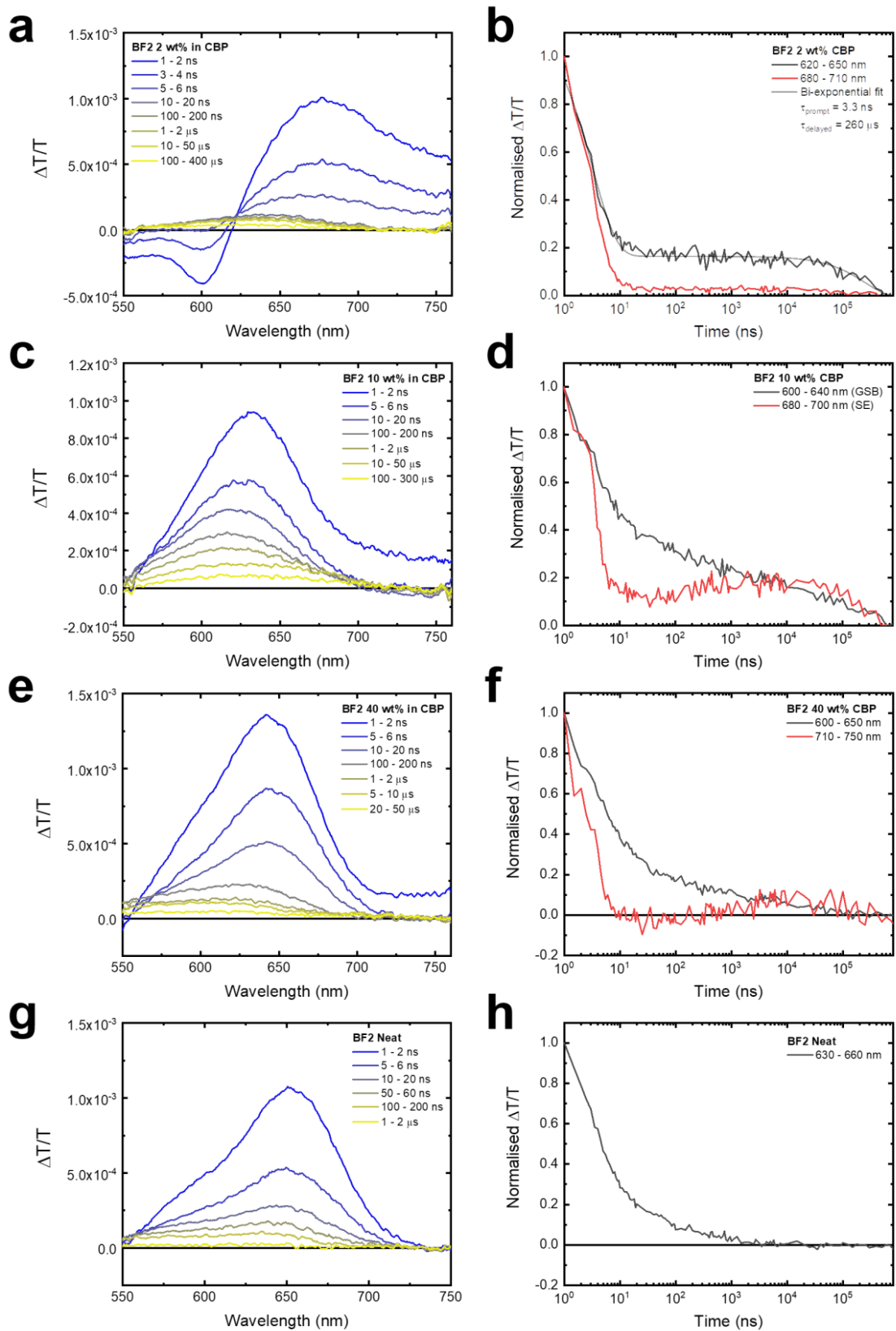

**Figure S10:** The nanosecond TA spectra and kinetics of BF2 at 2, 10 and 40 wt% in CBP and a neat BF2 film. 532 nm excitation with fluences of 692, 346, 59.7 and 47.6 μJ cm$^{-2}$ were used, respectively.

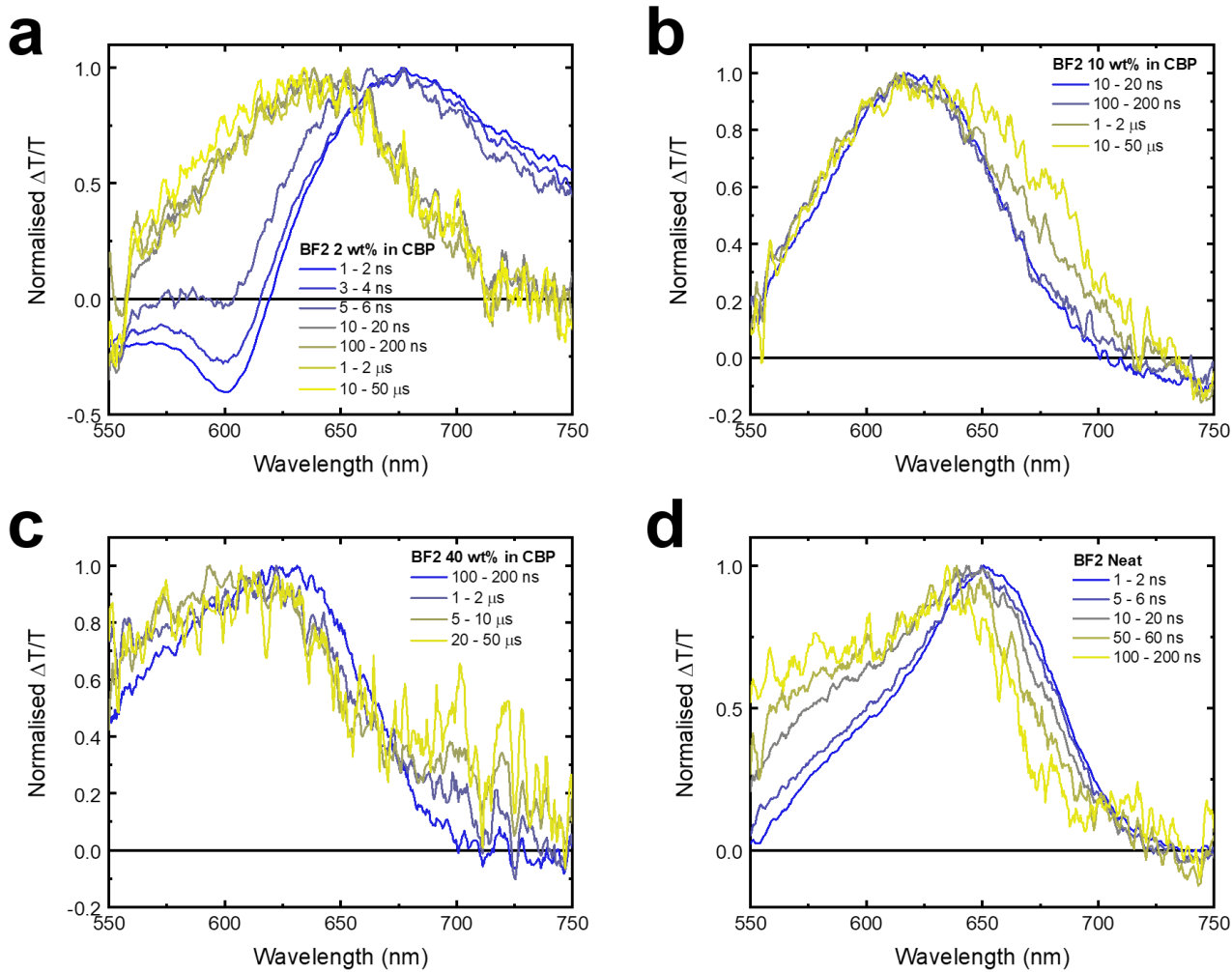

**Figure S11:** The normalised nanosecond TA spectra of BF2 at 2, 10 and 40 wt% in CBP and a neat BF2 film. 532 nm excitation with fluences of 692, 346, 59.7 and 47.6 µJ cm$^{-2}$ were used, respectively. The regrowth of the SE at the lower energy edge of the GSB can clearly be seen in the 10 and 40 wt% films but is absent in the 2 wt% and neat films.

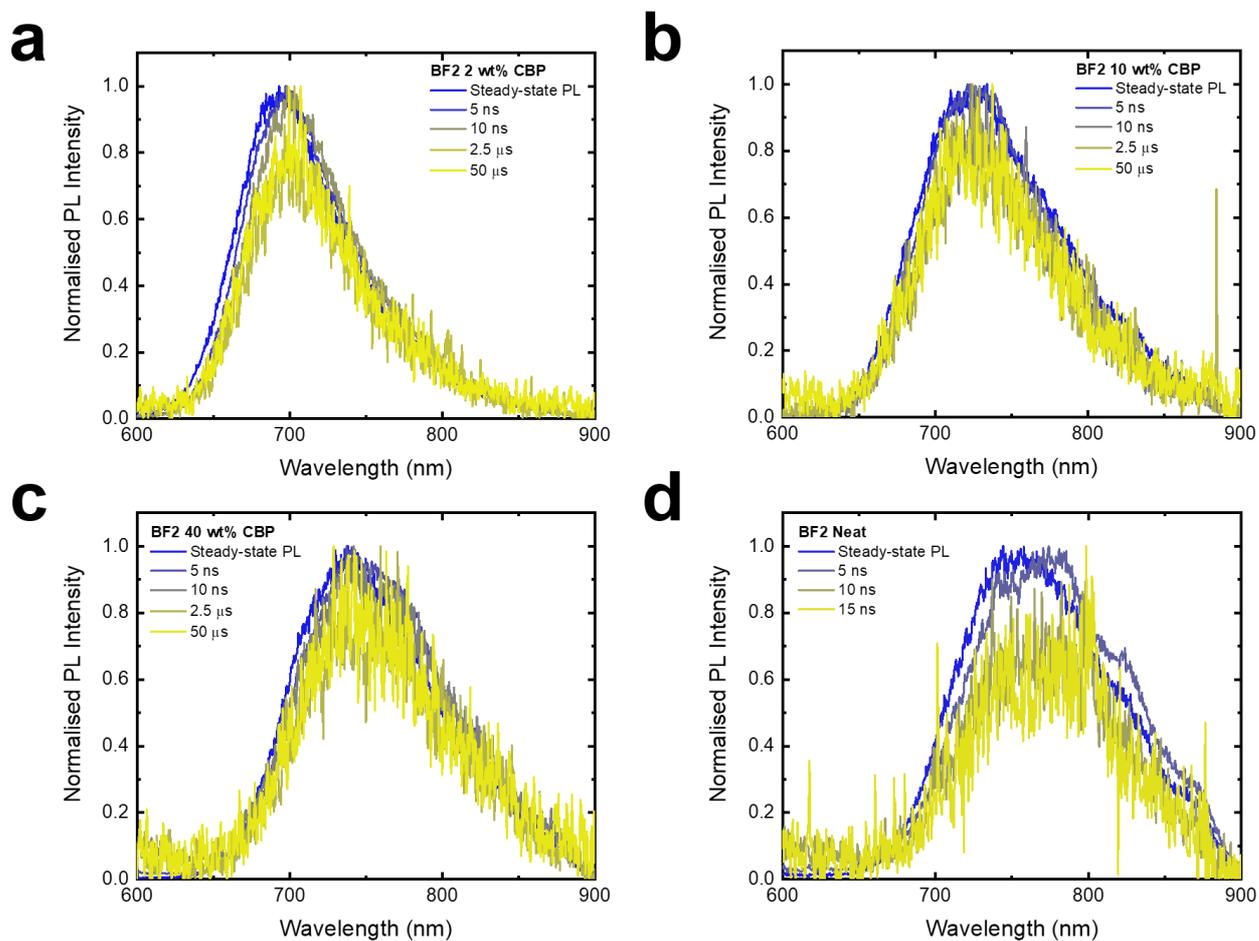

**Figure S12:** The transient PL spectra of BF2 at 2, 10 and 40 wt% in CBP and a neat BF2 film. 532 nm excitation with a fluence 27.8 μJ cm$^{-2}$ was used for all samples.

## PLDMR spectroscopy and hyperfine interactions

In PLDMR spectroscopy, an external magnetic field ($B_0$ ~330 mT) energetically splits the $T_+$ and $T_-$ triplet sublevels from the S and $T_0$ through the Zeeman effect[11]. By sweeping $B_0$, Zeeman sublevel transitions are induced in the triplet manifold when the applied X-band microwaves are resonant with the transition energy. To enable detection with a high sensitivity, the sample is illuminated with a 405 nm continuous-wave laser, which initially creates singlet Frenkel excitons, and the resulting PL is monitored. When the fixed-frequency microwaves become resonant with a Zeeman transition, the steady-state triplet sublevel populations will be modified and the PL yield will be enhanced or quenched, depending on whether the new triplet sublevel occupancies can better couple to the emissive singlet manifold than the unperturbed population[12]. Thus, PLDMR is well-suited to detecting and identifying triplet states in organic materials.

The HFI originates from the magnetic coupling between the electron spin and the nuclear spin (primarily of $^1$H and $^{14}$N nuclei[13]); as a result, every organic molecule possesses a local hyperfine field ($B_{HF}$) about which the paramagnetic spin of an unpaired electron, known as a radical, will precess[14]. HFI-ISC is typically observed in spin-correlated radical pairs[11,13,15], which consist of two weakly exchange-coupled radicals and can be considered analogous to the loosely-bound inter-CT states present in BF2. Here, local variances in $B_{HF}$, including the different $B_{HF}$ experienced by the electron and hole in the inter-CT state[16], results in the spins of individual radicals precessing with slightly different frequencies and, in absence of an applied external magnetic field ($B_0$), also about randomly-distributed axis[11,17]. Thus, the HFI can promote ISC transitions with a rate of ~$10^8$–$10^6$ s$^{-1}$ via a periodic oscillation between the $m_s$=0 singlet (S) and the $m_s$=-1,0,+1 triplet sublevels (T-, T$_0$, T$_+$) of the inter-CT state (Fig. 3a)[13,18,19]; this is achieved by inducing a phase-lag between the precession of the radical spins[11]. We note similar spin-mixing effects can be achieved through the $\Delta g$ mechanism, whereby small variations in the *g*-factor of the positive and negative polarons comprising the inter-CT state can also lead to dephasing between the radical spins[11]. However, the $\Delta g$ mechanism exhibits a strong dependence on the strength of **B$_0$** and is only able to mediate spin-mixing between the inter-$^1$CT$_0$ and inter-$^3$CT$_0$ with a rate of ~10$^7$-10$^6$ s$^{-1}$ at fields of ~1 T[11,15,18,20]. Consequently, at the field values studied here (and under OLED device operation at zero-field), it is likely that only the HFI can contribute significantly to the ISC processes between inter-CT states.

To estimate the size of the EPR **D**-parameter (corresponding to the strength of the dipolar interaction between two unpaired spins), and thus the electron-hole separation in the BF2 film, we can use the linewidth of the PLDMR signal. Whilst broadening mechanisms will also contribute to the observed linewidth, an upper estimate of 2**D** can be obtained from the full width at half maximum (FWHM) of the signal. In neat BF2, the FWHM of the inter-CT state signal is ~1.5 mT; this corresponds to **D** ≤0.75 mT. As we now know the strength of the dipolar interaction (in mT), we can estimate an lower bound for the electron-hole separation (in nm) from the following equation[21,22]:

$$r_{e-h} = \sqrt[3]{\frac{2.785}{D}} \quad (S1)$$

From this, we obtain a lower bound for the electron-hole separation for the inter-CT state in the neat BF2 film of ≥1.5 nm.

To ascertain the singlet-triplet interconversion timescale enabled by the HFI, we turn to the spin-locking data (Fig. 3d). In the PLDMR measurements, the active magnetic field (**B**) is determined by:

$$\boldsymbol{B} = \boldsymbol{B_0} + \boldsymbol{B_{HF}} \qquad (S2)$$

In the presence of a large **B₀**, **B** is overwhelmingly determined by **B₀**; this results in the two unpaired spins precessing around **B₀**. As previously discussed, this prevents HFI-interaction mediated transitions between $S_0$ and the $T_+$ and $T_-$. However, the spin precession frequency difference resulting from the **B_HF** component parallel to **B₀** ($\Delta\omega_{HF}$) is retained; thus, $S_0$ to $T_0$ HFI-induced transitions can still occur. In this case, the singlet-triplet mixing frequency ($\Omega$) is $\sim\Delta\omega_{HF}$. However, when **B₁** becomes comparable to the **B_HF** component parallel to **B₀** (with **B₁** is perpendicular to **B₀**), the spins reorient and begin to precess around **B₁** instead (as viewed in a rotating co-ordinate frame fixed with respect to **B₀**). As a result, $\Omega$ now becomes[23]:

$$\Omega \approx \frac{\Delta\omega_{HF}^2}{2\omega_1} \qquad (S3)$$

Where $\omega_1$ is the frequency of the spin precession around **B₁**. From this, we can estimate the effective **B_HF** ($|\boldsymbol{B_{HF}}|^{eff}$) to be:

$$|\boldsymbol{B_{HF}}|^{eff} \sim 2\boldsymbol{B_1^*} \qquad (S4)$$

Here, $\boldsymbol{B_1^*}$ represents the microwave field at which the W-shaped resonance starts to develop:

$$\boldsymbol{B}_1^* = \frac{\hbar \omega_{B_1^*}}{g\mu_B} \tag{S5}$$

Where ℏ is the reduced Planck's constant, $g$ is the electron $g$-factor and $\mu_B$ is the Bohr magneton. For BF2, $\boldsymbol{B}_1^* = 0.38$ mT, giving $|\boldsymbol{B}_{HF}|^{eff} \sim 0.76$ mT. This can be converted into a frequency (ν) through the following formula:

$$\nu\ (MHz) = 10^{-9} \frac{g\mu_B}{h} B\ (mT) \tag{S6}$$

Where h is Planck's constant. Thus, 0.76 mT corresponds to an effective $\Delta\omega_{HF}$ of 21.3 MHz. As $\Delta\omega_{HF}$ represents the time taken for a complete $S_0$-$T_0$-$S_0$ cycle, the frequency of one singlet to triplet conversion step is 10.7 MHz, equal to an ISC (and rISC) timescale of ~24 ns.

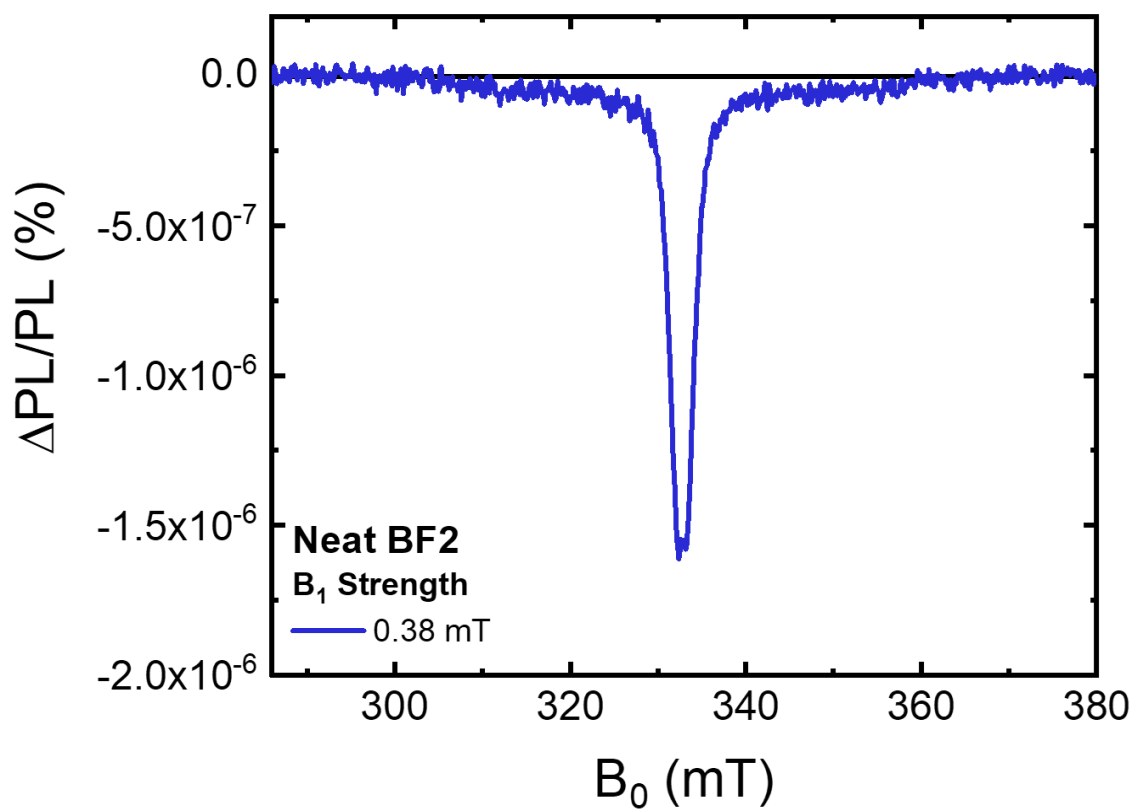

**Figure S13:** The full field range PLDMR response of a neat BF2 film at 293 K with 405 nm excitation (30 mW). The sharp peak at ~333 mT is a signature of inter-CT states, with the weak and broader resonance between 300 – 360 mT originating from triplet excitons localised on a single BF2 molecule.

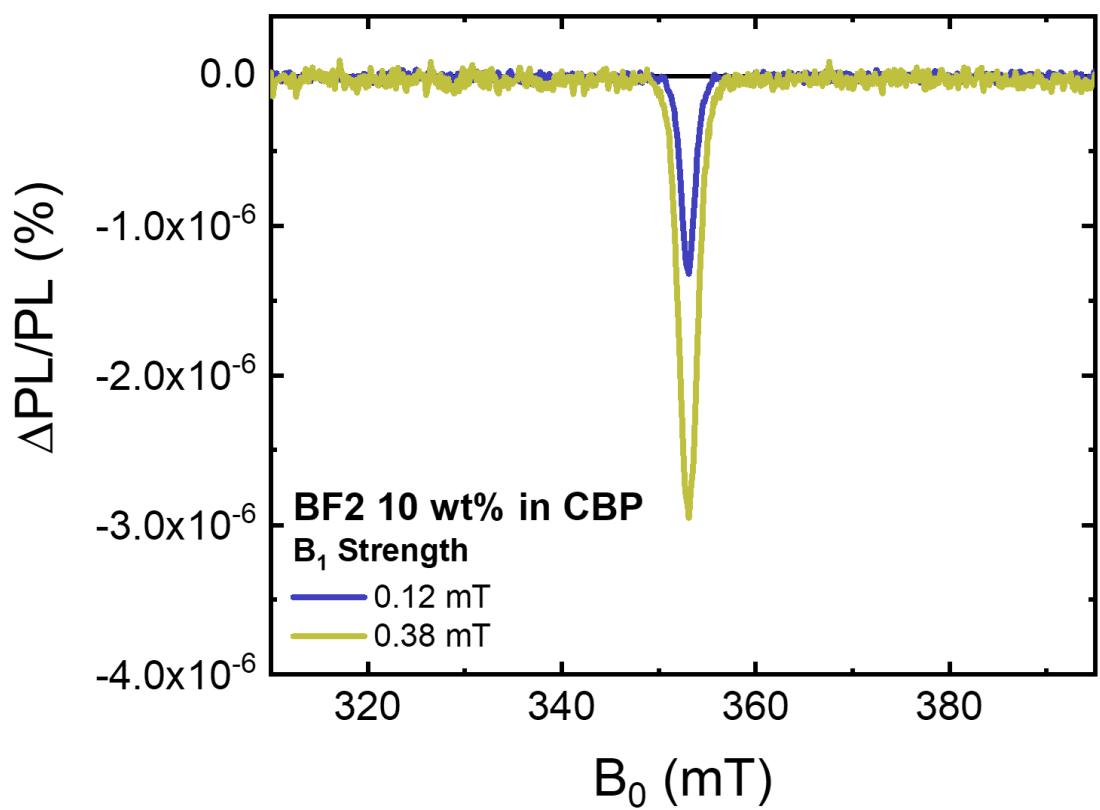

**Figure S14:** The full field range PLDMR response of BF2 diluted at 10 wt% in CBP film at 293 K with 405 nm excitation (30 mW). The sharp peak at ~353 mT is a signature of inter-CT states, with no broader resonance observed that could be assigned to triplet excitons localised on a single BF2 molecule. The inter-CT peak is shifted by ~20 mT compared to the neat film due to a small change in the frequency of the X-band microwaves between measurements. The electron *g*-factor remains unchanged.

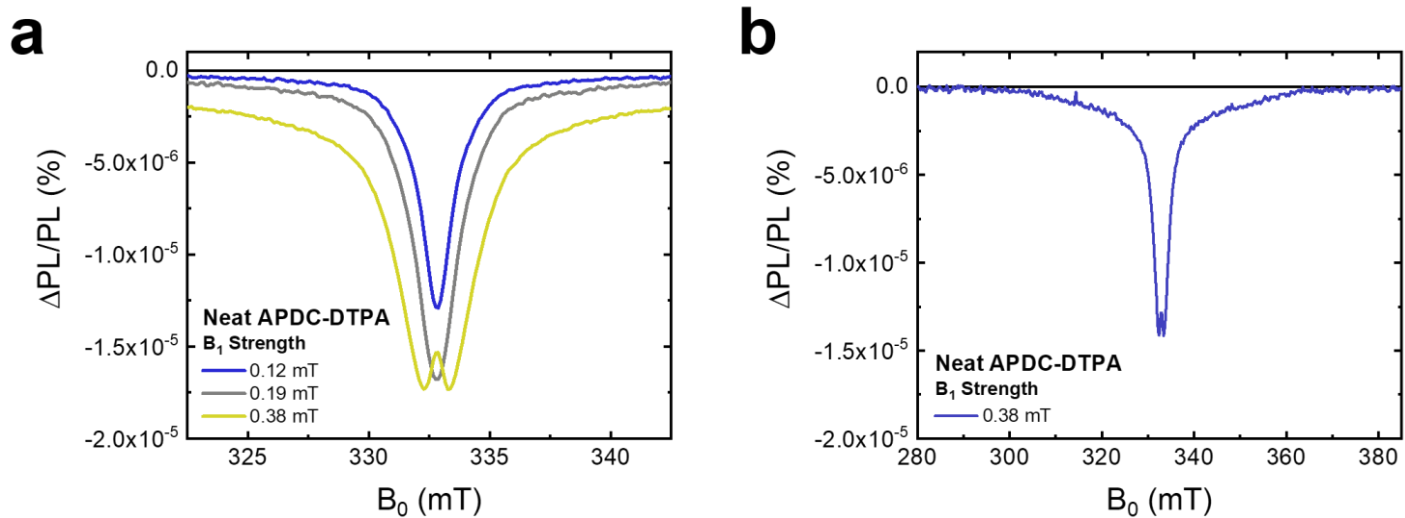

**Figure S15:** (a) The PLDMR response of a neat APDC-DTPA film at 293 K with 405 nm excitation (30 mW). As $B_1$ increases, a "W"-shaped peak begins to form, confirming the spin-locking behaviour and the presence of HFI-ISC processes between inter-CT states in a neat APDC-DTPA film. (b) The full field range PLDMR response of the neat APDC-DTPA film. The broader resonance between 300 – 360 mT originates from triplet excitons localised on a single APDC-DTPA molecule.

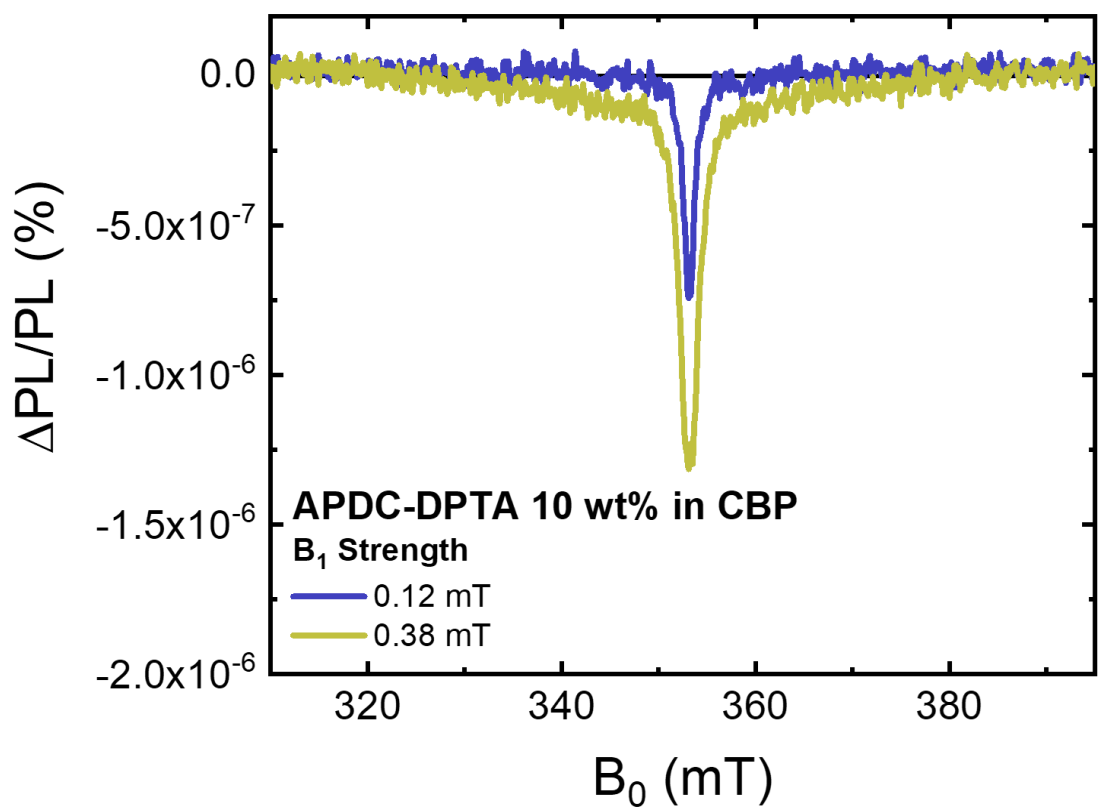

**Figure S16:** The PLDMR response of APDC-DTPA diluted at 10 wt % in CBP at 293 K with 405 nm excitation (30 mW). The sharp peak at ~353 mT is a signature of inter-CT states, with the weak and broader resonance between 300 – 360 mT originating from triplet excitons localised on a single APDC-DTPA molecule. The inter-CT peak is shifted by ~20 mT compared to the neat film due to a small change in the frequency of the X-band microwaves between measurements. The electron *g*-factor remains unchanged.

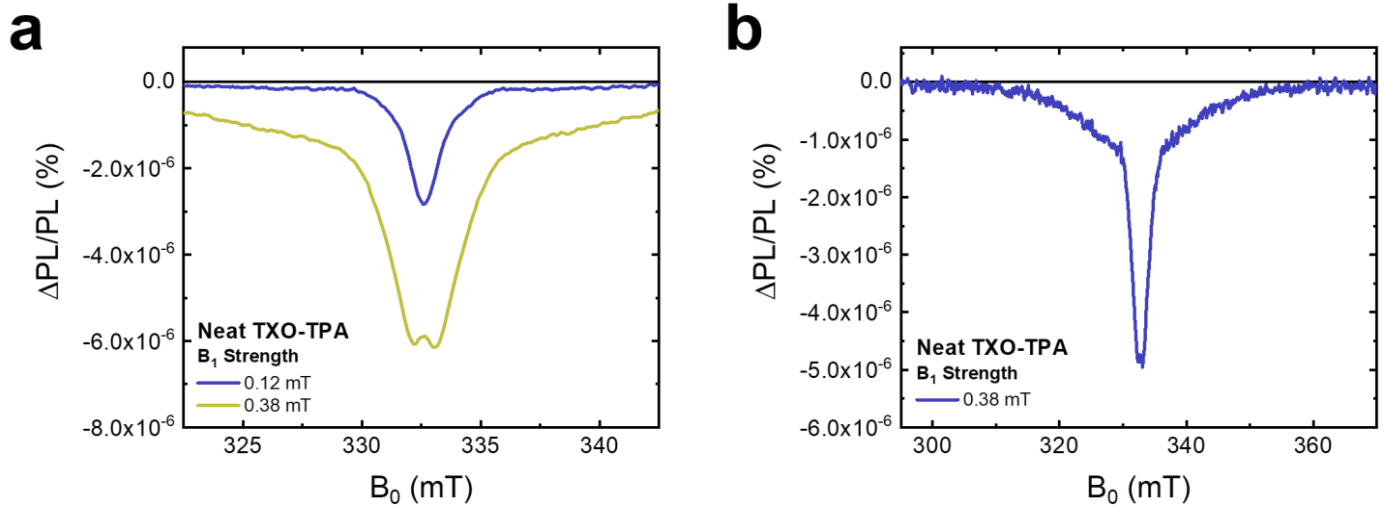

**Figure S17:** The PLDMR response of a neat TXO-TPA film at 293 K with 405 nm excitation (30 mW). As **B$_1$** increases through a reduced microwave attenuation, a "W"-shaped peak begins to form, confirming the spin-locking behaviour and the presence of HFI-ISC processes between inter-CT states in a neat TXO-TPA film. **(b)** The full field range PLDMR response of the neat TXO-TPA film. The broader resonance between 300 – 360 mT originates from triplet excitons localised on a single TXO-TPA molecule.

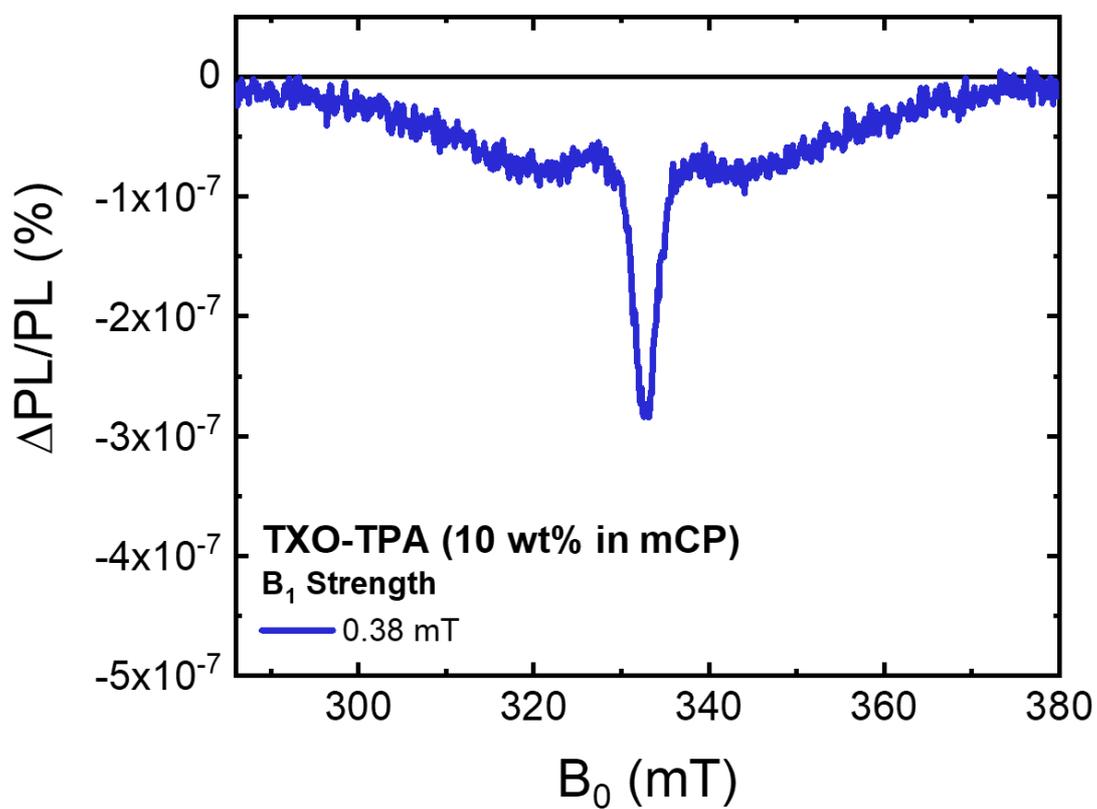

**Figure S18:** The PLDMR response of TXO-TPA diluted in mCP at 10 wt% at 293 K with 405 nm excitation (30 mW). The sharp peak at ~353 mT is a signature of inter-CT states, with the broader resonance between 290 – 370 mT originating from triplet excitons localised on a single TXO-TPA molecule.

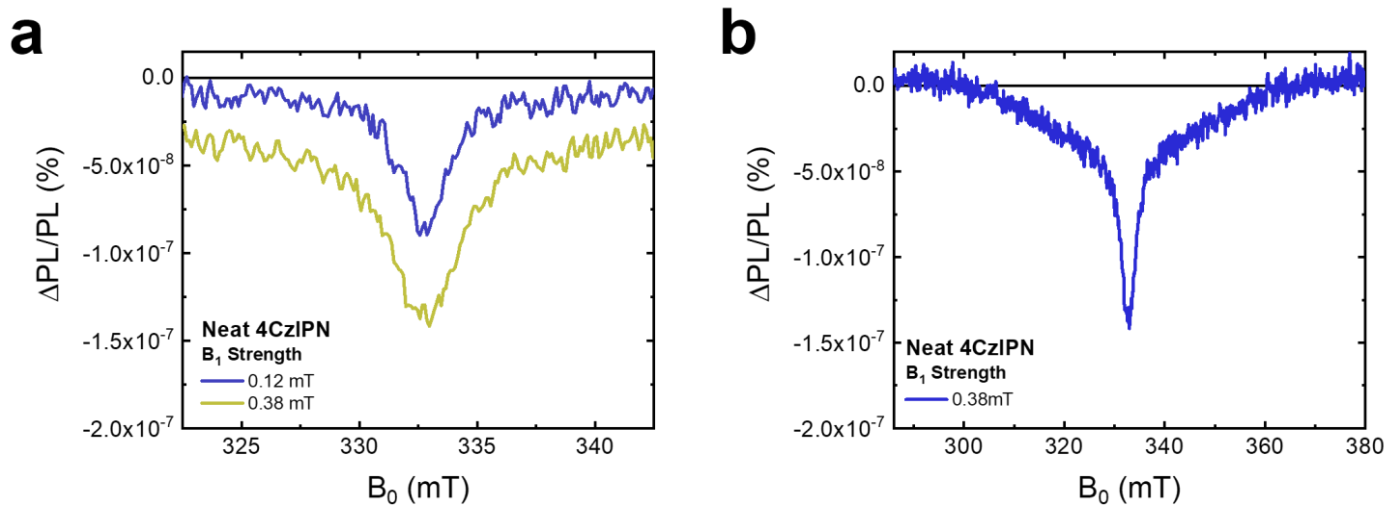

**Figure S19:** The PLDMR response of a neat 4CzIPN film at 293 K with 405 nm excitation (30 mW). The PLDMR response is significantly weaker than in the other DF materials included in this study, suggesting a reduced ability of 4CzIPN to form intermolecular CT states where HFI-ISC processes can take place. As $B_1$ increases, there is no formation of a "W"-shaped peak, indicating that spin-locking is not observed in the neat 4CzIPN film. **(b)** The full field range PLDMR response of the neat 4CzIPN film. The broader resonance between 300 – 360 mT originates from triplet excitons localised on a single 4CzIPN molecule.

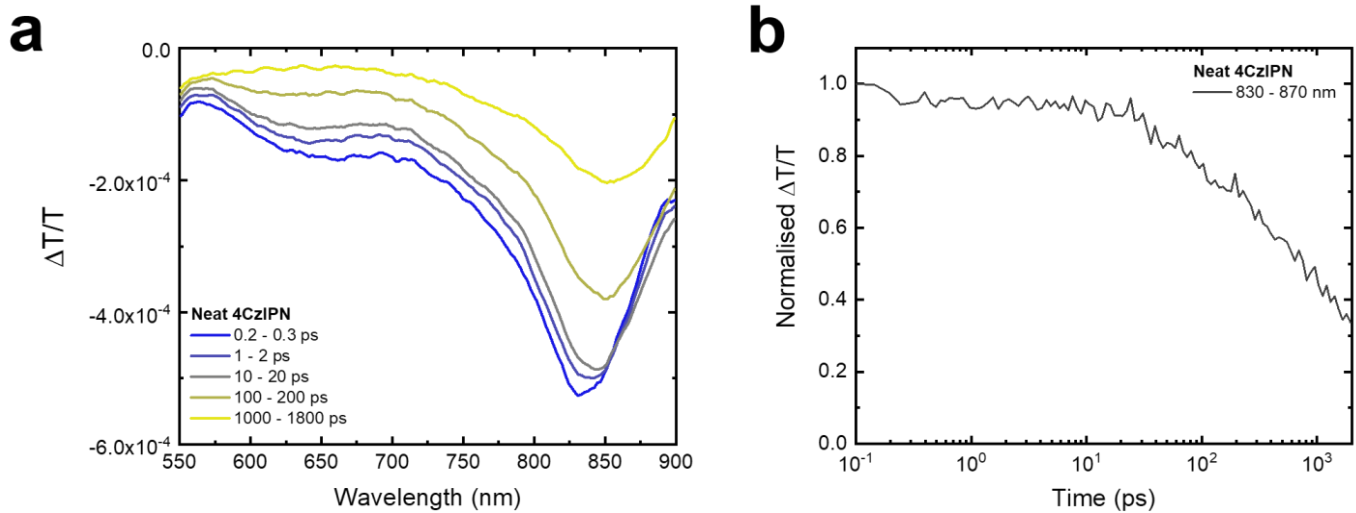

**Figure S20:** (a) The TA spectra of a neat 4CzIPN film, excited at 450 nm with a fluence of 31.8 µJ cm$^{-2}$. We see a small red shift in the 4CzIPN intra-$^1$CT PIA in the first few ps, potentially due to molecular relaxation, but no further spectral evolution on longer timescales. The 4CzIPN intra-$^1$CT PIA at 850 nm begins to decrease in intensity beyond 10 ps as recombination back to the ground state takes place. The absence of significant spectral evolution or the formation of new PIA bands that can be attributed to electron or hole polarons, as observed in a neat BF2 film (Fig. 2a), suggests that intermolecular charge transfer does not readily occur in a neat 4CzIPN film. (b) The TA kinetics of the intra-$^1$CT PIA (830 – 870 nm).

# Computational details

**Screened range separated hybrid approach**

Isolated BF2 and 4CzIPN molecules (monomers) were optimised at the DFT level with the exchange-correlation B3LYP[24] functional and the 6-31G(d) basis set for all the atomic species. To model dimers, we took advantage of the crystallographic unit cell of a curcuminoid derivative closely-related to BF2[25] and 4CzIPN[26], reported elsewhere in the literature. In the former, the hydrogen atom of the central dioxaborine ring in the parental curcuminoid molecule was replaced with an ethoxycarbonyl group to match the BF2 molecular structure. Dimer configurations were then optimised at the ground state with DFT, using the same level of theory as above, further including the D3 version of Grimme's empirical dispersion with the Becke-Johnson damping[27]. The DFT optimised BF2 dimer was then used to build a larger aggregate, namely a BF2 tetramer, whose constituting monomers were found to be 3.5 Å apart after the optimisation. Fig. S21 shows the BF2 and 4CzIPN dimer structures, whose optimisation yielded an intermolecular distance $d_\perp$ ~3.5 Å, along with the optimised BF2 tetramer.

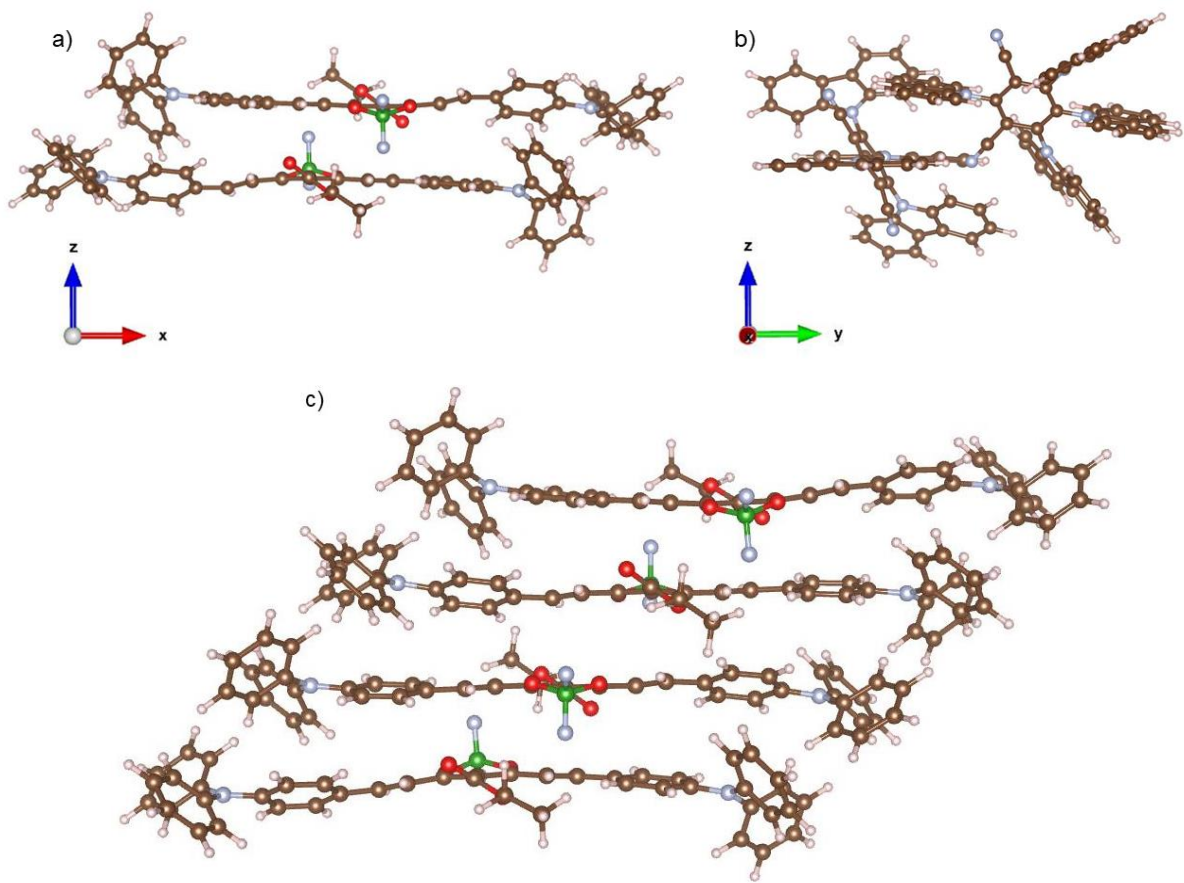

**Figure S21:** Ground state DFT optimised structure of the a) BF2 dimer, b) 4CzIPN dimer and c) BF2 tetramer.

Optoelectronic properties of the dimers were computed at the Time Dependent (TD) DFT level, resorting to the Tamm-Dancoff approximation (TDA)[28]. Singlet and triplet energies, as well as oscillator strengths, were computed as a function of $d_\perp$ by fixing the position of one fragment and shifting the other along the *z* axis. To take into account solid-state screening effects of the surrounding polarizable environment in our calculations, we exploited a *screened* range separated hybrid (SRSH) functional approach[29,30]. In this state-of-the-art method, the Coulomb potential is partitioned into a short-range (SR) and a long-range (LR) contributions, where the asymptotic behaviour of the inter-electron potential tends to $\frac{1}{\varepsilon r}$, so that in the LR domain it is properly screened by the condensed phase scalar dielectric constant ε. In addition, $\frac{1}{\varepsilon} = \alpha + \beta$, where $\alpha + \beta$ quantifies the fraction of exact Hartree-Fock (HF) exchange included in the LR part, while *α* quantifies the fraction of HF exchange included in the SR domain. In these calculations, as implemented in Gaussian16[31] and Q-Chem[32] programs, the LC-ωhPBE functional[33] was used along with the 6-311+G(d,p) basis set. Initially, the range-separation parameter ω was optimally tuned in gas-phase for each previously optimized monomer and dimer. The BF2 monomer yielded an ω value of 0.113 Bohr$^{-1}$, while the dimer gave an ω value of 0.086 Bohr$^{-1}$. The 4CzIPN monomer yielded an ω value of 0.107 Bohr$^{-1}$, while the dimer gave an ω value of 0.094 Bohr$^{-1}$. Then, by retaining the α value at 0.2 (typical for organic molecules) and the gas-phase ω value, a dielectric constant of ε = 3.00 was chosen and, thus, the β parameter was set at 0.1333, according to the above equation. As regards the BF2 tetramer, because of computational costs of such a larger aggregate, SRSH TDA TDDFT calculations were carried out for singlets and triplets at the LC-ωhPBE/6-31G(d) level, using the same parameters as for the BF2 dimer (*i.e.*, ω = 0.086 Bohr$^{-1}$, α = 0.2 and β = 0.1333).

**Diabatization of low-lying states**

Diabatization of the electronic transitions was performed for singlet and triplet excited states computed as described above by means of the Boys localization scheme, as implemented in the Q-Chem program. For monomers and dimers, the 6-311G(d,p) basis set was used, while the tetramer was treated with the 6-31G(d) basis set. The number of adiabatic states considered corresponds to the number of intra- and inter-CT contributions for each system. Namely, diabatization of BF2 and 4CzIPN monomers was obtained from two and four adiabatic states, respectively. For the molecular dimers, eight adiabatic states were considered for BF2 (four intra-CT and four inter-CT) and sixteen for 4CzIPN. For the BF2 tetramer, a total of twenty adiabatic states were considered (eight intra-CT and twelve inter-CT).

**Spin-Component Scaling second-order approximate Coupled-Cluster (SCS-CC2) excited states calculations**

The isolated BF2 monomer was optimised at the ground state at the DFT-PBE0 level of theory using the def2-TZVP basis set[34]. Subsequently, excited state energies were obtained at the SCS-CC2/def2-TZVP level of theory using the spin-adapted formulation of the linear response theory. All SCS-CC2 calculations were performed with Turbomole/7.4 package[35].

**BF2 dimer structure**

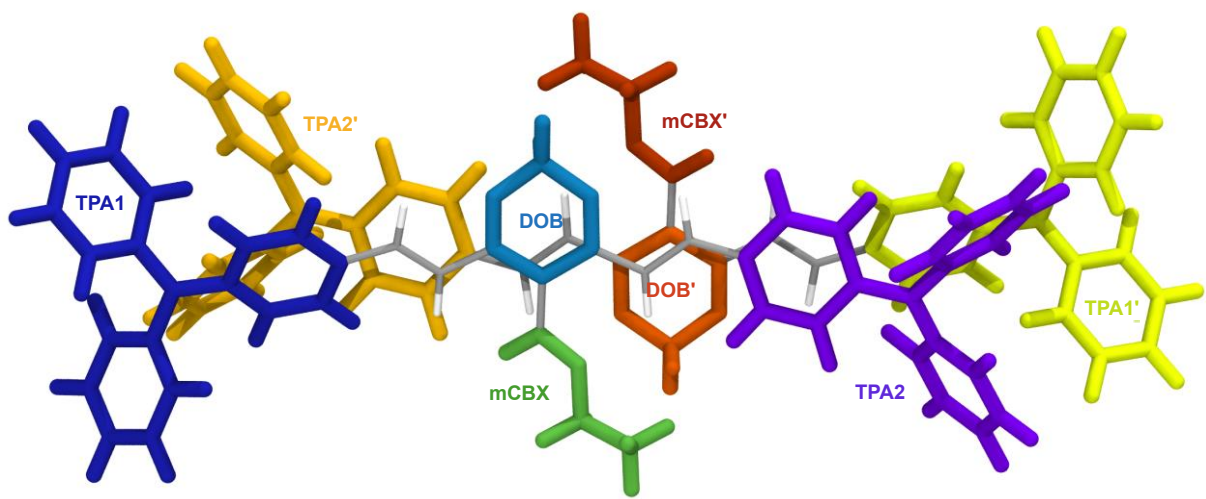

**Figure S22:** Optimised structure of the ground state BF2 dimer and nomenclature of the fragments considered in the diabatization analysis.

**BF2 monomer: singlet states**

Diabatization of singlet excited states of the BF2 monomer: $Z_1$ and $Z_2$ are $D_1 \rightarrow A$ and $D_2 \rightarrow A$ diabatic contributions, respectively. $D_1$ = TPA1; $D_2$ = TPA2; A = DOB.

| i | $S_i$ | $Z_i$ | | $\omega$ | $Z_1$ | $Z_2$ |
|---|---|---|---|---|---|---|
| 1 | 2.54 (2.22) | 2.79 | | $S_1$ | 54 | 46 |
| 2 | 2.94 (0.07) | 2.83 | | $S_2$ | 46 | 54 |

**Table S3:** (left) Energies of adiabatic ($S_i$) and diabatic ($Z_i$) states (in eV, oscillator strength in parenthesis); (right) Diabatic contributions $\omega$ (in %) as obtained from Boys diabatization for the two lowest singlet states of the BF2 monomer.

| State | TPA1 | DOB | TPA2 | mCBX |
|---|---|---|---|---|
| $Z_1$ | **0.632** | **-0.333** | -0.101 | 0.003 |
| $Z_2$ | -0.086 | **-0.326** | **0.604** | 0.003 |

**Table S4:** Relative Mulliken fragment charges of diabatic states of the BF2 monomer with respect to the ground state charge distribution.

The two lowest excited states $S_1$ and $S_2$ are the in-phase/out-of-phase combinations of two quasi-degenerate zwitterionic states $Z_1$ and $Z_2$ that correspond to the intra-CT from each TPA to the central DOB unit.

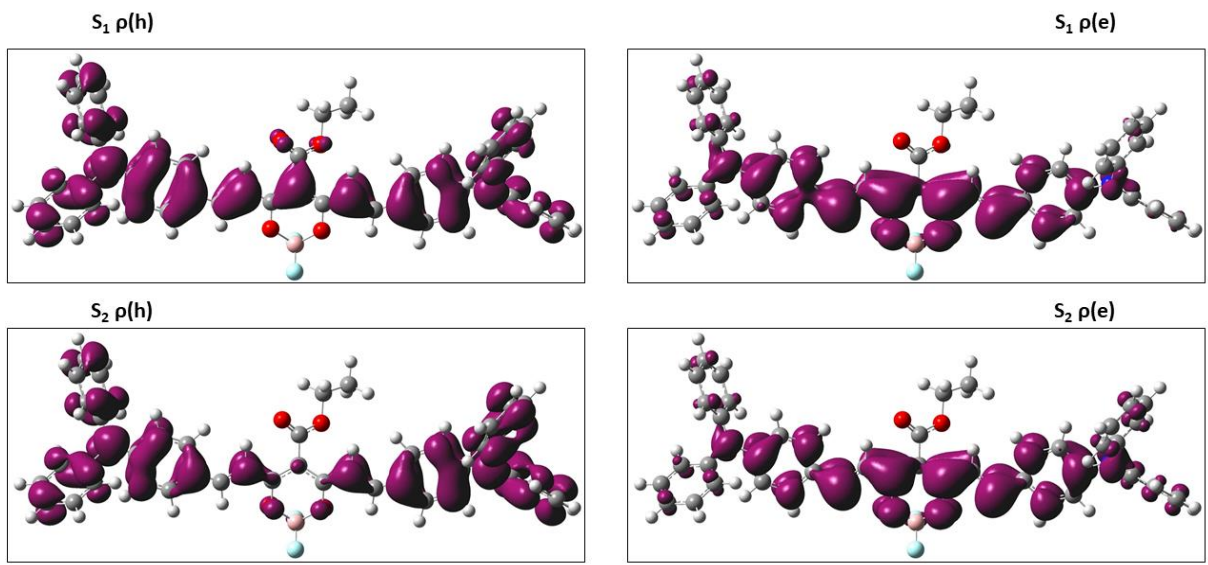

**Figure S23:** Hole ($\rho(h)$) and electron ($\rho(e)$) density plots of the two lowest excited states $S_1$ and $S_2$ of the BF2 monomer.

**BF2 monomer: $\Delta E_{ST}$**

|  | $T_1$ (eV) | $T_2$ (eV) | $S_1$ (eV) | $S_2$ (eV) |
|---|---|---|---|---|
| **SRSH TDA TDDFT** | 1.83 | 2.01 | 2.35 | 2.72 |
| **SCS-CC2** | 2.48 | 2.63 | 2.76 | 3.19 |

**Table S5:** $T_1$, $T_2$, $S_1$ and $S_2$ energies computed at the SRSH TDA TDDFT/6-311G+(d,p) and SCS-CC2/def2-TZVP levels of theory.

In addition, to obtain a more robust estimation of the singlet-triplet energy gap ($\Delta E_{ST}$), higher-level of theory SCS-CC2/def2-TZVP calculations were performed on the BF2 monomer at its ground state optimised structure. If, on the one hand, SRSH TDA TDDFT calculations yielded a $\Delta E_{ST}$ of 0.52 eV (with $S_1$ at 2.35 eV and $T_1$ at 1.83 eV), on the other hand, SCS-CC2 ones gave a lower $\Delta E_{ST}$ of 0.28 eV (with $S_1$ at 2.76 eV and $T_1$ at 2.48 eV), in close agreement with experiment.

**BF2 dimer, $d_\perp = 3.5$ Å: singlet states**

| i | $S_i$ | $Z_{intra}$ | $Z_{inter}$ | $Z_1$ | $Z_2$ | $Z_3$ | $Z_4$ | $Z_5$ | $Z_6$ | $Z_7$ | $Z_8$ |
|---|---|---|---|---|---|---|---|---|---|---|---|
| 1 | 1.98 (0.00) | 41 | 59 | 24 | 23 | 6 | 6 | 10 | 10 | 10 | 10 |
| 2 | 2.07 (0.03) | 4 | 96 | 45 | 45 | 3 | 3 | 1 | 1 | 1 | 1 |
| 3 | 2.23 (0.00) | 51 | 49 | 24 | 24 | 0 | 0 | 15 | 15 | 11 | 11 |
| 4 | 2.38 (2.04) | 66 | 34 | 0 | 0 | 17 | 17 | 5 | 5 | 28 | 28 |
| 5 | 2.44 (2.42) | 57 | 43 | 0 | 0 | 21 | 21 | 27 | 27 | 2 | 2 |
| 6 | 2.49 (0.00) | 16 | 84 | 2 | 2 | 40 | 40 | 8 | 8 | 0 | 0 |
| 7 | 2.69 (0.03) | 73 | 27 | 4 | 4 | 9 | 9 | 17 | 17 | 20 | 19 |
| 8 | 2.73 (0.00) | 92 | 8 | 0 | 0 | 4 | 4 | 17 | 17 | 29 | 30 |

**Table S6:** Energies (in eV) of adiabatic singlet ($S_i$) states (oscillator strength in parenthesis) and their diabatic composition in terms of intra (in blue) and intermolecular (in red) contributions and $Z_i$ states: $Z_{inter} = Z_1 + Z_2 + Z_3 + Z_4$ and $Z_{intra} = Z_5 + Z_6 + Z_7 + Z_8$.

The four lowest-energy diabatic states correspond to two pairs of degenerate states ($Z_1/Z_2$ and $Z_3/Z_4$) with inter-CT character. Symmetry breaking between the two TPA groups of BF2 because of the slip of one monomer with respect to the other along the long molecular axis lifts the degeneracy between $Z_{1/2}$ and $Z_{3/4}$. The nearly degenerate $Z_5$-$Z_8$ states on the other hand correspond to intra-CT excitations. The computed adiabatic excited states all exhibit mixing of these inter- and intra-CT diabatic states. Significant inter-CT and strong inter/intra coupling suggests that Kasha's model breaks down here.

| state | ΔE | TPA1 | DOB | TPA2 | mCBX | TPA1' | DOB' | TPA2' | mCBX' |
|---|---|---|---|---|---|---|---|---|---|
| $Z_1$ | 2.12 | 0.001 | 0.071 | **0.770** | 0.004 | -0.158 | **-0.409** | -0.092 | -0.002 |
| $Z_2$ | 2.13 | -0.159 | **-0.409** | -0.091 | -0.002 | 0.001 | 0.071 | **0.770** | 0.004 |
| $Z_3$ | 2.44 | -0.101 | **-0.406** | -0.160 | -0.002 | **0.765** | 0.072 | -0.004 | 0.004 |
| $Z_4$ | 2.45 | **0.765** | 0.072 | -0.004 | 0.004 | -0.100 | **-0.406** | -0.161 | -0.002 |
| $Z_5$ | 2.45 | -0.128 | **-0.328** | **0.654** | 0.002 | 0.006 | -0.001 | -0.018 | 0.000 |
| $Z_6$ | 2.45 | 0.006 | -0.001 | -0.018 | 0.000 | -0.126 | **-0.328** | **0.652** | 0.002 |
| $Z_7$ | 2.48 | -0.003 | -0.003 | 0.005 | -0.002 | **0.644** | **-0.348** | -0.100 | 0.001 |
| $Z_8$ | 2.48 | **0.643** | **-0.348** | -0.099 | 0.001 | -0.003 | -0.003 | 0.005 | -0.002 |

**Table S7:** Excitation energy (in eV) and relative Mulliken fragment charges of diabatic states of the BF2 dimer with respect to the ground state charge distribution.

Although states $S_1$ and $S_3$ involve the same diabatic states with similar weight, they differ by the sign of the amplitude of $Z_1$ and $Z_2$ contributions. $S_2$ shows a quasi-pure inter-CT character.

Bright states $S_4$ and $S_5$ are expressed as a combination of inter-CT excitations (from TPA1' to DOB and TPA1 to DOB', $Z_3$ and $Z_4$ respectively) with a mixing of two intra-CT on each molecule, namely $Z_5/Z_6$ and $Z_7/Z_8$ (mixing of $Z_1$ and $Z_2$ of the monomer).

|       | e-h radius |       | e-h radius | $Z_{intra}$ | $Z_{inter}$ |
|-------|------------|-------|------------|-------------|-------------|
| $Z_1$ | 4.589      | $S_1$ | 6.178      | 41          | 59          |
| $Z_2$ | 4.579      | $S_2$ | 4.973      | 4           | 96          |
| $Z_3$ | 9.776      | $S_3$ | 5.867      | 51          | 49          |
| $Z_4$ | 9.755      | $S_4$ | 7.818      | 66          | 34          |
| $Z_5$ | 7.230      | $S_5$ | 8.246      | 57          | 43          |
| $Z_6$ | 7.200      | $S_6$ | 9.146      | 16          | 84          |
| $Z_7$ | 6.800      | $S_7$ | 7.315      | 73          | 27          |
| $Z_8$ | 6.774      | $S_8$ | 7.161      | 92          | 8           |

**Table S8:** Electron-hole (e-h) radius (in Å) for the diabatic (left) and adiabatic (right) singlet states. Intra and intermolecular contributions are indicated in blue and red, respectively.

**BF2 dimer, $d_\perp$ = 3.5 Å: intra-$^1$CT/inter-$^1$CT couplings**

In order to stablish the intra-$^1$CT/inter-$^1$CT relationship in the BF2 dimer, we transform the $Z_i$ basis of diabatic states into a new basis of intra and inter diabats by diagonalizing the respective matrix blocks. Table S9 shows diabatic energies and electronic couplings between the two lowest intra-$^1$CT and inter-$^1$CT states.

|                | intra-$^1$CT$_1$ | intra-$^1$CT$_2$ | inter-$^1$CT$_1$ | inter-$^1$CT$_2$ |
|----------------|------------------|------------------|------------------|------------------|
| intra-$^1$CT$_1$ | 2.151          | 0                | -131             | 17               |
| intra-$^1$CT$_2$ | 0              | 2.402            | 1                | 3                |
| inter-$^1$CT$_1$ | -131           | 1                | 2.089            | 0                |
| inter-$^1$CT$_2$ | 17             | 3                | 0                | 2.093            |

**Table S9:** Diabatic energies (diagonal values in eV) and electronic couplings (off-diagonal values in meV) between the lowest intra-$^1$CT and inter-$^1$CT states.

Gap between the lowest intra-$^1$CT and inter-$^1$CT states: $\Delta E$ = 2.151-2.089 = 0.062 eV.

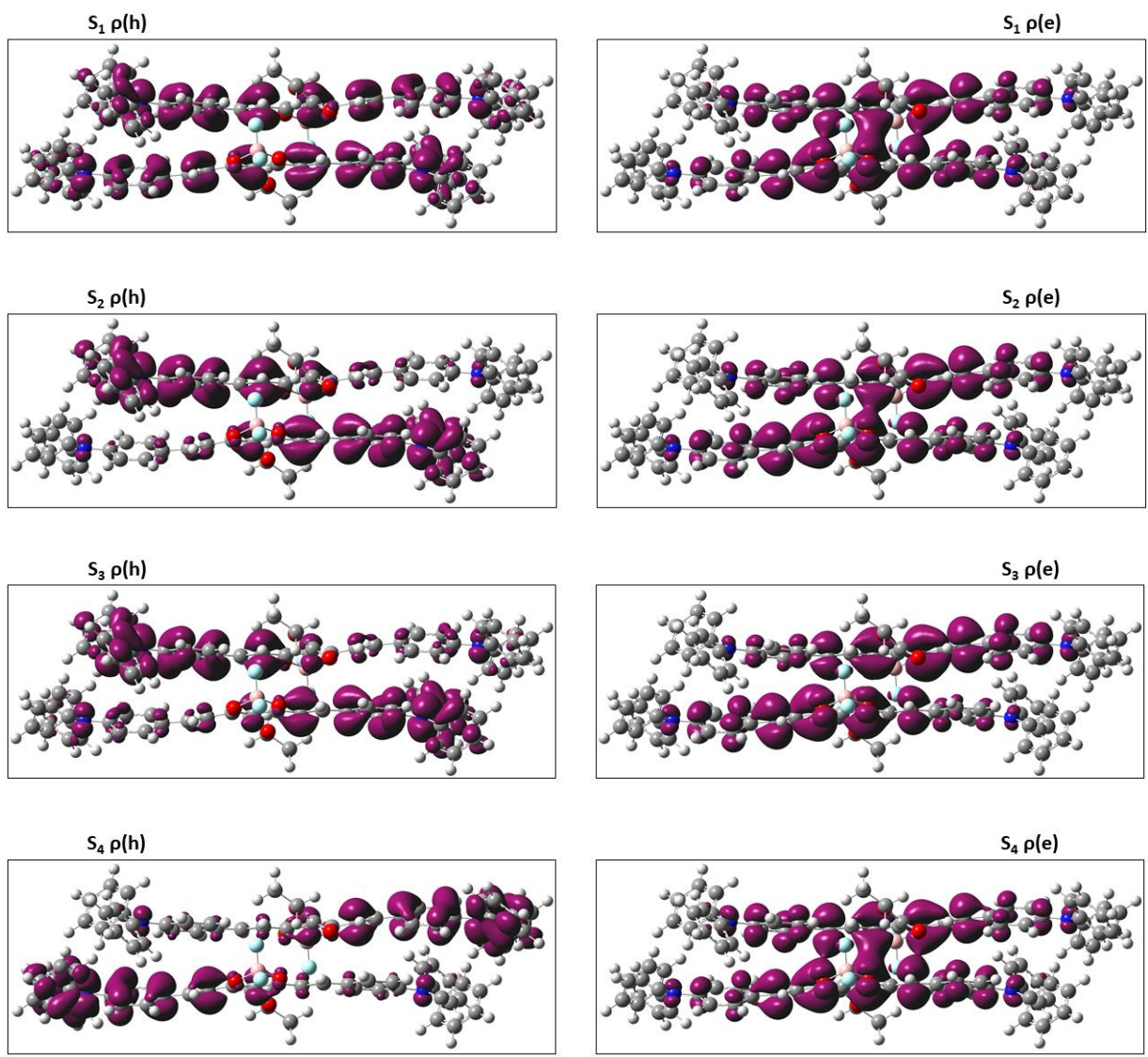

**Figure S24:** Hole (ρ(h)) and electron (ρ(e)) density plots from $S_1$ to $S_4$ of the BF2 dimer at $d_\perp$ = 3.5 Å.

**BF2 dimer, $d_\perp = 3.5$ Å: triplet states**

| i | $T_i$ | $Z_{intra}$ | $Z_{inter}$ | $Z_1$ | $Z_2$ | $Z_3$ | $Z_4$ | $Z_5$ | $Z_6$ | $Z_7$ | $Z_8$ |
|---|------|-----|-----|----|----|----|----|----|----|----|----|
| 1 | 1.65 | 81 | 19 | 32 | 31 | 9 | 9 | 7 | 7 | 2 | 2 |
| 2 | 1.75 | 98 | 2 | 39 | 40 | 10 | 10 | 1 | 1 | 0 | 0 |
| 3 | 1.91 | 62 | 38 | 4 | 4 | 27 | 27 | 19 | 19 | 0 | 0 |
| 4 | 2.01 | 98 | 2 | 12 | 12 | 37 | 37 | 0 | 0 | 1 | 1 |
| 5 | 2.12 | 34 | 66 | 6 | 6 | 11 | 11 | 25 | 24 | 9 | 9 |
| 6 | 2.16 | 17 | 83 | 7 | 7 | 2 | 2 | 39 | 41 | 2 | 2 |
| 7 | 2.45 | 4 | 96 | 0 | 0 | 2 | 2 | 3 | 3 | 46 | 44 |
| 8 | 2.50 | 5 | 95 | 0 | 0 | 2 | 2 | 6 | 6 | 40 | 42 |

**Table S10:** Energies (in eV) of adiabatic triplet ($T_i$) states and their diabatic composition in terms of intra (in blue) and intermolecular (in red) contributions and $Z_i$ states: $Z_{intra} = Z_1 + Z_2 + Z_3 + Z_4$ and $Z_{inter} = Z_5 + Z_6 + Z_7 + Z_8$.

| State | TPA1 | DOB | TPA2 | mCBX | TPA1' | DOB' | TPA2' | mCBX' |
|-------|------|------|------|------|-------|------|-------|-------|
| $Z_1$ | 0.006 | 0.005 | -0.006 | -0.001 | -0.052 | **-0.258** | **0.392** | 0.005 |
| $Z_2$ | -0.052 | **-0.259** | **0.395** | 0.005 | 0.006 | 0.005 | -0.006 | -0.001 |
| $Z_3$ | -0.002 | -0.003 | -0.002 | -0.001 | **0.386** | **-0.228** | -0.039 | 0.000 |
| $Z_4$ | **0.386** | **-0.228** | -0.039 | 0.000 | -0.002 | -0.003 | -0.002 | -0.001 |
| $Z_5$ | 0.001 | 0.070 | **0.753** | 0.003 | -0.155 | **-0.403** | -0.089 | -0.002 |
| $Z_6$ | -0.155 | **-0.402** | -0.089 | -0.002 | 0.001 | 0.070 | **0.753** | 0.003 |
| $Z_7$ | -0.086 | **-0.378** | -0.126 | -0.002 | **0.701** | 0.062 | -0.018 | 0.005 |
| $Z_8$ | **0.700** | 0.062 | -0.017 | 0.005 | -0.085 | **-0.378** | -0.126 | -0.002 |

**Table S11:** Relative Mulliken fragment charges of diabatic states of the BF2 dimer with respect to the ground state charge distribution.

| | e-h radius | | e-h radius | $Z_{intra}$ | $Z_{inter}$ |
|---|---|---|---|---|---|
| $Z_1$ | 3.763 | $T_1$ | 4.085 | 81 | 19 |
| $Z_2$ | 3.786 | $T_2$ | 3.764 | 98 | 2 |
| $Z_3$ | 3.537 | $T_3$ | 3.913 | 62 | 38 |
| $Z_4$ | 3.527 | $T_4$ | 3.701 | 98 | 2 |
| $Z_5$ | 4.479 | $T_5$ | 4.991 | 34 | 66 |
| $Z_6$ | 4.466 | $T_6$ | 4.494 | 17 | 83 |
| $Z_7$ | 9.108 | $T_7$ | 8.596 | 4 | 96 |
| $Z_8$ | 9.085 | $T_8$ | 8.208 | 5 | 95 |

**Table S12:** Electron-hole (e-h) radius (in Å) for the diabatic (left) and adiabatic (right) triplet states. Intra and intermolecular contributions are indicated in blue and red, respectively.

**BF2 dimer, $d_\perp$ = 3.5 Å: intra-$^3$CT/inter-$^3$CT couplings**

To establish the intra-$^3$CT/inter-$^3$CT relationship in the BF2 dimer, we transform the $Z_i$ basis of diabatic states into a new basis of intra and inter diabats by diagonalizing the respective matrix blocks. Table S13 shows diabatic energies and electronic couplings between the two lowest intra-$^3$CT and inter-$^3$CT states.

|  | intra-$^3$CT$_1$ | intra-$^3$CT$_2$ | inter-$^3$CT$_1$ | inter-$^3$CT$_2$ |
|---|---|---|---|---|
| intra-$^3$CT$_1$ | 1.751 | 0 | 0 | 194 |
| intra-$^3$CT$_2$ | 0 | 1.755 | 32 | 24 |
| inter-$^3$CT$_1$ | 0 | 32 | 2.040 | 0 |
| inter-$^3$CT$_2$ | 194 | 24 | 0 | 2.070 |

**Table S13:** Diabatic energies (diagonal values in eV) and electronic couplings (off-diagonal values in meV) between the lowest intra-$^3$CT and inter-$^3$CT states.

Gap between the lowest intra-$^3$CT and inter-$^3$CT states: $\Delta E$ = 2.040-1.751 = 0.289 eV

## BF2 dimer, distance dependence

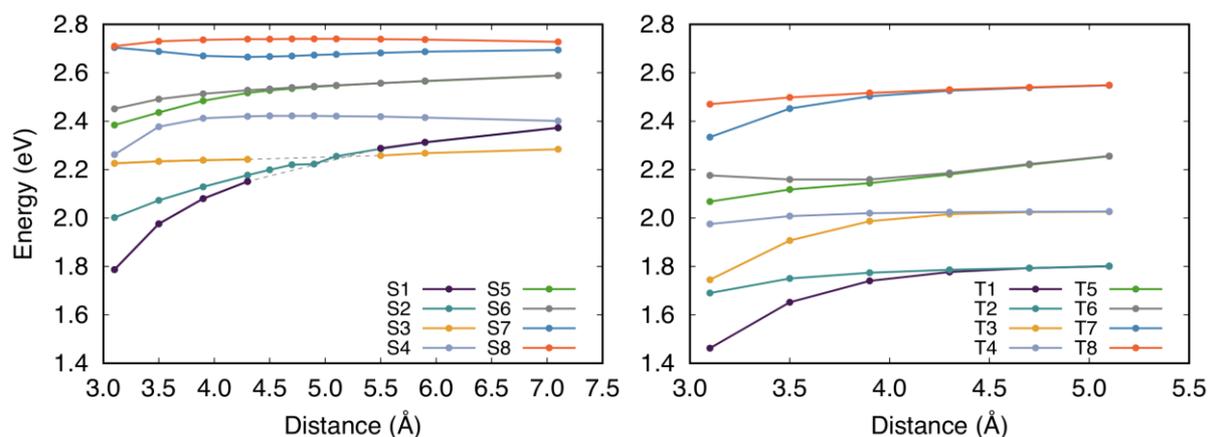

**Figure S25:** Adiabatic singlet (left) and triplet (right) state energies as a function of the distance between the BF2 molecules.

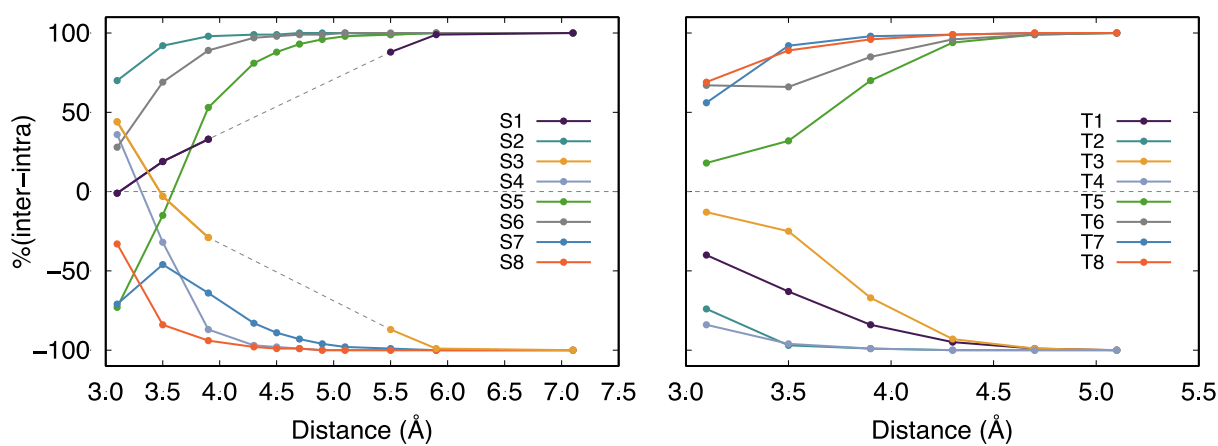

**Figure S26** Adiabatic singlet (left) and triplet (right) state wave function as a function of the distance between the BF2 molecules.

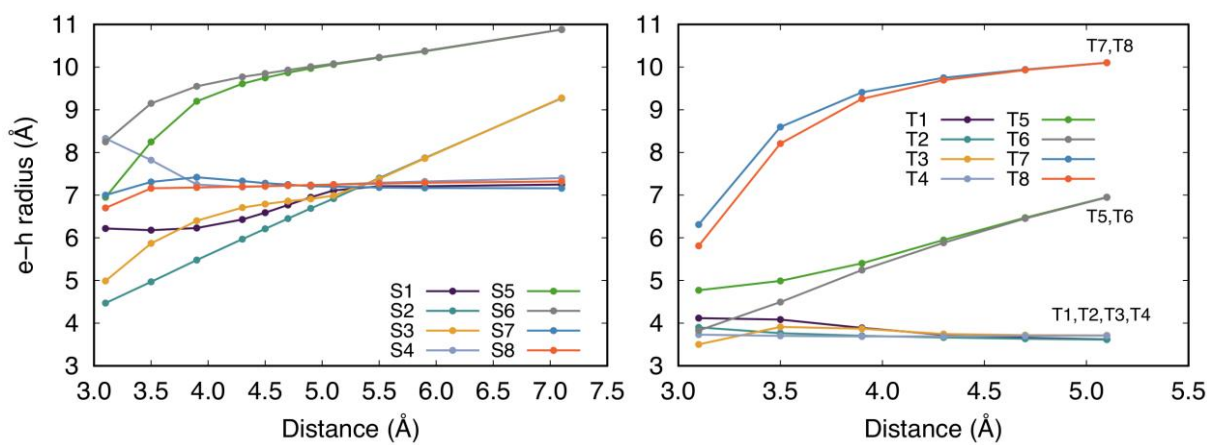

**Figure S27:** Adiabatic singlet (left) and triplet (right) electron-hole (e-h) radius (in Å) as a function of the distance between the BF2 molecules.

**BF2 tetramer structure**

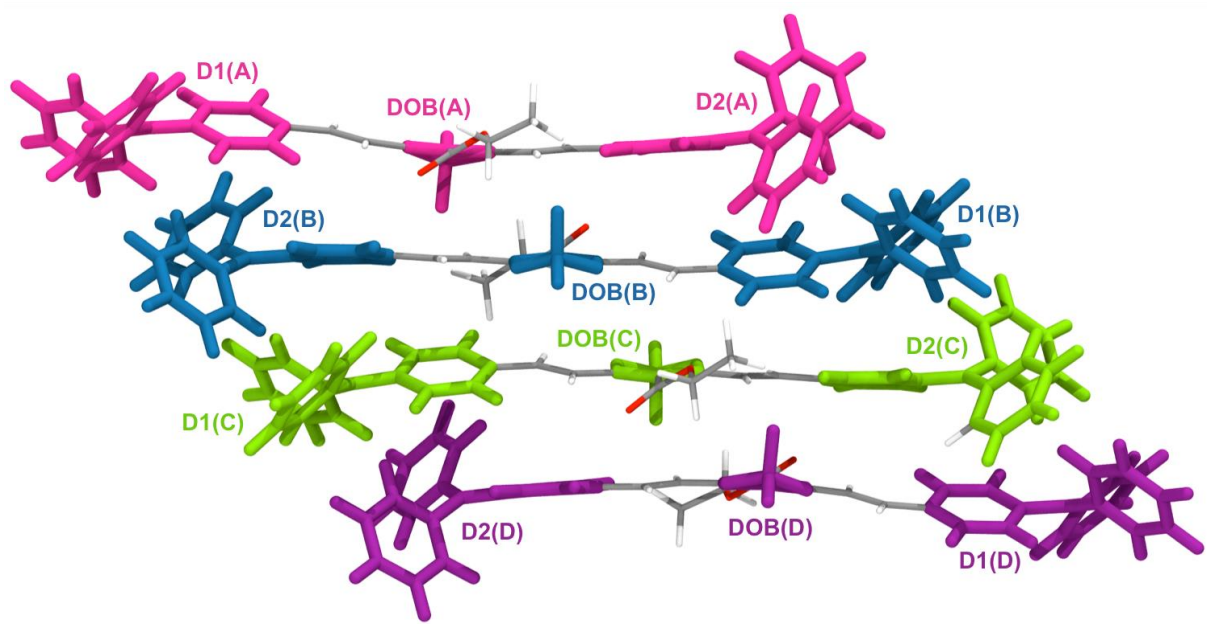

**Figure S28:** Optimised structure of the ground state BF2 tetramer and nomenclature of the fragments considered in the diabatization analysis (D1=TPA1, D2=TPA2).

**BF2 tetramer: singlet states**

| i | $S_i$ | $Z_{intra}$ | $Z_{inter}$ |
|---|---|---|---|
| 1 | 1.87 (0.00) | 17 | 83 |
| 2 | 1.99 (0.02) | 8 | 92 |
| 3 | 2.03 (0.01) | 5 | 95 |
| 4 | 2.05 (0.00) | 2 | 98 |
| 5 | 2.13 (0.04) | 5 | 95 |
| 6 | 2.13 (0.09) | 6 | 94 |
| 7 | 2.19 (0.00) | 36 | 64 |
| 8 | 2.20 (0.03) | 53 | 47 |
| 9 | 2.28 (0.00) | 58 | 42 |
| 10 | 2.29 (0.00) | 1 | 99 |
| 11 | 2.33 (0.02) | 7 | 93 |
| 12 | 2.35 (0.00) | 24 | 76 |
| 13 | 2.36 (0.03) | 9 | 91 |
| 14 | 2.36 (0.13) | 8 | 92 |
| 15 | 2.39 (0.00) | 19 | 81 |
| 16 | 2.44 (0.71) | 20 | 80 |
| 17 | 2.47 (0.01) | 13 | 87 |
| 18 | 2.49 (0.80) | 12 | 88 |
| 19 | 2.51 (0.17) | 19 | 81 |
| 20 | 2.53 (4.68) | 77 | 23 |

**Table S14:** Energies (in eV) of adiabatic singlet ($S_i$) states (oscillator strength in parenthesis) and their diabatic composition in terms of intra (in blue) and intermolecular (in red) contributions and $Z_i$ states: $Z_{inter} = Z_{1-8} + Z_{11-16} + Z_{19-20}$ and $Z_{intra} = Z_9 + Z_{10} + Z_{17} + Z_{18}$.

| | e-h radius | | e-h radius | $Z_{intra}$ | $Z_{inter}$ |
|---|---|---|---|---|---|
| $Z_1$ | 4.242 | $S_1$ | 5.309 | 17 | 83 |
| $Z_2$ | 4.123 | $S_2$ | 5.774 | 8 | 92 |
| $Z_3$ | 4.082 | $S_3$ | 4.909 | 5 | 95 |
| $Z_4$ | 4.630 | $S_4$ | 5.565 | 2 | 98 |
| $Z_5$ | 4.271 | $S_5$ | 4.739 | 5 | 95 |
| $Z_6$ | 4.651 | $S_6$ | 4.690 | 6 | 94 |
| $Z_7$ | 9.891 | $S_7$ | 5.288 | 36 | 64 |
| $Z_8$ | 9.827 | $S_8$ | 3.783 | 53 | 47 |
| $Z_9$ | 1.307 | $S_9$ | 3.469 | 58 | 42 |
| $Z_{10}$ | 1.424 | $S_{10}$ | 7.360 | 1 | 99 |
| $Z_{11}$ | 6.922 | $S_{11}$ | 8.680 | 7 | 93 |
| $Z_{12}$ | 6.917 | $S_{12}$ | 5.398 | 24 | 76 |
| $Z_{13}$ | 6.979 | $S_{13}$ | 7.348 | 9 | 91 |
| $Z_{14}$ | 6.979 | $S_{14}$ | 6.767 | 8 | 92 |
| $Z_{15}$ | 9.011 | $S_{15}$ | 7.661 | 19 | 81 |
| $Z_{16}$ | 8.974 | $S_{16}$ | 6.971 | 20 | 80 |
| $Z_{17}$ | 1.243 | $S_{17}$ | 7.441 | 13 | 87 |
| $Z_{18}$ | 1.632 | $S_{18}$ | 7.906 | 12 | 88 |
| $Z_{19}$ | 10.970 | $S_{19}$ | 7.004 | 19 | 81 |
| $Z_{20}$ | 10.927 | $S_{20}$ | 2.937 | 77 | 23 |

**Table S15:** Electron-hole (e-h) radius (in Å) for the diabatic (left) and (right) adiabatic singlet states. Intra and intermolecular contributions are indicated in blue and red, respectively.

| state | monomer | ΔE | DOB | TPA1 | TPA2 | mCBX |
|---|---|---|---|---|---|---|
| $Z_9$ | A | 2.28 | **-0.282** | **0.129** | **0.279** | 0.006 |
| $Z_{10}$ | D | 2.28 | **-0.300** | **0.142** | **0.293** | 0.007 |
| $Z_{17}$ | B | 2.36 | **-0.278** | **0.264** | **0.137** | 0.007 |
| $Z_{18}$ | C | 2.37 | **-0.259** | **0.246** | **0.125** | 0.006 |

**Table S16:** Intra-CT diabatic states excitation energies (in eV) and relative Mulliken fragment charges of diabatic states of the BF2 tetramer with respect to the ground state charge distribution.

| state | monomer | ΔE | DOB | TPA1 | TPA2 | mCBX |
|---|---|---|---|---|---|---|
| $Z_1$ | C | 2.14 | 0.057 | 0.001 | **0.647** | 0.005 |
| | D | | **-0.379** | -0.105 | -0.093 | -0.001 |
| $Z_2$ | A | 2.14 | 0.064 | 0.029 | **0.651** | 0.006 |
| | B | | **-0.386** | -0.106 | -0.083 | 0.000 |
| $Z_3$ | D | 2.14 | 0.065 | 0.032 | **0.663** | 0.006 |
| | C | | **-0.396** | -0.116 | -0.082 | 0.000 |
| $Z_4$ | B | 2.14 | 0.037 | **0.666** | 0.004 | 0.004 |
| | C | | **-0.374** | -0.094 | -0.087 | -0.001 |
| $Z_5$ | B | 2.14 | 0.057 | 0.001 | **0.657** | 0.004 |
| | A | | **-0.383** | -0.106 | -0.094 | -0.001 |
| $Z_6$ | C | 2.15 | 0.042 | **0.674** | 0.006 | 0.004 |
| | B | | **-0.378** | -0.094 | -0.091 | -0.001 |
| $Z_7$ | C | 2.26 | 0.030 | **0.458** | -0.025 | 0.004 |
| | D | | **-0.357** | -0.086 | -0.146 | -0.001 |
| $Z_8$ | B | 2.27 | 0.038 | **0.478** | -0.023 | 0.003 |
| | A | | **-0.365** | -0.088 | -0.149 | -0.001 |
| $Z_{11}$ | B | 2.29 | 0.057 | **0.694** | 0.066 | 0.007 |
| | D | | **-0.412** | -0.140 | -0.105 | -0.002 |
| $Z_{12}$ | C | 2.30 | 0.057 | **0.696** | 0.065 | 0.007 |
| | A | | **-0.413** | -0.139 | -0.106 | -0.002 |
| $Z_{13}$ | A | 2.31 | 0.064 | 0.034 | **0.747** | 0.006 |
| | C | | **-0.418** | -0.121 | -0.119 | -0.002 |
| $Z_{14}$ | D | 2.32 | 0.064 | 0.034 | **0.745** | 0.006 |
| | B | | **-0.418** | -0.121 | -0.119 | -0.002 |
| $Z_{15}$ | B | 2.35 | 0.017 | -0.028 | **0.520** | 0.007 |
| | C | | **-0.258** | 0.008 | -0.056 | -0.001 |
| $Z_{16}$ | C | 2.35 | 0.016 | -0.029 | **0.538** | 0.007 |
| | B | | **-0.264** | -0.009 | -0.057 | -0.001 |
| $Z_{19}$ | A | 2.39 | 0.061 | 0.045 | **0.745** | 0.006 |
| | D | | **-0.420** | -0.135 | -0.122 | -0.002 |
| $Z_{20}$ | D | 2.40 | 0.060 | 0.044 | **0.744** | 0.006 |
| | A | | **-0.420** | -0.134 | -0.123 | -0.002 |

**Table S17:** Inter-CT diabatic states excitation energies (in eV) and relative Mulliken fragment charges of diabatic states of the BF2 tetramer with respect to the ground state charge distribution.

| ω | $Z_1$ | $Z_2$ | $Z_3$ | $Z_4$ | $Z_5$ | $Z_6$ | $Z_7$ | $Z_8$ | $Z_9$ | $Z_{10}$ | $Z_{11}$ | $Z_{12}$ | $Z_{13}$ | $Z_{14}$ | $Z_{15}$ | $Z_{16}$ | $Z_{17}$ | $Z_{18}$ | $Z_{19}$ | $Z_{20}$ |
|---|---|---|---|---|---|---|---|---|---|---|---|---|---|---|---|---|---|---|---|---|
| $S_1$ | 2 | 6 | 5 | 20 | 2 | 18 | 6 | 6 | 0 | 0 | 0 | 0 | 3 | 3 | 5 | 6 | 8 | 8 | 0 | 0 |
| $S_2$ | 27 | 1 | 1 | 2 | 23 | 2 | 12 | 11 | 4 | 4 | 5 | 5 | 0 | 0 | 0 | 0 | 0 | 0 | 1 | 1 |
| $S_3$ | 0 | 22 | 23 | 14 | 0 | 14 | 0 | 0 | 2 | 2 | 1 | 1 | 9 | 9 | 1 | 1 | 0 | 0 | 1 | 1 |
| $S_4$ | 29 | 2 | 3 | 2 | 30 | 1 | 5 | 7 | 0 | 0 | 6 | 6 | 2 | 2 | 2 | 2 | 1 | 1 | 1 | 1 |
| $S_5$ | 2 | 49 | 6 | 22 | 1 | 0 | 0 | 0 | 5 | 0 | 0 | 0 | 1 | 4 | 8 | 1 | 0 | 0 | 0 | 0 |
| $S_6$ | 2 | 0 | 41 | 4 | 5 | 28 | 0 | 1 | 0 | 4 | 0 | 0 | 3 | 0 | 0 | 9 | 1 | 1 | 0 | 0 |
| $S_7$ | 0 | 0 | 0 | 4 | 0 | 4 | 3 | 3 | 1 | 3 | 12 | 11 | 8 | 9 | 0 | 0 | 16 | 16 | 5 | 4 |
| $S_8$ | 3 | 3 | 3 | 1 | 3 | 1 | 1 | 2 | 20 | 22 | 4 | 6 | 8 | 5 | 3 | 3 | 7 | 4 | 0 | 0 |
| $S_9$ | 9 | 3 | 2 | 0 | 7 | 0 | 3 | 2 | 30 | 27 | 8 | 4 | 0 | 0 | 1 | 1 | 0 | 1 | 3 | 1 |
| $S_{10}$ | 8 | 0 | 0 | 7 | 9 | 6 | 3 | 4 | 0 | 1 | 14 | 16 | 1 | 1 | 2 | 2 | 0 | 0 | 13 | 12 |
| $S_{11}$ | 0 | 0 | 0 | 0 | 1 | 0 | 24 | 26 | 4 | 2 | 8 | 5 | 0 | 0 | 10 | 11 | 0 | 0 | 4 | 3 |
| $S_{12}$ | 9 | 2 | 1 | 1 | 6 | 3 | 8 | 3 | 12 | 10 | 11 | 13 | 11 | 6 | 1 | 0 | 1 | 1 | 0 | 1 |
| $S_{13}$ | 0 | 7 | 1 | 5 | 3 | 2 | 6 | 10 | 1 | 6 | 11 | 0 | 21 | 1 | 0 | 1 | 2 | 0 | 19 | 4 |
| $S_{14}$ | 4 | 1 | 8 | 0 | 3 | 3 | 3 | 0 | 6 | 1 | 3 | 16 | 4 | 30 | 1 | 0 | 0 | 0 | 1 | 16 |
| $S_{15}$ | 2 | 0 | 0 | 4 | 1 | 4 | 15 | 15 | 0 | 1 | 1 | 2 | 0 | 0 | 9 | 6 | 10 | 8 | 10 | 11 |
| $S_{16}$ | 3 | 1 | 1 | 4 | 3 | 3 | 3 | 6 | 7 | 10 | 0 | 0 | 0 | 0 | 30 | 25 | 0 | 2 | 1 | 0 |
| $S_{17}$ | 0 | 0 | 0 | 8 | 0 | 7 | 4 | 3 | 3 | 1 | 0 | 0 | 3 | 1 | 21 | 31 | 6 | 3 | 6 | 2 |
| $S_{18}$ | 0 | 3 | 0 | 1 | 0 | 1 | 1 | 0 | 1 | 3 | 10 | 4 | 16 | 7 | 1 | 0 | 0 | 8 | 27 | 16 |
| $S_{19}$ | 0 | 2 | 3 | 1 | 0 | 2 | 0 | 1 | 1 | 0 | 4 | 11 | 7 | 15 | 2 | 1 | 16 | 2 | 8 | 23 |
| $S_{20}$ | 1 | 0 | 1 | 0 | 1 | 0 | 2 | 2 | 1 | 3 | 0 | 0 | 2 | 5 | 1 | 1 | 31 | 43 | 1 | 3 |

**Table S18:** Diabatic contributions ω (in %) as obtained from Boys diabatization.

**BF2 tetramer: triplet states**

| i | $T_i$ | $Z_{intra}$ | $Z_{inter}$ |
|---|------|------|------|
| 1 | 1.61 | 79 | 21 |
| 2 | 1.68 | 83 | 17 |
| 3 | 1.73 | 94 | 6 |
| 4 | 1.78 | 98 | 2 |
| 5 | 1.86 | 54 | 46 |
| 6 | 1.92 | 48 | 52 |
| 7 | 2.00 | 61 | 39 |
| 8 | 2.03 | 95 | 5 |
| 9 | 2.09 | 46 | 54 |
| 10 | 2.10 | 43 | 57 |
| 11 | 2.12 | 16 | 84 |
| 12 | 2.12 | 23 | 77 |
| 13 | 2.13 | 28 | 72 |
| 14 | 2.18 | 6 | 94 |
| 15 | 2.30 | 3 | 97 |
| 16 | 2.34 | 5 | 95 |
| 17 | 2.35 | 0 | 100 |
| 18 | 2.36 | 2 | 98 |
| 19 | 2.37 | 7 | 93 |
| 20 | 2.38 | 10 | 90 |

**Table S19:** Energies (in eV) of adiabatic triplet ($T_i$) states and their diabatic composition in terms of intra (in blue) and intermolecular (in red) contributions and $Z_i$ states: $Z_{inter} = Z_{9\text{-}20}$ and $Z_{intra} = Z_{1\text{-}8}$.

|  | e-h radius |  | e-h radius | $Z_{intra}$ | $Z_{inter}$ |
| --- | --- | --- | --- | --- | --- |
| $Z_1$ | 3.969 | $T_1$ | 3.918 | 79 | 21 |
| $Z_2$ | 3.972 | $T_2$ | 4.070 | 83 | 17 |
| $Z_3$ | 3.420 | $T_3$ | 3.895 | 94 | 6 |
| $Z_4$ | 3.438 | $T_4$ | 3.633 | 98 | 2 |
| $Z_5$ | 3.616 | $T_5$ | 3.938 | 54 | 46 |
| $Z_6$ | 3.707 | $T_6$ | 4.136 | 48 | 52 |
| $Z_7$ | 3.443 | $T_7$ | 3.887 | 61 | 39 |
| $Z_8$ | 3.455 | $T_8$ | 3.569 | 95 | 5 |
| $Z_9$ | 4.045 | $T_9$ | 5.459 | 46 | 54 |
| $Z_{10}$ | 4.026 | $T_{10}$ | 5.321 | 43 | 57 |
| $Z_{11}$ | 4.170 | $T_{11}$ | 4.263 | 16 | 84 |
| $Z_{12}$ | 3.886 | $T_{12}$ | 4.644 | 23 | 77 |
| $Z_{13}$ | 3.877 | $T_{13}$ | 4.352 | 28 | 72 |
| $Z_{14}$ | 4.171 | $T_{14}$ | 4.635 | 6 | 94 |
| $Z_{15}$ | 9.313 | $T_{15}$ | 7.904 | 3 | 97 |
| $Z_{16}$ | 9.476 | $T_{16}$ | 6.168 | 5 | 95 |
| $Z_{17}$ | 8.870 | $T_{17}$ | 7.487 | 0 | 100 |
| $Z_{18}$ | 8.765 | $T_{18}$ | 6.888 | 2 | 98 |
| $Z_{19}$ | 6.934 | $T_{19}$ | 7.947 | 7 | 93 |
| $Z_{20}$ | 6.905 | $T_{20}$ | 7.344 | 10 | 90 |

**Table S20:** Electron-hole (e-h) radius (in Å) for the diabatic (left) and adiabatic (right) triplet states. Intra and intermolecular contributions are indicated in blue and red, respectively.

|  |  | A | | B | | C | | D | |
|---|---|---|---|---|---|---|---|---|---|
| state | $\Delta E$ | e- | h+ | e- | h+ | e- | h+ | e- | h+ |
| $Z_1$ | 1.81 | 0.000 | 0.000 | -0.004 | 0.002 | -0.021 | 0.028 | **-0.973** | **0.970** |
| $Z_2$ | 1.81 | **-0.973** | **0.969** | -0.021 | 0.028 | -0.004 | 0.002 | 0.000 | 0.000 |
| $Z_3$ | 1.85 | -0.021 | 0.018 | **-0.925** | **0.932** | -0.051 | 0.048 | -0.003 | 0.002 |
| $Z_4$ | 1.85 | -0.003 | 0.002 | -0.050 | 0.052 | **-0.926** | **0.929** | -0.020 | 0.017 |
| $Z_5$ | 1.91 | -0.008 | 0.004 | -0.057 | 0.032 | **-0.910** | **0.924** | -0.023 | 0.039 |
| $Z_6$ | 1.92 | -0.023 | 0.043 | **-0.900** | **0.920** | -0.067 | 0.033 | -0.008 | 0.003 |
| $Z_7$ | 2.00 | 0.000 | 0.000 | -0.001 | 0.000 | -0.026 | 0.020 | **-0.971** | **0.979** |
| $Z_8$ | 2.00 | **-0.969** | **0.977** | -0.028 | 0.021 | -0.001 | 0.000 | 0.000 | 0.000 |
| $Z_9$ | 2.10 | -0.034 | 0.816 | **-0.783** | 0.141 | -0.135 | 0.041 | -0.046 | 0.001 |
| $Z_{10}$ | 2.10 | -0.035 | 0.001 | -0.141 | 0.031 | **-0.787** | 0.151 | -0.036 | **0.816** |
| $Z_{11}$ | 2.11 | -0.004 | 0.027 | -0.037 | 0.061 | -0.075 | **0.887** | **-0.883** | 0.025 |
| $Z_{12}$ | 2.11 | -0.088 | 0.003 | **-0.773** | 0.068 | -0.128 | **0.727** | -0.010 | 0.201 |
| $Z_{13}$ | 2.11 | -0.011 | 0.193 | -0.119 | **0.744** | **-0.782** | 0.059 | -0.085 | 0.002 |
| $Z_{14}$ | 2.11 | **-0.875** | 0.027 | -0.087 | **0.887** | -0.033 | 0.068 | -0.004 | 0.017 |
| $Z_{15}$ | 2.27 | **-0.857** | 0.036 | -0.115 | **0.711** | -0.024 | 0.240 | -0.003 | 0.013 |
| $Z_{16}$ | 2.27 | -0.002 | 0.013 | -0.016 | 0.225 | -0.103 | **0.725** | **-0.878** | 0.036 |
| $Z_{17}$ | 2.27 | -0.001 | **0.777** | -0.015 | **0.214** | **-0.296** | 0.008 | **-0.687** | 0.000 |
| $Z_{18}$ | 2.28 | **-0.648** | 0.000 | **-0.333** | 0.007 | -0.016 | **0.197** | -0.002 | **0.795** |
| $Z_{19}$ | 2.29 | -0.001 | **0.318** | -0.011 | **0.648** | **-0.314** | 0.033 | **-0.672** | 0.000 |
| $Z_{20}$ | 2.30 | **-0.701** | 0.001 | **-0.285** | 0.035 | -0.010 | **0.679** | -0.002 | **0.284** |

**Table S21:** Diabatic states excitation energies (in eV) and e-/h+ contributions to the relative Mulliken fragment charges of diabatic states of the BF2 tetramer with respect to the ground state charge distribution.

| $\omega$ | $Z_1$ | $Z_2$ | $Z_3$ | $Z_4$ | $Z_5$ | $Z_6$ | $Z_7$ | $Z_8$ | $Z_9$ | $Z_{10}$ | $Z_{11}$ | $Z_{12}$ | $Z_{13}$ | $Z_{14}$ | $Z_{15}$ | $Z_{16}$ | $Z_{17}$ | $Z_{18}$ | $Z_{19}$ | $Z_{20}$ |
|---|---|---|---|---|---|---|---|---|---|---|---|---|---|---|---|---|---|---|---|---|
| $T_1$ | 4 | 5 | 19 | 18 | 16 | 16 | 1 | 1 | 2 | 1 | 2 | 5 | 5 | 2 | 2 | 2 | 0 | 0 | 0 | 0 |
| $T_2$ | 24 | 25 | 7 | 8 | 5 | 5 | 5 | 5 | 3 | 3 | 3 | 0 | 0 | 3 | 2 | 2 | 0 | 0 | 0 | 0 |
| $T_3$ | 34 | 31 | 4 | 4 | 3 | 3 | 8 | 7 | 1 | 1 | 0 | 1 | 1 | 0 | 0 | 1 | 0 | 0 | 0 | 0 |
| $T_4$ | 12 | 14 | 22 | 22 | 11 | 10 | 3 | 4 | 1 | 1 | 0 | 0 | 0 | 0 | 0 | 0 | 0 | 0 | 0 | 0 |
| $T_5$ | 1 | 1 | 12 | 12 | 13 | 12 | 1 | 1 | 1 | 1 | 9 | 11 | 11 | 10 | 0 | 0 | 1 | 1 | 0 | 0 |
| $T_6$ | 2 | 3 | 9 | 8 | 9 | 9 | 4 | 4 | 3 | 3 | 18 | 2 | 2 | 18 | 1 | 2 | 0 | 0 | 2 | 2 |
| $T_7$ | 5 | 5 | 3 | 3 | 2 | 2 | 21 | 20 | 13 | 12 | 3 | 3 | 3 | 3 | 1 | 1 | 0 | 0 | 0 | 0 |
| $T_8$ | 4 | 4 | 9 | 9 | 10 | 9 | 24 | 26 | 2 | 2 | 0 | 0 | 0 | 0 | 0 | 0 | 0 | 0 | 0 | 0 |
| $T_9$ | 0 | 1 | 0 | 1 | 11 | 2 | 21 | 9 | 4 | 1 | 13 | 0 | 0 | 3 | 6 | 14 | 8 | 2 | 3 | 1 |
| $T_{10}$ | 0 | 0 | 2 | 0 | 5 | 17 | 4 | 14 | 0 | 2 | 9 | 3 | 0 | 12 | 12 | 2 | 2 | 9 | 4 | 5 |
| $T_{11}$ | 0 | 0 | 3 | 2 | 5 | 1 | 1 | 3 | 2 | 0 | 5 | 23 | 35 | 10 | 0 | 0 | 5 | 1 | 2 | 3 |
| $T_{12}$ | 4 | 3 | 3 | 6 | 0 | 0 | 3 | 3 | 20 | 26 | 0 | 11 | 4 | 0 | 1 | 1 | 4 | 7 | 1 | 1 |
| $T_{13}$ | 4 | 5 | 0 | 0 | 5 | 7 | 3 | 3 | 26 | 26 | 2 | 1 | 1 | 2 | 1 | 0 | 0 | 0 | 6 | 6 |
| $T_{14}$ | 1 | 1 | 2 | 2 | 0 | 0 | 0 | 0 | 15 | 15 | 1 | 23 | 21 | 2 | 1 | 1 | 5 | 5 | 2 | 3 |
| $T_{15}$ | 0 | 0 | 0 | 0 | 1 | 1 | 0 | 0 | 0 | 0 | 4 | 1 | 2 | 2 | 5 | 5 | 34 | 24 | 11 | 8 |
| $T_{16}$ | 0 | 0 | 0 | 0 | 1 | 2 | 0 | 0 | 1 | 1 | 10 | 8 | 7 | 10 | 0 | 1 | 7 | 16 | 17 | 17 |
| $T_{17}$ | 0 | 0 | 0 | 0 | 0 | 0 | 0 | 0 | 4 | 1 | 0 | 1 | 3 | 2 | 7 | 7 | 20 | 10 | 35 | 10 |
| $T_{18}$ | 0 | 0 | 0 | 0 | 1 | 1 | 0 | 0 | 1 | 4 | 6 | 1 | 1 | 5 | 1 | 6 | 5 | 13 | 14 | 41 |
| $T_{19}$ | 1 | 1 | 2 | 2 | 0 | 0 | 0 | 0 | 0 | 1 | 7 | 4 | 2 | 4 | 22 | 36 | 6 | 10 | 0 | 1 |
| $T_{20}$ | 1 | 1 | 2 | 1 | 2 | 2 | 0 | 1 | 1 | 1 | 6 | 2 | 3 | 11 | 37 | 18 | 3 | 2 | 2 | 3 |

**Table S22:** Diabatic contributions ω (in %) as obtained from Boys diabatization.

## BF2 tetramer: intra-CT/inter-CT energy diagram

To simplify the above information, we partially diagonalize the singlet and triplet $Z_i$ diabatic states obtained for the BF2 tetramer within the intra-CT, inter-CT involving 1st neighbours and inter-CT beyond 1st neighbours.

| states | intra | intra | intra | intra | inter(1st) | inter(1st) | inter(1st) | inter(1st) | inter(1st) | inter(1st) | inter(1st) | inter(1st) | inter(1st) | inter(1st) | inter | inter | inter | inter | inter | inter |
|---|---|---|---|---|---|---|---|---|---|---|---|---|---|---|---|---|---|---|---|---|
| intra | 2400 | 0 | 0 | 0 | 3 | -5 | -3 | 2 | 1 | 5 | 1 | 15 | 23 | -16 | -84 | 35 | 5 | -50 | -31 | 20 |
| intra | 0 | 2388 | 0 | 0 | -3 | -5 | 4 | 2 | -2 | -6 | -2 | 12 | -23 | -19 | 22 | 83 | 16 | 42 | -45 | -6 |
| intra | 0 | 0 | 2364 | 0 | 36 | 0 | -31 | -1 | -17 | -78 | 8 | -3 | -27 | -13 | -40 | -18 | -2 | -3 | -6 | -40 |
| intra | 0 | 0 | 0 | 2335 | 5 | -44 | 3 | 8 | 109 | -18 | -2 | 73 | -8 | 143 | 15 | -34 | 12 | -4 | -33 | 7 |
| inter(1st) | 3 | -3 | 36 | 5 | 2499 | 0 | 0 | 0 | 0 | 0 | 0 | 0 | 0 | 0 | -8 | 1 | -8 | 14 | 18 | 75 |
| inter(1st) | -5 | -5 | 0 | -44 | 0 | 2397 | 0 | 0 | 0 | 0 | 0 | 0 | 0 | 0 | -23 | 47 | -6 | -5 | -39 | 6 |
| inter(1st) | -3 | 4 | -31 | 3 | 0 | 0 | 2323 | 0 | 0 | 0 | 0 | 0 | 0 | 0 | -15 | -9 | -4 | -6 | 14 | 67 |
| inter(1st) | 2 | 2 | -1 | 8 | 0 | 0 | 0 | 2310 | 0 | 0 | 0 | 0 | 0 | 0 | -47 | 95 | -4 | -3 | -21 | 5 |
| inter(1st) | 1 | -2 | -17 | 109 | 0 | 0 | 0 | 0 | 2182 | 0 | 0 | 0 | 0 | 0 | -4 | -7 | -22 | 0 | 8 | -1 |
| inter(1st) | 5 | -6 | -78 | -18 | 0 | 0 | 0 | 0 | 0 | 2170 | 0 | 0 | 0 | 0 | -29 | -13 | 0 | -19 | 6 | 38 |
| inter(1st) | 1 | -2 | 8 | -2 | 0 | 0 | 0 | 0 | 0 | 0 | 2121 | 0 | 0 | 0 | 134 | 61 | 1 | 15 | 0 | 18 |
| inter(1st) | 15 | 12 | -3 | 73 | 0 | 0 | 0 | 0 | 0 | 0 | 0 | 2119 | 0 | 0 | -16 | 36 | 87 | 3 | 61 | -11 |
| inter(1st) | 23 | -23 | -27 | -8 | 0 | 0 | 0 | 0 | 0 | 0 | 0 | 0 | 2055 | 0 | -9 | -5 | 8 | 106 | -22 | -54 |
| inter(1st) | -16 | -19 | -13 | 143 | 0 | 0 | 0 | 0 | 0 | 0 | 0 | 0 | 0 | 1948 | -33 | 66 | 49 | -7 | -42 | 5 |
| inter | -84 | 22 | -40 | 15 | -8 | -23 | -15 | -47 | -4 | -29 | 134 | -16 | -9 | -33 | 2329 | 0 | 0 | 0 | 0 | 0 |
| inter | 35 | 83 | -18 | -34 | 1 | 47 | -9 | 95 | -7 | -13 | 61 | 36 | -5 | 66 | 0 | 2317 | 0 | 0 | 0 | 0 |
| inter | 5 | 16 | -2 | 12 | -8 | -6 | -4 | -4 | -22 | 0 | 1 | 87 | 8 | 49 | 0 | 0 | 2312 | 0 | 0 | 0 |
| inter | -50 | 42 | -3 | -4 | 14 | -5 | -6 | -3 | 0 | -19 | 15 | 3 | 106 | -7 | 0 | 0 | 0 | 2295 | 0 | 0 |
| inter | -31 | -45 | -6 | -33 | 18 | -39 | 14 | -21 | 8 | 6 | 0 | 61 | -22 | -42 | 0 | 0 | 0 | 0 | 2278 | 0 |
| inter | 20 | -6 | -40 | 7 | 75 | 6 | 67 | 5 | -1 | 38 | 18 | -11 | -54 | 5 | 0 | 0 | 0 | 0 | 0 | 2248 |

**Table S23:** Diabatic energies (diagonal) and electronic couplings (off-diagonal) between the lowest intra-$^1$CT (blue), 1st neighbours inter-$^1$CT (red) and inter-$^1$CT beyond 1st neighbours (green). All values are expressed in meV.

| states | intra | intra | intra | intra | intra | intra | intra | intra | inter(1st) | inter(1st) | inter(1st) | inter(1st) | inter(1st) | inter(1st) | inter(1st) | inter(1st) | inter | inter | inter | inter |
|---|---|---|---|---|---|---|---|---|---|---|---|---|---|---|---|---|---|---|---|---|
| intra | 2052 | 0 | 0 | 0 | 0 | 0 | 0 | 0 | 37 | -25 | -11 | 12 | 25 | -43 | -37 | -24 | 5 | -10 | 5 | -1 |
| intra | 0 | 2050 | 0 | 0 | 0 | 0 | 0 | 0 | -30 | -32 | 12 | 12 | -35 | -32 | 39 | -25 | 11 | -3 | -2 | -4 |
| intra | 0 | 0 | 2022 | 0 | 0 | 0 | 0 | 0 | 45 | -10 | 16 | 2 | 32 | -2 | -64 | -15 | -3 | -2 | 9 | 5 |
| intra | 0 | 0 | 0 | 2005 | 0 | 0 | 0 | 0 | -13 | 93 | 3 | -29 | 5 | -34 | 0 | 125 | 4 | -3 | 4 | -3 |
| intra | 0 | 0 | 0 | 0 | 1788 | 0 | 0 | 0 | 3 | -18 | 3 | 61 | -8 | -56 | 5 | 43 | 8 | -7 | 4 | -7 |
| intra | 0 | 0 | 0 | 0 | 0 | 1752 | 0 | 0 | 45 | 1 | -65 | 22 | 56 | -12 | 56 | 9 | -5 | -7 | -18 | -7 |
| intra | 0 | 0 | 0 | 0 | 0 | 0 | 1751 | 0 | 0 | 49 | -8 | -112 | 16 | 37 | 5 | -65 | -5 | 2 | 4 | -13 |
| intra | 0 | 0 | 0 | 0 | 0 | 0 | 0 | 1723 | -48 | -7 | -110 | -9 | -165 | 10 | -105 | 0 | -12 | -15 | -23 | -11 |
| inter(1st) | 37 | -30 | 45 | -13 | 3 | 45 | 0 | -48 | 2344 | 0 | 0 | 0 | 0 | 0 | 0 | 0 | 3 | 2 | 2 | 0 |
| inter(1st) | -25 | -32 | -10 | 93 | -18 | 1 | 49 | -7 | 0 | 2341 | 0 | 0 | 0 | 0 | 0 | 0 | 0 | 0 | -2 | 4 |
| inter(1st) | -11 | 12 | 16 | 3 | 3 | -65 | -8 | -110 | 0 | 0 | 2177 | 0 | 0 | 0 | 0 | 0 | 4 | 1 | 52 | 24 |
| inter(1st) | 12 | 12 | 2 | -29 | 61 | 22 | -112 | -9 | 0 | 0 | 0 | 2111 | 0 | 0 | 0 | 0 | 61 | -58 | 8 | -13 |
| inter(1st) | 25 | -35 | 32 | 5 | -8 | 56 | 16 | -165 | 0 | 0 | 0 | 0 | 2079 | 0 | 0 | 0 | -57 | -59 | 36 | 17 |
| inter(1st) | -43 | -32 | -2 | -34 | -56 | -12 | 37 | 10 | 0 | 0 | 0 | 0 | 0 | 2069 | 0 | 0 | -10 | 21 | 2 | -3 |
| inter(1st) | -37 | 39 | -64 | 0 | 5 | 56 | 5 | -105 | 0 | 0 | 0 | 0 | 0 | 0 | 2032 | 0 | 18 | 22 | 87 | 39 |
| inter(1st) | -24 | -25 | -15 | 125 | 43 | 9 | -65 | 0 | 0 | 0 | 0 | 0 | 0 | 0 | 0 | 2019 | -11 | 8 | 45 | -97 |
| inter | 5 | 11 | -3 | 4 | 8 | -5 | -5 | -12 | 3 | 0 | 4 | 61 | -57 | -10 | 18 | -11 | 2326 | 0 | 0 | 0 |
| inter | -10 | -3 | -2 | -3 | -7 | -7 | 2 | -15 | 2 | 0 | 1 | -58 | -59 | 21 | 22 | 8 | 0 | 2319 | 0 | 0 |
| inter | 5 | -2 | 9 | 4 | 4 | -18 | 4 | -23 | 2 | -2 | 52 | 8 | 36 | 2 | 87 | 45 | 0 | 0 | 2252 | 0 |
| inter | -1 | -4 | 5 | -3 | -7 | -7 | -13 | -11 | 0 | 4 | 24 | -13 | 17 | -3 | 39 | -97 | 0 | 0 | 0 | 2247 |

**Table S24:** Diabatic energies (diagonal) and electronic couplings (off-diagonal) between the lowest intra-$^3$CT (blue), 1st neighbours inter-$^3$CT (red) and inter-$^3$CT beyond 1st neighbours (green). All values are expressed in meV.

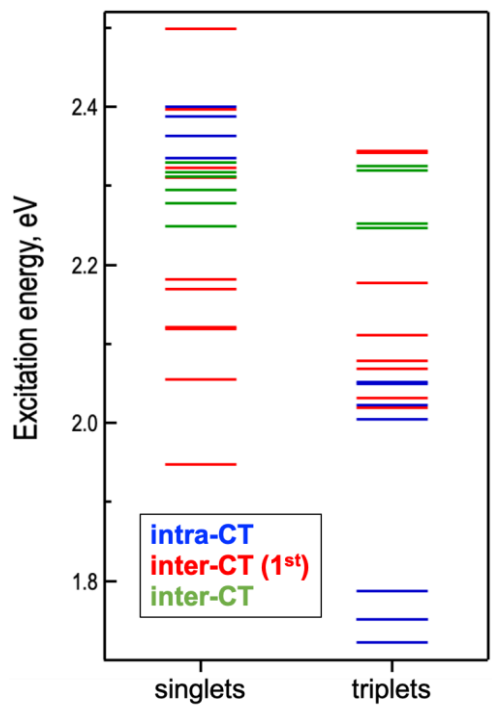

**Figure S29:** Singlet and triplet intra-CT and inter-CT diabatic state energy distribution for the BF2 tetramer. Blue: intra-CT; red: inter-CT (1st neighbours); green: inter-CT (beyond 1st neighbours).

**4CzIPN dimer structure**

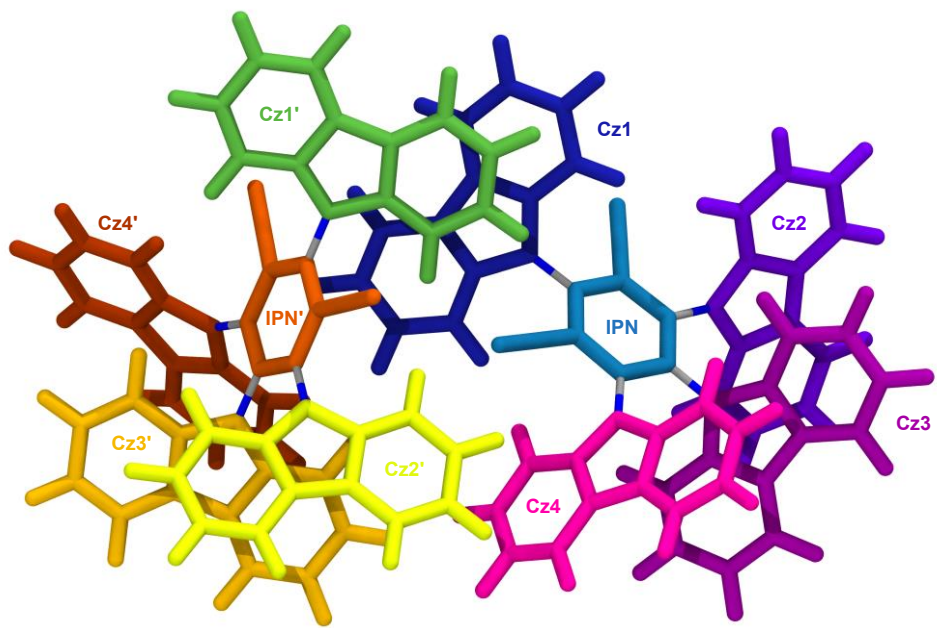

**Figure S30:** Optimised structure of the ground state 4CzIPN dimer and nomenclature of the fragments considered in the diabatization analysis.

**4CzIPN monomer: singlet states**

Diabatization of singlet excited states of the 4CzIPN monomer: $Z_i$ are D →A diabatic contributions, respectively. D = Cz; A = IPN.

| i | $S_i$ | $Z_i$ | $\omega$ | $Z_1$ | $Z_2$ | $Z_3$ | $Z_4$ |
|---|---|---|---|---|---|---|---|
| 1 | 2.88 (0.09) | 3.03 | $S_1$ | 28 | 27 | 20 | 25 |
| 2 | 3.07 (0.04) | 3.03 | $S_2$ | 51 | 49 | 0 | 0 |
| 3 | 3.09 (0.01) | 3.05 | $S_3$ | 7 | 9 | 79 | 4 |
| 4 | 3.13 (0.02) | 3.06 | $S_4$ | 14 | 14 | 0 | 71 |

**Table S25:** (left) Energies of adiabatic ($S_i$) and diabatic ($Z_i$) states (in eV, oscillator strength in parenthesis); (right) Diabatic contributions $\omega$ (in %) as obtained from Boys diabatization for the four lowest singlet states of the 4CzIPN monomer.

| State | IPN | Cz1 | Cz2 | Cz3 | Cz4 |
|-------|-----|-----|-----|-----|-----|
| $Z_1$ | **-0.836** | -0.002 | -0.029 | 0.058 | **0.808** |
| $Z_2$ | **-0.836** | -0.002 | **0.808** | 0.059 | -0.029 |
| $Z_3$ | **-0.845** | -0.002 | 0.028 | **0.792** | 0.027 |
| $Z_4$ | **-0.851** | **0.918** | -0.033 | -0.002 | -0.033 |

**Table S26:** Relative Mulliken fragment charges of diabatic states of the 4CzIPN monomer with respect to the ground state charge distribution.

The four lowest excited states $S_{1-4}$ correspond to combinations of the intra-CT diabatic states (from each Cz to the central IPN unit).

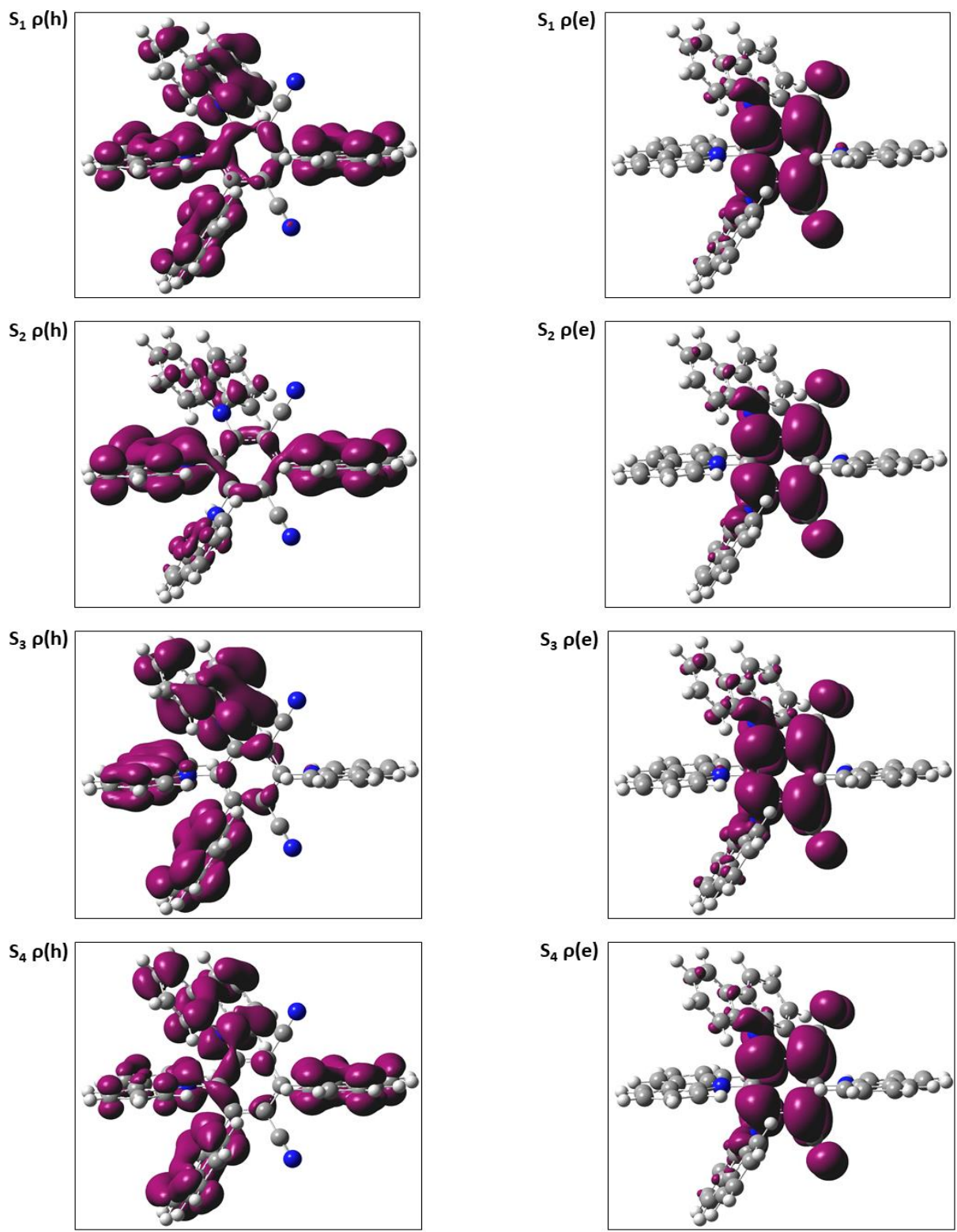

**Figure S31:** Hole (ρ(h)) and electron (ρ(e)) density plots of the four lowest excited states S$_{1-4}$ of the 4CzIPN monomer.

**4CzIPN dimer, $d_\perp = 3.5$ Å**

| i | $S_i$ | intra | inter | $T_i$ | intra | inter |
|---|---|---|---|---|---|---|
| 1 | 2.52 (0.04) | 78 | 22 | 2.36 | 98 | 2 |
| 2 | 2.59 (0.05) | 100 | 0 | 2.41 | 100 | 0 |
| 3 | 2.64 (0.01) | 16 | 84 | 2.56 | 94 | 6 |
| 4 | 2.72 (0.02) | 56 | 44 | 2.59 | 100 | 0 |
| 5 | 2.77 (0.02) | 100 | 0 | 2.62 | 50 | 50 |
| 6 | 2.80 (0.01) | 49 | 51 | 2.66 | 57 | 43 |
| 7 | 2.84 (0.06) | 99 | 1 | 2.69 | 91 | 9 |
| 8 | 2.85 (0.00) | 81 | 19 | 2.70 | 99 | 1 |
| 9 | 2.85 (0.07) | 81 | 19 | 2.73 | 94 | 6 |
| 10 | 2.88 (0.00) | 10 | 90 | 2.76 | 100 | 0 |
| 11 | 2.90 (0.01) | 61 | 39 | 2.79 | 100 | 0 |
| 12 | 2.91 (0.02) | 100 | 0 | 2.82 | 19 | 81 |
| 13 | 2.92 (0.18) | 98 | 2 | 2.86 | 82 | 18 |
| 14 | 2.97 (0.06) | 89 | 11 | 2.88 | 26 | 74 |
| 15 | 2.98 (0.01) | 93 | 7 | 2.89 | 90 | 10 |
| 16 | 2.98 (0.09) | 88 | 12 | 2.90 | 100 | 0 |

**Table S27:** Energies (in eV) of adiabatic singlet ($S_i$) (oscillator strength in parenthesis) and triplet ($T_i$) states. Intra and intermolecular contributions are indicated in blue and red, respectively.

| i | $S_i$ | $T_i$ |
|---|---|---|
| 1 | 4.329 | 3.742 |
| 2 | 3.636 | 3.497 |
| 3 | 7.612 | 3.138 |
| 4 | 5.469 | 2.966 |
| 5 | 3.759 | 5.422 |
| 6 | 5.688 | 5.704 |
| 7 | 3.290 | 2.730 |

**Table S28:** Electron-hole (e-h) radius (in Å) for the adiabatic singlet ($S_i$) and triplet ($T_i$) states.

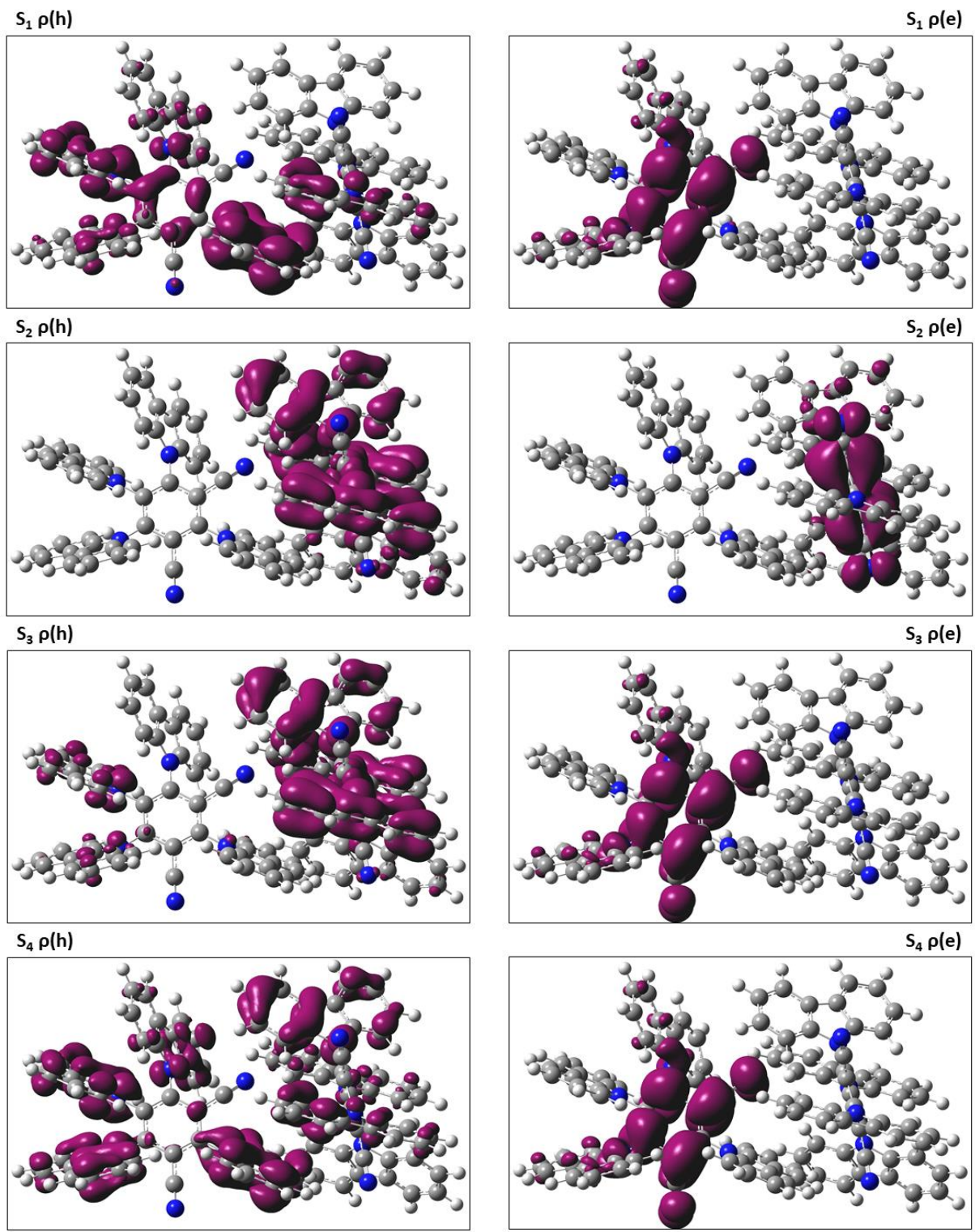

**Figure S32:** Hole (ρ(h)) and electron (ρ(e)) density plots from $S_1$ to $S_4$ of the 4CzIPN dimer at $d_\perp = 3.5$ Å.

**4CzIPN diabatization at $d_\perp$ = 3.5 Å: singlet states**

|                | IPN   | Cz1   | Cz2   | Cz3  | Cz4   | IPN'  | Cz1'  | Cz2'  | Cz3'  | Cz4'  |
|----------------|-------|-------|-------|------|-------|-------|-------|-------|-------|-------|
| $Z_1$ (2.72)   | **-0.88** | 0.01  | -0.06 | 0.00 | -0.05 | 0.07  | **0.83**  | 0.01  | 0.06  | 0.01  |
| $Z_2$ (2.74)   | 0.00  | 0.01  | 0.00  | 0.00 | 0.00  | **-0.81** | **0.90**  | -0.05 | 0.00  | -0.04 |
| $Z_3$ (2.75)   | **-0.83** | **0.92**  | -0.05 | 0.00 | -0.04 | 0.00  | 0.01  | 0.00  | 0.00  | 0.00  |
| $Z_4$ (2.77)   | **-0.83** | 0.00  | -0.02 | **0.70** | **0.16**  | 0.00  | 0.00  | 0.00  | 0.00  | 0.00  |
| $Z_5$ (2.77)   | 0.00  | 0.00  | 0.00  | 0.00 | 0.00  | **-0.81** | 0.00  | **0.77**  | 0.04  | 0.00  |
| $Z_6$ (2.78)   | **-0.87** | 0.00  | -0.06 | 0.00 | -0.04 | 0.06  | 0.01  | **0.79**  | 0.07  | 0.03  |
| $Z_7$ (2.79)   | **-0.81** | 0.00  | **0.67**  | 0.08 | 0.05  | 0.00  | 0.00  | 0.00  | 0.00  | 0.00  |
| $Z_8$ (2.79)   | **-0.84** | 0.00  | **0.25**  | **0.51** | 0.07  | 0.00  | 0.00  | 0.00  | 0.00  | 0.00  |
| $Z_9$ (2.80)   | 0.00  | 0.00  | 0.00  | 0.00 | 0.00  | **-0.86** | 0.00  | **0.53**  | **0.34**  | 0.00  |
| $Z_{10}$ (2.81) | **-0.87** | -0.01 | -0.06 | 0.00 | -0.05 | 0.01  | 0.00  | **0.71**  | **0.27**  | 0.00  |
| $Z_{11}$ (2.82) | 0.00  | 0.00  | 0.00  | 0.00 | 0.00  | **-0.82** | 0.00  | -0.03 | **0.65**  | **0.21**  |
| $Z_{12}$ (2.85) | **-0.86** | **0.23**  | -0.04 | 0.00 | -0.01 | 0.00  | **0.67**  | 0.00  | 0.01  | 0.00  |
| $Z_{13}$ (2.88) | **-0.81** | **0.52**  | -0.02 | 0.00 | **0.19**  | 0.00  | 0.06  | 0.01  | 0.04  | 0.01  |
| $Z_{14}$ (2.92) | **-0.76** | **0.47**  | 0.03  | **0.10** | -0.01 | 0.00  | **0.16**  | 0.00  | 0.01  | 0.01  |
| $Z_{15}$ (2.96) | 0.01  | **0.16**  | 0.00  | 0.01 | 0.00  | **-0.88** | **0.80**  | -0.05 | 0.00  | -0.05 |
| $Z_{16}$ (2.97) | -0.03 | 0.01  | 0.00  | 0.00 | 0.01  | **-0.73** | **0.36**  | **0.18**  | **0.18**  | 0.01  |

**Table S29:** Relative Mulliken fragment charges of diabatic states (energy in eV in parenthesis) of the 4CzIPN dimer with respect to the ground state charge distribution.

The lowest diabatic state $Z_1$ corresponds to inter-CT excitation from Cz1' to IPN. The $Z_2$/$Z_3$ quasi-degenerate states correspond to intra-CT from Cz1 (Cz1') to IPN (IPN'). The next state with inter-CT character $Z_6$ involves a transfer of the electron density from Cz2' to IPN. Diabatic states $Z_{12}$-$Z_{15}$ have mixed inter-/intra-CT character with $Z_{12}$ being mostly inter-CT while $Z_{13}$-$Z_{15}$ correspond primarily to local excitations.

| ω | $Z_1$ | $Z_2$ | $Z_3$ | $Z_4$ | $Z_5$ | $Z_6$ | $Z_7$ | $Z_8$ | $Z_9$ | $Z_{10}$ | $Z_{11}$ | $Z_{12}$ | $Z_{13}$ | $Z_{14}$ | $Z_{15}$ | $Z_{16}$ |
|---|---|---|---|---|---|---|---|---|---|---|---|---|---|---|---|---|
| $S_1$ | 10 | 0 | 31 | 17 | 0 | 1 | 9 | 5 | 0 | 1 | 0 | 9 | 14 | 1 | 0 | 0 |
| $S_2$ | 0 | 32 | 0 | 0 | 25 | 0 | 0 | 0 | 14 | 0 | 27 | 0 | 0 | 0 | 2 | 0 |
| $S_3$ | 37 | 0 | 0 | 5 | 0 | 30 | 4 | 5 | 0 | 17 | 0 | 0 | 1 | 0 | 0 | 0 |
| $S_4$ | 6 | 0 | 13 | 14 | 0 | 13 | 20 | 8 | 0 | 11 | 0 | 13 | 1 | 0 | 0 | 0 |
| $S_5$ | 0 | 49 | 0 | 0 | 25 | 0 | 0 | 1 | 20 | 0 | 0 | 0 | 0 | 0 | 5 | 0 |
| $S_6$ | 31 | 0 | 1 | 2 | 0 | 5 | 2 | 33 | 1 | 10 | 0 | 5 | 9 | 0 | 0 | 0 |
| $S_7$ | 0 | 0 | 0 | 31 | 1 | 0 | 57 | 7 | 2 | 0 | 0 | 0 | 1 | 0 | 0 | 0 |
| $S_8$ | 6 | 0 | 0 | 15 | 18 | 2 | 0 | 13 | 33 | 3 | 0 | 9 | 0 | 1 | 0 | 0 |
| $S_9$ | 4 | 0 | 0 | 9 | 15 | 0 | 4 | 25 | 27 | 7 | 0 | 7 | 0 | 0 | 0 | 0 |
| $S_{10}$ | 0 | 0 | 6 | 0 | 0 | 37 | 0 | 0 | 0 | 47 | 0 | 5 | 2 | 1 | 0 | 0 |
| $S_{11}$ | 6 | 0 | 47 | 2 | 0 | 11 | 0 | 0 | 0 | 3 | 0 | 20 | 7 | 5 | 0 | 0 |
| $S_{12}$ | 0 | 8 | 0 | 0 | 12 | 0 | 0 | 0 | 3 | 0 | 69 | 0 | 0 | 2 | 5 | 0 |
| $S_{13}$ | 0 | 0 | 1 | 0 | 2 | 0 | 0 | 0 | 0 | 0 | 1 | 2 | 5 | 85 | 1 | 2 |
| $S_{14}$ | 0 | 1 | 0 | 3 | 0 | 0 | 2 | 1 | 0 | 0 | 1 | 11 | 15 | 4 | 10 | 51 |
| $S_{15}$ | 0 | 7 | 0 | 1 | 0 | 0 | 1 | 0 | 0 | 0 | 2 | 7 | 15 | 0 | 66 | 0 |
| $S_{16}$ | 0 | 1 | 0 | 1 | 0 | 0 | 0 | 0 | 0 | 0 | 0 | 12 | 29 | 0 | 11 | 45 |

**Table S30:** Diabatic contributions ω (in %) as obtained from Boys diabatization.

**4CzIPN diabatization at $d_\perp$ = 3.5 Å: triplet states**

| | IPN | Cz1 | Cz2 | Cz3 | Cz4 | IPN' | Cz1' | Cz2' | Cz3' | Cz4' |
|---|---|---|---|---|---|---|---|---|---|---|
| $Z_1$ (2.56) | 0.00 | 0.00 | 0.00 | 0.00 | 0.00 | **-0.75** | 0.00 | **0.78** | 0.00 | -0.04 |
| $Z_2$ (2.60) | **-0.73** | 0.00 | **0.77** | 0.00 | -0.04 | 0.00 | 0.00 | 0.00 | 0.00 | 0.00 |
| $Z_3$ (2.62) | **-0.79** | **0.87** | -0.05 | 0.00 | -0.04 | 0.00 | 0.01 | 0.00 | 0.00 | 0.00 |
| $Z_4$ (2.65) | **-0.65** | **0.46** | 0.03 | **0.12** | 0.02 | 0.00 | 0.03 | 0.00 | 0.00 | 0.00 |
| $Z_5$ (2.65) | 0.00 | 0.02 | 0.00 | 0.00 | 0.00 | **-0.79** | **0.85** | -0.05 | 0.00 | -0.04 |
| $Z_6$ (2.65) | **-0.72** | 0.00 | -0.05 | 0.01 | **0.75** | 0.00 | 0.00 | 0.00 | 0.00 | 0.00 |
| $Z_7$ (2.66) | **-0.78** | 0.00 | -0.04 | **0.76** | 0.06 | 0.00 | 0.00 | 0.00 | 0.00 | 0.00 |
| $Z_8$ (2.70) | 0.00 | 0.00 | 0.00 | 0.00 | 0.00 | **-0.64** | **0.39** | 0.08 | **0.15** | 0.02 |
| $Z_9$ (2.72) | **-0.88** | 0.00 | -0.06 | 0.00 | -0.05 | 0.07 | **0.82** | 0.02 | 0.06 | 0.02 |
| $Z_{10}$ (2.73) | 0.00 | 0.00 | 0.00 | 0.00 | 0.00 | **-0.77** | 0.00 | -0.04 | **0.78** | 0.04 |
| $Z_{11}$ (2.75) | 0.00 | 0.00 | 0.00 | 0.00 | 0.00 | **-0.71** | 0.00 | -0.04 | 0.01 | **0.74** |
| $Z_{12}$ (2.77) | **-0.84** | 0.00 | **0.42** | **0.43** | 0.00 | 0.00 | 0.00 | 0.00 | 0.00 | 0.00 |
| $Z_{13}$ (2.79) | 0.00 | 0.00 | 0.00 | 0.00 | 0.00 | **-0.85** | 0.00 | **0.66** | **0.23** | -0.03 |
| $Z_{14}$ (2.79) | **-0.87** | 0.00 | -0.06 | 0.00 | -0.05 | 0.06 | 0.01 | **0.81** | 0.07 | 0.04 |
| $Z_{15}$ (2.80) | **-0.87** | -0.01 | -0.06 | 0.00 | -0.05 | 0.01 | 0.00 | **0.71** | **0.27** | 0.00 |
| $Z_{16}$ (2.82) | **-0.87** | **0.68** | -0.06 | 0.00 | -0.05 | 0.00 | **0.30** | 0.00 | 0.00 | 0.00 |

**Table S31:** Relative Mulliken fragment charges of diabatic states (energy in eV in parenthesis) of the 4CzIPN dimer with respect to the ground state charge distribution.

| ω | $Z_1$ | $Z_2$ | $Z_3$ | $Z_4$ | $Z_5$ | $Z_6$ | $Z_7$ | $Z_8$ | $Z_9$ | $Z_{10}$ | $Z_{11}$ | $Z_{12}$ | $Z_{13}$ | $Z_{14}$ | $Z_{15}$ | $Z_{16}$ |
|---|---|---|---|---|---|---|---|---|---|---|---|---|---|---|---|---|
| $T_1$ | 0 | 15 | 35 | 0 | 0 | 7 | 27 | 0 | 1 | 0 | 0 | 8 | 0 | 0 | 0 | 5 |
| $T_2$ | 38 | 0 | 0 | 0 | 25 | 0 | 0 | 1 | 0 | 25 | 3 | 0 | 8 | 0 | 0 | 0 |
| $T_3$ | 0 | 36 | 2 | 23 | 0 | 28 | 1 | 0 | 2 | 0 | 0 | 4 | 0 | 3 | 1 | 1 |
| $T_4$ | 37 | 0 | 0 | 0 | 18 | 0 | 0 | 14 | 0 | 7 | 22 | 0 | 1 | 0 | 0 | 0 |
| $T_5$ | 0 | 3 | 8 | 20 | 0 | 5 | 7 | 0 | 34 | 0 | 0 | 3 | 0 | 10 | 6 | 3 |
| $T_6$ | 0 | 24 | 14 | 8 | 0 | 4 | 1 | 0 | 8 | 0 | 0 | 1 | 0 | 20 | 15 | 6 |
| $T_7$ | 1 | 1 | 7 | 45 | 1 | 27 | 3 | 4 | 6 | 0 | 0 | 1 | 0 | 2 | 2 | 2 |
| $T_8$ | 10 | 0 | 0 | 3 | 16 | 1 | 1 | 64 | 1 | 0 | 2 | 0 | 2 | 0 | 0 | 0 |
| $T_9$ | 0 | 16 | 5 | 0 | 0 | 17 | 33 | 0 | 2 | 0 | 0 | 19 | 0 | 2 | 2 | 3 |
| $T_{10}$ | 13 | 0 | 0 | 0 | 9 | 0 | 0 | 1 | 0 | 19 | 8 | 0 | 50 | 0 | 0 | 0 |
| $T_{11}$ | 0 | 0 | 0 | 0 | 27 | 0 | 0 | 17 | 0 | 4 | 51 | 0 | 1 | 0 | 0 | 0 |
| $T_{12}$ | 0 | 0 | 3 | 0 | 0 | 5 | 1 | 0 | 43 | 0 | 0 | 0 | 0 | 10 | 29 | 8 |
| $T_{13}$ | 0 | 2 | 1 | 0 | 0 | 4 | 17 | 0 | 0 | 0 | 0 | 49 | 0 | 6 | 12 | 9 |
| $T_{14}$ | 0 | 1 | 0 | 0 | 0 | 2 | 7 | 0 | 2 | 0 | 0 | 13 | 0 | 41 | 31 | 3 |
| $T_{15}$ | 0 | 1 | 24 | 0 | 0 | 0 | 3 | 0 | 3 | 0 | 0 | 1 | 0 | 6 | 2 | 60 |
| $T_{16}$ | 0 | 0 | 0 | 0 | 4 | 0 | 0 | 0 | 0 | 44 | 13 | 0 | 39 | 0 | 0 | 0 |

**Table S32:** Diabatic contributions ω (in %) as obtained from Boys diabatization.

## Calculation of the hyperfine coupling in the BF2 dimer

In this section we present the approach developed to compute the hyperfine coupling (HFC) in the excited states of the BF2 dimer. By considering an inter-CT state involving a monomer A and a monomer B, we can depict two scenarios:

1) The electron density is fully transferred from monomer A(B) to monomer B(A) giving rise to a pure inter-CT state, where the monomer A(B) acquires a cationic nature and monomer B(A) an anionic nature.

2) The electron density is partially transferred from monomer A(B) to monomer B(A), giving rise to a hybrid intra-inter CT state, where the hole (electron) density, which we associate with the cation (anion), is delocalized over the entire dimer.

In the first case, the hole and the electron are well localized on different molecules and the local magnetic field they experience can be easily evaluated by computing the HFC on the cation and anion species, respectively. In the second case, the hole and the electron are delocalized on the two monomers and consequently the local magnetic field competing to each monomer arises from a mixed cationic and anionic character. It follows that the evaluation of the effective HFC characterizing each monomer requires the knowledge of the amount of the hole (cationic character) and the electron (anionic character) localized on monomer A and monomer B. This can be done by resorting to the *attachment/detachment* formalism, weighting the HFC of the electron (anion) and hole (cation) on the amount of attachment and detachment density localized on each monomer.

Our approach encompasses three steps:

1) The calculation of the HFC of the cation and anion of the isolated monomers by accounting for both the isotropic and dipolar part, according to the following equations[36]:

$$B_{iso} = \left[ \sum_{\alpha}^{nuclei} [A_{\alpha}^2 I_{\alpha}(I_{\alpha} + 1)] N_{\alpha} \right]^{\frac{1}{2}} \quad (S7a)$$

$$B_{dip} = \frac{1}{3} \left[ \sum_{\alpha}^{nuclei} [\boldsymbol{T}_{\alpha}:\boldsymbol{T}_{\alpha} I_{\alpha}(I_{\alpha} + 1)] N_{\alpha} \right]^{\frac{1}{2}} \quad (S7b)$$

where $A_{\alpha}$ is the isotropic constant describing the coupling between the nuclear spin of atom α and the electron spin, $I_{\alpha}$ is the nuclear spin, $\boldsymbol{T}_{\alpha}$ is the second-rank tensor describing the dipolar coupling between the nuclear spin of atom α and the electron spin, and $N_{\alpha}$ is the natural abundance of the α-th nucleus. The term $\boldsymbol{T}_{\alpha}:\boldsymbol{T}_{\alpha}$ is the inner product of the tensor with itself, computed as:

$$\boldsymbol{T}_{\alpha}:\boldsymbol{T}_{\alpha} = (T_{\alpha}^x)^2 + (T_{\alpha}^y)^2 + (T_{\alpha}^z)^2 \quad (S8)$$

being $T_{\alpha}^i$ the elements of the tensor in the diagonal form.

The total effective magnetic field arising from the isotropic and dipolar HFC is then:

$$B_{HF} = \sqrt{B_{iso}^2 + B_{dip}^2} \quad (S9)$$

This is done for both the monomers in the anionic and cationic form. The isotropic constant, $A_\alpha$, and the dipolar tensor, $\boldsymbol{T}_\alpha$, were computed at DFT level, with the LC-ωhPBE functional using the previously mentioned parameters for the BF2 monomer and the EPR-III basis set, providing the following results:

|  | $B_{iso}$ [mT] | $B_{dip}$ [mT] | $B_{HF}$ [mT] |
|---|---|---|---|
| *Anion* | 0.899 (104) | 0.292 (34) | 0.945 (110) |
| *Cation* | 0.701 (81) | 0.354 (41) | 0.786 (91) |

**Table S33:** Isotropic, dipolar and total HFC computed for the anion and cation of the BF2 monomer. The values of $B_{iso}$ in parentheses are expressed in neV.

In our case, because the BF2 dimer is made up of two identical molecules, the HFC calculation was carried out only for one monomer. As natural abundance, we considered 0.0107 for $^{13}C$, 1 for $^{1}H$, 1 for $^{19}F$, 0.00038 for $^{17}O$, 0.99 for $^{14}N$ and 0.80 for $^{11}B$.

2) The calculation of the attachment and detachment densities associated with the excited state of interest are carried out using the NANCY_EX package[37] from SRSH TDDFT calculations with the LC-ωhPBE functional using the previously mentioned parameters for the BF2 dimer and the 6-311+G(d,p) basis set.

3) Weighting of the HFC of the cation and anion on the percentage of detachment and attachment density localized on each monomer:

$$B_{HF,anion}^{m1\%} = B_{HF,anion}^{m1} \times Att^{m1} \quad (S10a)$$

$$B_{HF,anion}^{m2\%} = B_{HF,anion}^{m2} \times Att^{m2} \quad (S10b)$$

$$B_{HF,anion}^{dimer} = \sqrt{\left(B_{HF,anion}^{m1\%}\right)^2 + \left(B_{HF,anion}^{m2\%}\right)^2} \qquad (S10c)$$

$$B_{HF,cation}^{m1\%} = B_{HF,cation}^{m1} \times Det^{m1} \qquad (S11a)$$

$$B_{HF,cation}^{m2\%} = B_{HF,cation}^{m2} \times Det^{m2} \qquad (S11b)$$

$$B_{HF,cation}^{dimer} = \sqrt{\left(B_{HF,cation}^{m1\%}\right)^2 + \left(B_{HF,cation}^{m2\%}\right)^2} \qquad (S11c)$$

where $Att^{mi} = \sum_{j}^{atoms(mi)} \rho_j^{att}$ and $Det^{mi} = \sum_{j}^{atoms(mi)} \rho_j^{det}$ are the amount of the attachment and detachment density, respectively, localized on the *mi*-th monomer.

| $d_\perp = 3.5\text{Å}$ | $T_1$ | $T_2$ | $T_3$ | $T_4$ | $T_5$ | $T_6$ | $T_7$ | $T_8$ | $T_9$ | $T_{10}$ |
|---|---|---|---|---|---|---|---|---|---|---|
| $Att^{m1}$ | 0.499 | 0.501 | 0.500 | 0.499 | 0.498 | 0.502 | 0.496 | 0.504 | 0.499 | 0.501 |
| $Att^{m2}$ | 0.501 | 0.499 | 0.500 | 0.500 | 0.502 | 0.498 | 0.504 | 0.496 | 0.501 | 0.499 |
| $Det^{m1}$ | 0.499 | 0.501 | 0.501 | 0.499 | 0.502 | 0.498 | 0.505 | 0.495 | 0.499 | 0.501 |
| $Det^{m2}$ | 0.501 | 0.498 | 0.499 | 0.500 | 0.498 | 0.502 | 0.495 | 0.505 | 0.501 | 0.499 |
| $B_{HF,anion}^{m1\%}[mT]$ | 0.472 | 0.473 | 0.473 | 0.472 | 0.471 | 0.474 | 0.469 | 0.476 | 0.472 | 0.473 |
| $B_{HF,anion}^{m2\%}[mT]$ | 0.473 | 0.472 | 0.473 | 0.473 | 0.474 | 0.471 | 0.476 | 0.469 | 0.473 | 0.472 |
| $B_{HF,cation}^{m1\%}[mT]$ | 0.392 | 0.394 | 0.394 | 0.392 | 0.395 | 0.391 | 0.397 | 0.389 | 0.392 | 0.394 |
| $B_{HF,cation}^{m2\%}[mT]$ | 0.394 | 0.391 | 0.392 | 0.393 | 0.391 | 0.395 | 0.389 | 0.397 | 0.394 | 0.392 |
| $B_{HF,anion}^{dimer}[mT]$ | 0.668 | 0.668 | 0.668 | 0.668 | 0.668 | 0.668 | 0.668 | 0.668 | 0.668 | 0.668 |
| $B_{HF,cation}^{dimer}[mT]$ | 0.556 | 0.555 | 0.556 | 0.555 | 0.556 | 0.556 | 0.556 | 0.556 | 0.556 | 0.556 |
| $B_{HF,anion}^{dimer}[neV]$ | 77 | 77 | 77 | 77 | 77 | 77 | 77 | 77 | 77 | 77 |
| $B_{HF,cation}^{dimer}[neV]$ | 64 | 64 | 64 | 64 | 64 | 64 | 64 | 64 | 64 | 64 |

**Table S34:** Attachment and detachment density localized on each BF2 monomer, the weighted hole (cation) and electron (anion) HFC according to equations S10a-S10b and S11a-S11b and the total hole (cation) and electron (anion) HFC in the dimer, obtained using equations S10c and S11c.

| $d_\perp = 6.1\text{Å}$ | $T_1$ | $T_2$ | $T_3$ | $T_4$ | $T_5$ | $T_6$ | $T_7$ | $T_8$ | $T_9$ | $T_{10}$ |
|---|---|---|---|---|---|---|---|---|---|---|
| $Att^{m1}$ | 0.999 | 0.001 | 1.000 | 0.000 | 1.000 | 0.000 | 0.999 | 0.001 | 0.994 | 0.006 |
| $Att^{m2}$ | 0.001 | 0.999 | 0.000 | 1.000 | 0.000 | 1.000 | 0.001 | 0.999 | 0.006 | 0.994 |
| $Det^{m1}$ | 0.999 | 0.001 | 1.000 | 0.000 | 0.000 | 1.000 | 0.001 | 0.999 | 0.994 | 0.006 |
| $Det^{m2}$ | 0.001 | 0.999 | 0.000 | 1.000 | 1.000 | 0.000 | 0.999 | 0.001 | 0.006 | 0.994 |
| $B_{HF,anion}^{m1\%}[mT]$ | 0.944 | 0.001 | 0.945 | 0.000 | 0.945 | 0.000 | 0.944 | 0.001 | 0.939 | 0.006 |
| $B_{HF,anion}^{m2\%}[mT]$ | 0.001 | 0.944 | 0.000 | 0.945 | 0.000 | 0.945 | 0.001 | 0.944 | 0.006 | 0.939 |
| $B_{HF,cation}^{m1\%}[mT]$ | 0.785 | 0.001 | 0.786 | 0.000 | 0.000 | 0.786 | 0.001 | 0.785 | 0.781 | 0.005 |
| $B_{HF,cation}^{m2\%}[mT]$ | 0.001 | 0.785 | 0.000 | 0.786 | 0.786 | 0.000 | 0.785 | 0.001 | 0.005 | 0.781 |
| $B_{HF,anion}^{dimer}[mT]$ | 0.944 | 0.944 | 0.945 | 0.945 | 0.945 | 0.945 | 0.944 | 0.944 | 0.939 | 0.939 |
| $B_{HF,cation}^{dimer}[mT]$ | 0.785 | 0.785 | 0.786 | 0.786 | 0.786 | 0.786 | 0.785 | 0.785 | 0.781 | 0.781 |
| $B_{HF,anion}^{dimer}[neV]$ | 109 | 109 | 110 | 110 | 110 | 110 | 109 | 109 | 109 | 109 |
| $B_{HF,cation}^{dimer}[neV]$ | 91 | 91 | 91 | 91 | 91 | 91 | 91 | 91 | 91 | 91 |

**Table S35:** Attachment and detachment density localized on each BF2 monomer, the weighted hole (cation) and electron (anion) HFC according to the equations S10a-S10b and S11a-S11b and the total hole (cation) and electron (anion) HFC in the dimer, obtained using equations S10c and S11c.

# Intersystem crossing rate between the singlet and triplet spin correlated pairs

Manolopoulos and co. have proposed an incoherent model to compute the rate of intersystem crossing between the singlet and triplet spin correlated radical pair ($^1$SCRP and $^3$SCRP)[38,39], which for these purposes can be considered analogous to the inter-CT states presented here. In their model, the $^1$SCRP is formed after photoexcitation to a locally excited singlet state ($^1$LE). Then, interconversion between $^1$SCRP and $^3$SCRP occurs through a hyperfine interaction-driven intersystem crossing (HFI-ISC) mechanism.

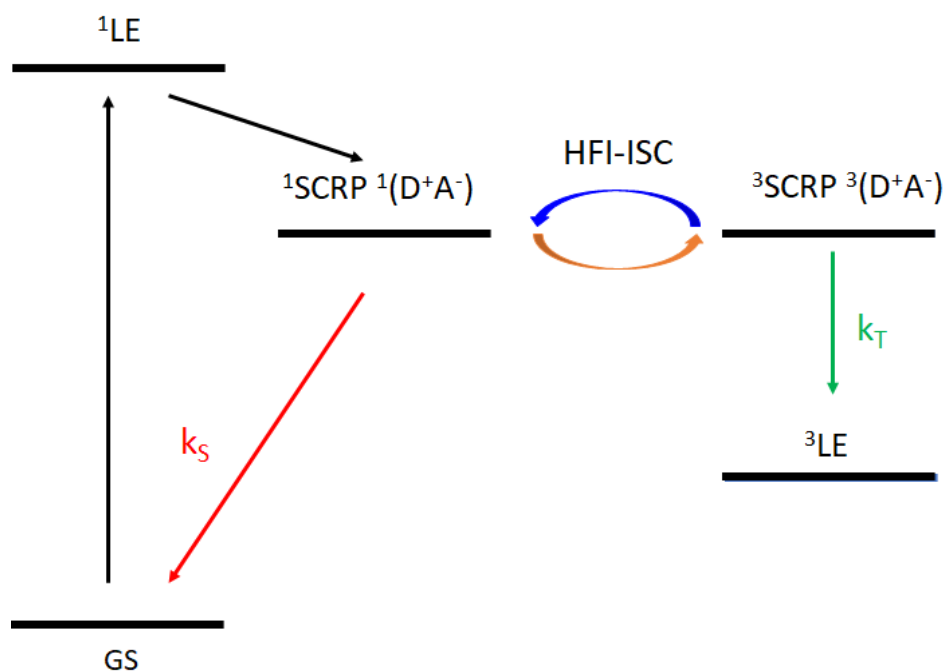

**Figure S33:** Electronic structure of the different states involved in the model proposed by Manolopoulos and co. in references 36 and 37.

The HFI-ISC is limited by both the decay rate to the ground state (GS) from the $^1$SCRP ($k_S$), the decay rate from $^3$SCRP to the $^3$LE ($k_T$) and the exchange energy J between the $^1$SCRP and $^3$SCRP.

In the incoherent, Markovian, limit, the HFI-ISC rate takes the following expression:

$$k_{HFI-ISC} = \frac{\omega_{D^+}^2 + \omega_{A^-}^2}{6} \frac{\gamma}{\gamma^2 + J^2} \qquad (S12)$$

where $\omega_{D^+}$ and $\omega_{A^-}$ are the pulsation associated with the hyperfine interaction corresponding to the electron donor and acceptor in their cationic and anionic states, respectively; $\gamma$ is the dephasing rate evaluated as the average decay rate $\left(\gamma = \frac{k_S + k_T}{2}\right)$.

The $k_{HFI-ISC}$ rate expression is obtained through a second-order perturbation treatment and is thus valid within a range of parameters such as $\frac{\omega_{D^+}^2 + \omega_{A^-}^2}{6} \ll \gamma^2 + J^2$. $\eta = \frac{\frac{\omega_{D^+}^2 + \omega_{A^-}^2}{6}}{\gamma^2 + J^2}$ must therefore be much lower than 1 for the perturbation treatment to remain valid.

The pulsations associated with the cation and anion hyperfine interaction as well as the exchange energy are obtained from first principles DFT and TDDFT calculations. In Figure 3e, HFI-ISC rate is obtained as a function of the exchange energy J and the dephasing rate $\gamma$ considering the distance dependence of the exchange energy J and $\gamma$ as a parameter. Here, the J distance dependence in Fig. S34 is fitted with a decaying exponential:

$$J = \alpha \, exp(-\beta d_\perp)$$

where $d_\perp$ is the intermolecular distance between the $^1$SCRP and $^3$SCRP and α and β are fitting parameters which amount to 4.76·10$^{-3}$ eV and 0.364 Å$^{-1}$, respectively.

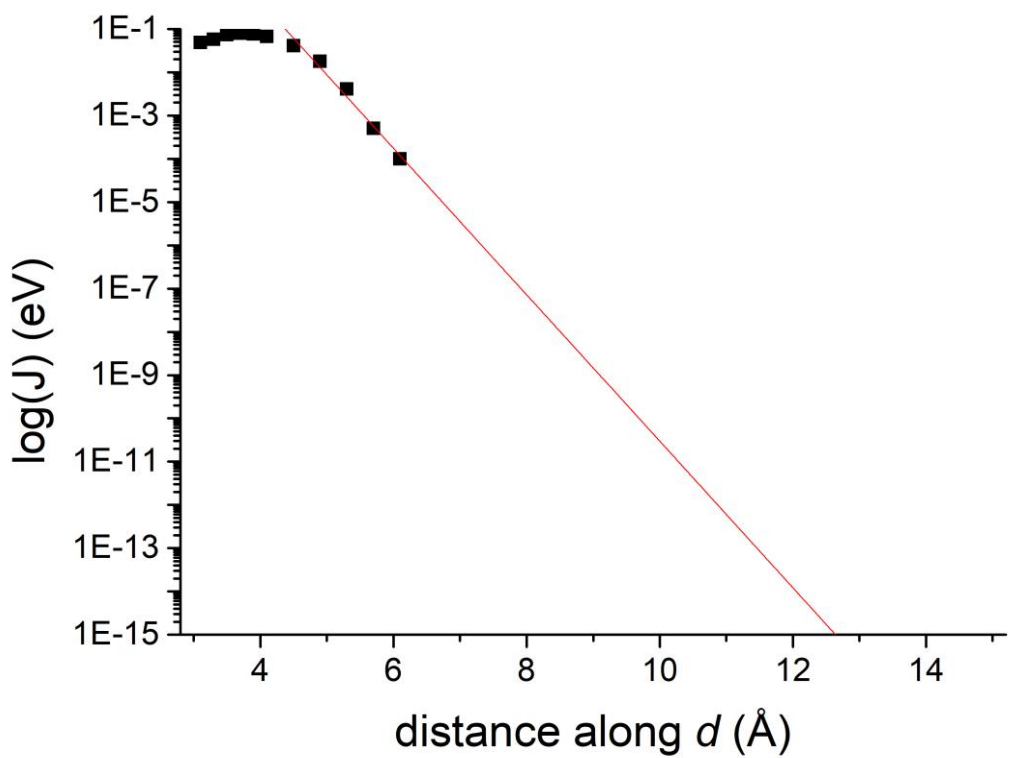

**Figure S34:** Distance dependence of the exchange energy J between the $^1$SCRP and $^3$SCRP.

In Fig. S35, we plot η as a function of J and γ. The dephasing rate γ is estimated to be 40 ns – 1 μs from the regrowth of the SE signal in the TA of BF2 at 10 wt% in CBP (Fig. 2d). Here, we note that the regrowth of the SE represents the reformation of intra-$^1$CT states, in which HFI-ISC cannot occur, from the HFI-ISC active inter-CT states. Thus, we take the onset of SE regrowth in Fig. 2d (40 ns) to provide a lower bound and the point at which SE regrowth ends (1 μs) as an upper bound for the dephasing time. In the range of dephasing rates inferred from experiments and exchange energies predicted by theory, η is much larger than 1 (see highlighted blue ellipsoid area in Fig. S35) suggesting that the singlet-triplet interconversion mechanism is likely to be coherent.

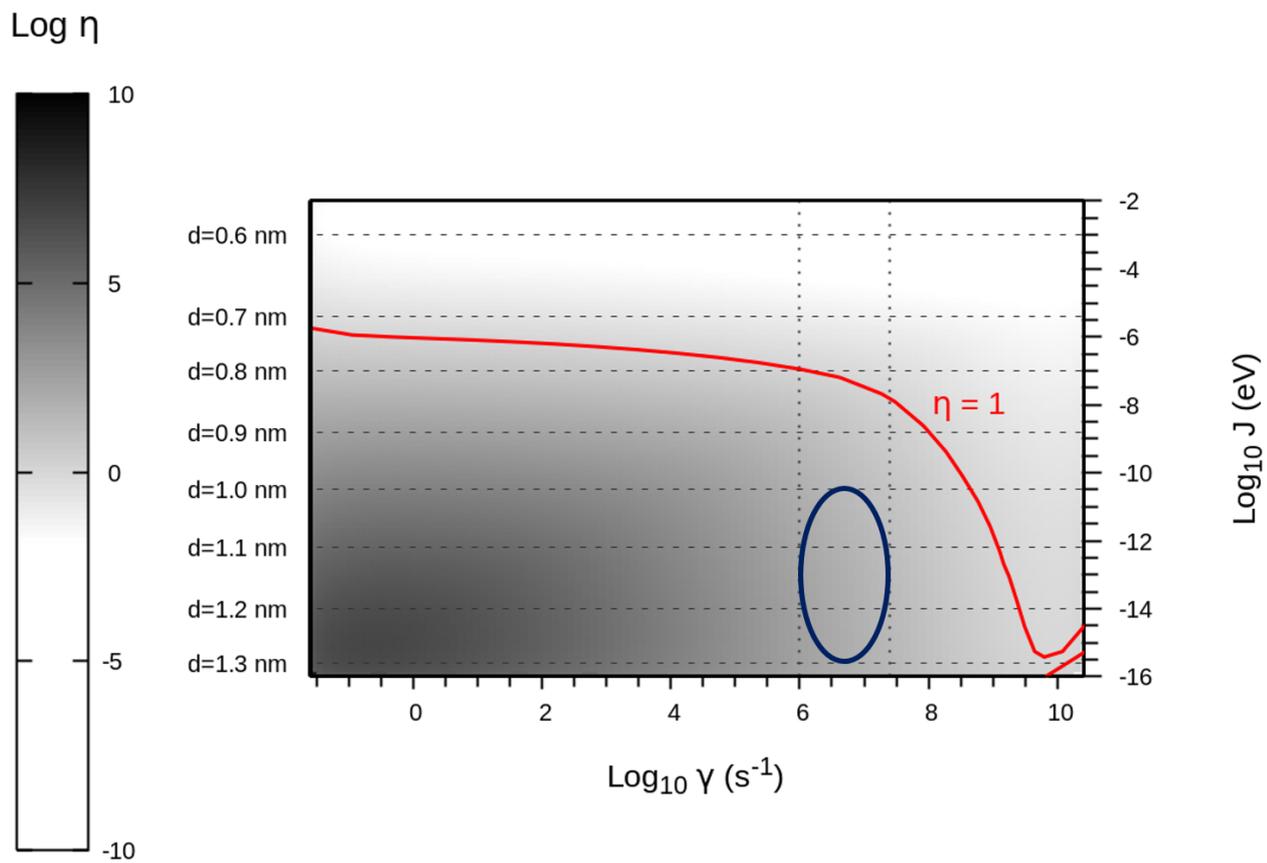

**Figure S35:** η as a function of the J and γ. The circled area represents the region of interest for the BF2 system explored in our work.